\documentclass[reprint, superscriptaddress, amsmath,amssymb, aps,]{revtex4-2}
\usepackage[utf8]{inputenc}
\usepackage{amsmath,esint}
\usepackage{hyperref}
\hypersetup{colorlinks=true,linkcolor=blue,citecolor=blue,urlcolor=blue}
\usepackage{amsfonts}
\usepackage{amssymb}
\usepackage{bm}
\usepackage{textcomp}
\usepackage{empheq}
\usepackage{siunitx}
\usepackage[titletoc,title]{appendix}
\usepackage{graphicx}
\usepackage{dcolumn}
\usepackage{bm}
\usepackage{subcaption}
\usepackage{xcolor}
\usepackage{physics}

\renewcommand{\Re}{\operatorname{Re}}
\renewcommand{\Im}{\operatorname{Im}}

\newcommand{\citeasnoun}[1]{Ref.~\cite{#1}}

\newcommand{\Figref}[1]{Fig.~\ref{fig:#1}}
\newcommand{\figref}[1]{Fig.~\ref{fig:#1}}
\renewcommand{\eqref}[1]{Eq.~(\ref{eq:#1})}
\newcommand{\Eqref}[1]{Equation~(\ref{eq:#1})}

\newcommand{\hl}[1]{{\color{black}#1}}

\newcommand{\magn}[1]{\left| #1 \right|}
\renewcommand{\vec}[1]{\boldsymbol{#1}}

\newcommand\D{\mathrm{d}}

\newcommand{\mat}[1]{\begin{matrix}#1\end{matrix}} 
\newcommand{\bb}[1]{\left(#1\right)}

\renewcommand{\Re}{\operatorname{Re}}
\renewcommand{\Im}{\operatorname{Im}}

\newcommand*{\Udag}{U^{\dagger}}
\newcommand*{\epsmax}{\varepsilon_{\rm max}}
\newcommand*{\epsmin}{\varepsilon_{\rm min}}
\renewcommand*{\d}{\,{\rm d}}

\begin{document}
\preprint{APS/123-QED}

\title{Conservation-law-based global bounds to quantum optimal control}

\author{Hanwen Zhang}
\affiliation{Department of Applied Physics, Yale University, New Haven, Connecticut 06511, USA}
\affiliation{Energy Sciences Institute, Yale University, New Haven, Connecticut 06511, USA}
\author{Zeyu Kuang}
\affiliation{Department of Applied Physics, Yale University, New Haven, Connecticut 06511, USA}
\affiliation{Energy Sciences Institute, Yale University, New Haven, Connecticut 06511, USA}
\author{Shruti Puri}
\affiliation{Department of Applied Physics, Yale University, New Haven, Connecticut 06511, USA}
\affiliation{Yale Quantum Institute, Yale University, New Haven, Connecticut 06511, USA}
\author{Owen D. Miller}
\affiliation{Department of Applied Physics, Yale University, New Haven, Connecticut 06511, USA}
\affiliation{Energy Sciences Institute, Yale University, New Haven, Connecticut 06511, USA}

\date{\today}
\begin{abstract}
Active control of quantum systems enables diverse applications ranging from quantum computation to manipulation of molecular processes. Maximum speeds and related bounds have been identified from uncertainty principles and related inequalities, but such bounds utilize only coarse system information, and loosen significantly in the presence of constraints and complex interaction dynamics. We show that an integral-equation-based formulation of conservation laws in quantum dynamics leads to a systematic framework for identifying fundamental limits to any quantum control scenario. We demonstrate the utility of our bounds in three scenarios---three-level driving, decoherence suppression, and maximum-fidelity gate implementations---and show that in each case our bounds are tight or nearly so. Global bounds complement local-optimization-based designs, illuminating performance levels that may be possible as well as those that cannot be surpassed.
\end{abstract}
\pacs{Valid PACS appear here}

\maketitle

In this Letter, we develop a framework for computing fundamental limits to what is possible via control of quantum systems. \hl{We show that quantum control problems can be transformed to quadratically constrained quadratic programs (QCQPs), with generalized probability conservation laws as the constraints,} adapting a mathematical approach recently developed for light--matter interactions~\cite{Kuang2020n2,Molesky2020c}. \hl{The QCQP formulation enables} global bounds via \hl{relaxations to semidefinite programs}~\cite{Laurent2005,Luo2010}. We demonstrate the power and utility of our method on three \hl{prototype} systems: (1) three-level system driving, where our bounds incorporate sophisticated information about interference between levels, and can account for constraints on undesirable transitions (as needed in transmons \cite{motzoi2009simple}, for example), (2) upper bounds to the \hl{suppression of} decoherence, and (3) the maximum fidelity of a control-based implementation of a single-qubit Hadamard gate. In each case we supplement our bounds with many local-optimization-based solutions, showing that they come quite close to (and in some cases achieve) our bounds, suggesting that our bounds are tight or nearly so. Our framework applies to open and closed systems, can be extended to related domains in NMR~\cite{Khaneja2005,Nielsen2010,Tovsner2018} and quantum complexity~\cite{Nielsen2006,Jefferson2017,Chapman2018}, and should reveal the limits of what is possible with quantum control.

Quantum control~\cite{Peirce1988,Werschnik2007,dAlessandro2007,Brif2010,Glaser2015} refers to the design and synthesis of efficient control sequences that drive a quantum system to maximize a desired objective, such as maximizing overlap with a target state or minimizing error in the implementation of a gate operation. Recent experiments have demonstrated the power of optimal control for wide-ranging applications~\cite{Bason2012,VanFrank2016,Heeres2017,Goetz2019,Vepsalainen2019,Lam2021}. Because the wave function $\ket{\psi(t)}$ that represents a quantum state is nonlinear in the control parameter $\varepsilon(t)$, it is generically difficult to identify globally optimal controls. One strategy is to use local numerical optimization over the control parameters (e.g. GRAPE~\cite{Khaneja2005,DeFouquieres2011,Dolde2014,Anderson2015}, the Krotov method~\cite{Krotov1993,Somloi1993,Zhu1998,Ohtsuki2004,Schirmer2011,Goerz2019}, and CRAB~\cite{Caneva2011,Doria2011}), optimizing over many initial conditions in the hopes of identifying high-performance local optima. Yet, except in the simplest of systems, one is left uncertain about the best performance possible. Alternatively, there are a variety of global bounds~\cite{Mandelstam1945,Fleming1973,Vaidman1992,Margolus1998,Khaneja2001,Khaneja2002,Giovannetti2003,Giovannetti2003n2,Carlini2006,Carlini2007,Deffner2013,DelCampo2013,Poggi2013,Hegerfeldt2013,Hegerfeldt2014,Russell2014,Brody2015,Russell2015,Frey2016,Deffner2017,Arenz2017,Lee2018}; most famously, the Mandelstam--Tamm (MT) bound. The MT bound is a prototype of ``quantum speed limits,'' which more generally have varying levels of complexity but are essentially time-energy uncertainty relations~\cite{Mandelstam1945,Fleming1973,Vaidman1992,Margolus1998,Giovannetti2003,Deffner2013,DelCampo2013,Frey2016,Deffner2017,Arenz2017,Lee2018,burgarth2020quantum}. The energy measure is typically a matrix norm of the Hamiltonian, but more complex details of the system interactions are not captured. Another class of bounds is obtained by analytically solving Pontryagin's maximum principle~\cite{Knowles1981}, which is only possible in simple cases such as two-level systems~\cite{Khaneja2001,Khaneja2002,Carlini2006,Carlini2007,Poggi2013,Hegerfeldt2013,Hegerfeldt2014}. Consequently, meaningful, accurate bounds cannot be computed for most quantum control systems of interest.

\emph{Formulation}---We consider a Hamiltonian of the form $H_0(t) + H_c'(t) = H_0(t) + \varepsilon(t) H_c(t)$, where $H_0$ is the non-controllable part of the Hamiltonian, $H_c'$ is the controllable part, and $\varepsilon$ is the control parameter to be optimized. We assume the control parameter is bounded between 0 and $\epsmax$ (any other minimum value can be shifted to 0 by replacing $H_0$ with $H_0 + \epsmin H_c$). Our method generalizes to any number of control parameters (cf. SM), but for simplicity we assume one throughout this paper. Any smooth, continuous, bounded control can be approximated with arbitrary accuracy by a ``bang--bang'' binary control that only takes the values 0 and $\epsmax$ (cf. SM), so we use bang--bang controls in our formulation. Instead of the differential Schr\"odinger equation for the time-evolution operator $U(t,t_0)$ (for an initial time $t_0$), we instead start with an integral form (equivalent to the Dyson equation~\cite{Englert2006,Sakurai1994} in the interaction picture):
\begin{align}
    U(t,t_0) = U_0(t,t_0) - \frac{i}{\hbar}\int_{t_0}^T G_0^{+}(t,t') H_c'(t') U(t',t_0) \hl{\d{t}'},
    \label{eq:inteq}
\end{align}
where $U_0$ and $G_0^{+}$ are the time-evolution operator and retarded Green's function in the absence of controls (i.e., for $H_0(t)$), and $T$ is the final time. To derive conservation laws, we start by taking the product of \eqref{inteq} with $\Udag(t,t_0) H_c'(t) D_i(t)$ from the left and integrating from an initial time $t_0$ to $T$:
\begin{align}
                    &\int_{t_0}^T \Udag(t,t_0) H_c'(t) D_i(t) U(t,t_0) \d{t}\nonumber \\
                    &+ \frac{i}{\hbar}\int_{t_0}^T \int_{t_0}^T \Udag(t,t_0) H_c'(t) D_i(t) G_0^{+}(t,t') H_c'(t') U(t',t_0) \d{t} \d{t'} \nonumber \\
                    &= \int_{t_0}^T \Udag(t,t_0) H_c'(t) D_i(t) U_0(t,t_0) \d{t}.
                    \label{eq:preconstr}
\end{align}
The variable $D_i(t)$ can be any time-dependent operator and is an optimization hyperparameter below in \eqref{opt_prob}; intuitively, allowing different possible choices of $D_i$ enables the isolation of particular times and elements in Hilbert space for which \eqref{preconstr} should be satisfied. \hl{The variable $H_c$ is effectively a renormalization that simplifies the probabilistic interpretation below; equivalently, it can be omitted.} The constraint of \eqref{preconstr} depends on both the time-evolution degrees of freedom $U(t)$ and the control variable degrees of freedom $\varepsilon(t)$. However, if we define a new variable $\Phi(t) = \varepsilon(t) H_c(t) U(t,t_0)$, this variable (the analog of a polarization field in electrodynamics~\cite{Jackson1999,Kuang2020n2}) can subsume both. \hl{Crucially, we can replace any instance of $\varepsilon(t)$ with $\varepsilon_{\rm max}$. This can be thought of as a two-step simplification: one could restrict the domains of the integrals to only times in which the control is on, in which case such a replacement is trivial. Next, $\varepsilon(t)$ only appears in a term of the form $\Phi^\dagger \varepsilon^{-1} \Phi$, which is zero even when $\varepsilon(t) = 0$, due to the quadratic dependence on $\Phi$. Hence we can extend the domain of the integrals back to the entire time interval from $t_0$ to $T$. Such ``domain-obliviousness''~\cite{Kuang2020n2} arises from our inclusion of $\varepsilon(t)$ and $U^\dagger(t,t_0)$ in the product term. Finally, we have the constraints:}
\begin{align}
                    &\int_{t_0}^T \!\int_{t_0}^T \!\Phi^{\dagger}(t) D_i(t) \!\left[ \frac{H_c^{-1}(t)}{\epsmax} \delta(t-t') \!+\! \frac{i}{\hbar} G_0^{+}(t,t') \right] \!\Phi(t')\hl{\D t \,\D t'} \nonumber \\
                    &= \int_{t_0}^T \Phi^{\dagger}(t) D_i(t) U_0(t,t_0) \d{t},
                    \label{eq:Dconstr}
\end{align}
where $H_c^{-1}$ is taken to be the pseudo inverse if $H_c$ is not invertible. For any $D_i(t)$, \eqref{Dconstr} is a quadratic equation in the variable $\Phi(t)$; the set of all possible $D_i(t)$ imply an infinite number of quadratic constraints.

\Eqref{Dconstr}  can be interpreted as a generalization of probability conservation. At any time $t_1$, conservation of probability implies unitarity of the time-evolution operator $U(t_1,t_0)$, such that $U^\dagger U = \mathcal{I}$, where $\mathcal{I}$ is the identity operator. From the integral equation for $U$, \eqref{inteq}, the difference $U^\dagger U - \mathcal{I}$ can be written
\begin{align}
    &\Udag(t_1,t_0) U(t_1,t_0) - \mathcal{I} \nonumber\\
    &= \frac{1}{\hbar^2}\int_{t_0}^{t_1} \int_{t_0}^{t_1} \Phi^{\dagger}(t'',t_0) U_0(t'',t') \Phi(t',t_0) \d{t'} \d{t''}\nonumber\\
    &+ \frac{2}{\hbar} \Im \int_{t_0}^{t_1} U_0(t_0,t') \Phi(t',t_0) \d{t'}.
    \label{eq:UUdiff}
\end{align}
If we take the imaginary part of \eqref{Dconstr}, and choose $D_i(t)$ to be the identity operator from $t_0$ to $t_1$ (and zero otherwise), the resulting constraint is precisely the one that requires the right-hand side of \eqref{UUdiff} to be zero (cf. SM). In other words, a subset of the constraints of \eqref{Dconstr} are those which enforce unitary evolution at all times. (In an open system described by a density matrix, unitarity is not preserved and the corresponding constraints instead represent conservation of probability flow, cf. SM.) 
\begin{figure*}[htb]
    \centering
    \includegraphics[width=1.0\textwidth]{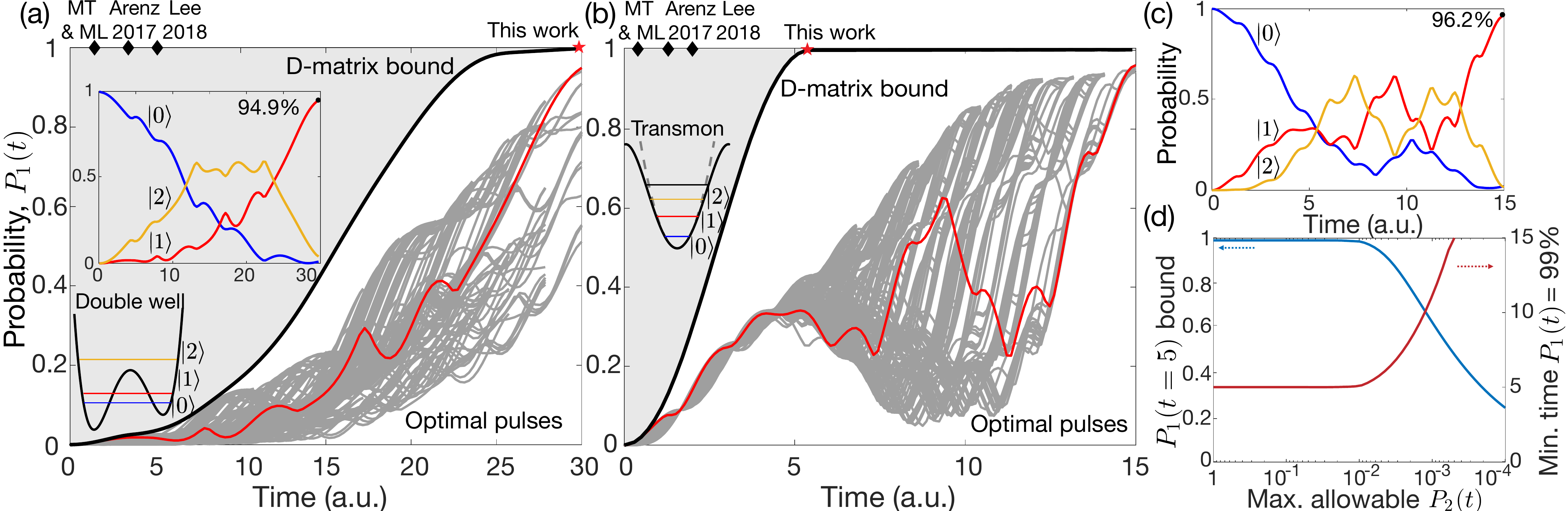}
    \caption{(a) Bounds on the maximum probability in state $\ket{1}$ as a function of time (solid black) for an asymmetric double-well potential, with shading above to indicate impossible values. Grey lines represent pulse evolutions optimized by gradient ascent, with the red line the very best evolution for final time 30. Inset: evolution of probabilities in states $\ket{0}, \ket{1}, \ket{2}$ for the optimal control, showing the complex dynamics captured by the bound. Black diamonds: evaluations of bounds of Mandelstam--Tamm, Margolus--Levitin, and Refs.~\cite{Arenz2017,Lee2018} for this problem. (b,c) Analogous to (a) but for a three-level model of a transmon qubit. (d) Incorporation of an additional constraint requiring small maximum allowable excitation probabilities of state 2. The bound on the maximum state-$\ket{1}$ probability (at time 5) decreases accordingly. The time to achieve $99\%$ state-$\ket{1}$ probability increases substantially with smaller allowed leakage rates.}
    \label{fig:figure1}
\end{figure*}

\hl{Although our derivation implies only that the conservation laws of \eqref{Dconstr} are necessary conditions for describing quantum evolution, one can show that they are sufficient as well: any $\Phi(t)$ that satisfies all possible versions of \eqref{Dconstr} implies a corresponding time-evolution operator $U(t,t_0)$ that satisfies \eqref{inteq} (cf. SM, \citeasnoun{Angeris2021}). Hence, we can replace the differential or integral dynamical equations with the conservation-law constraints of \eqref{Dconstr}. The optimal-control problem, for any objective $f$ that is a linear or quadratic function of the time-evolution operator $U$, and therefore a linear or quadratic function of $\Phi=\varepsilon H_c U$, is then the QCQP:}
\begin{equation}
    \begin{aligned}
        & \underset{\Phi}{\text{max.}} & & f(\Phi) \\
        & \text{s.t.} & & \textrm{\Eqref{Dconstr} satisfied for all } D_i(t).
    \end{aligned}
    \label{eq:opt_prob}
\end{equation}
We assume the problem has been discretized in any standard basis~\cite{Thijssen2012}. \hl{If we denote $\vec{\Phi}$ to be a single column vector containing the full discretization of $\Phi(t)$, \eqref{opt_prob} is a maximization of an objective of the form $\vec{\Phi}^\dagger A \vec{\Phi} + \Re(\vec{\alpha}^\dagger \vec{\Phi})$, where $A$ is Hermitian, subject to constraints of the form $\vec{\Phi}^\dagger B_i \vec{\Phi} +\Re(\vec{\beta}^\dagger_i \vec{\Phi}) = 0$ for all $i$. QCQPs are generically NP-hard to solve, but bounds on their solutions can be computed efficiently after semidefinite relaxation (SDR). SDRs transform quadratic forms $\Phi^\dagger A \Phi$ to equivalent linear forms $\Tr{AX}$, where $X$ is a rank-one positive semidefinite matrix, then drop the rank-one constraint. Finally, we are left with an objective of the form $\Tr{AX}+\Re(\vec{\alpha}^\dagger \vec{\Phi})$ subject to the constraints that $\Tr{B_i X}+\Re(\vec{\beta}_i^\dagger \vec{\Phi}) = 0$ for all $i$ and $X \succeq 0$, which is a semidefinite program whose global optimum can be computed via interior-point methods~\cite{Vandenberghe1996,Luo2010}.} As the bounds are computed over all possible matrices $D_i$, we label them ``D-matrix bounds.''  This framework applies broadly across quantum control; next, we demonstrate bounds for three prototypical systems.

\emph{Applications}---First, we compute bounds on driving three-level quantum systems. We consider two three-level systems described by Hamiltonians $H = \hbar \sum_{\hl{i}=1,2} \omega_j \ketbra{i}{i} - \varepsilon(t)\sum_{i,j =0,1,2} \mu_{ij} \ketbra{i}{j}$: one modeling an asymmetric double-well potential, with exact parameters from Sec.~2.8 of \citeasnoun{Werschnik2007} and given in the SM, and a second modeling a weakly nonlinear harmonic oscillator with nearest-level couplings, as is typically used to model a transmon qubit \cite{motzoi2009simple,krantz2019quantum}. (We consider both systems as they have different features: the first, couplings between all levels, and the second, small anharmonicity with hard-to-avoid leakage.) In each case we assume the system starts in the ground state, $\ket{0}$, and that we want to drive it to the first excited state, $\ket{1}$, as rapidly as possible. We denote the probability of occupying state $i$ at time $t$ by $P_i(t) = \magn{\braket{i}{\psi(t)}}^2$. There are two classes of bounds that we can compute: for a given amount of time $T$, the maximum probability in $\ket{1}$, $P_1(t)$; or, iteratively, the minimum amount of time to achieve near-unity probability in $\ket{1}$.

The black curve of \figref{figure1}(a) is the computed bound on $P_1(t)$ for the asymmetric-double-well model, for a bounded control field with $|\varepsilon(t)|\leq 0.15$. The shaded region of the figure is impossible to reach: our bounds indicate that any such evolution would necessarily violate at least one of the conservation laws. The grey lines are the results of local computational optimizations; we implemented a gradient-ascent optimization (similar to GRAPE) as described in the SM, for many different final times and initial pulse sequences. \hl{The gap between the local optimizations and the bounds arises from two sources---looseness in the bounds (from the SDR) or insufficient local exploration of the optimal pulses---though it is hard to pinpoint which source is more responsible.}  Also included in the figure are data points corresponding to evaluations of other bounds as applied to this problem: Mandelstam--Tamm (MT), Margolus--Levitin (ML), and Refs.~\cite{Arenz2017,Lee2018}. It takes some effort to map the various bounds to this problem, with varying degrees of looseness, which we discuss in detail in the SM. In particular, however, one can see that each of these bounds predicts minimal times an order of magnitude smaller than our approach. The inset provides a likely explanation: the optimal trajectory (highlighted in red) first populates the second excited state, then transitions to the first excited state through appropriate driving. Such complex dynamics cannot be captured by any previous bound approaches, but can be captured by our approach.

Parts (b--d) of \figref{figure1} show results for the transmon-qubit model, with $\omega_1 = 0.19$, $\omega_2 = 0.37$, $\mu_{10} = \mu_{01} = -1$, $\mu_{21} = \mu_{12} = -\sqrt{2}$ (all other $\mu_{ij} = 0$), and $|\varepsilon(t)| \leq 0.3$. \Figref{figure1}(b,c) are the transmon analogs of \figref{figure1}(a). The key novelty that is possible in this case is the addition of a constraint on the excitation probability of the second excited state, $\ket{2}$. Such ``leakage'' can be highly detrimental to the practical control of such systems, as they can open up additional decoherence channels~\cite{Wood2018}.  In our approach, we can simply add to \eqref{opt_prob} a (quadratic) constraint on the maximum allowed probability in $\ket{2}$. In \figref{figure1}(d), we show the bound for maximum $P_1(t)$ subject to varying constraints on the maximum allowed $P_2(t)$, at time $t=5$, which shows the dramatic reduction that is required if state-$\ket{2}$ transitions are to be avoided. Conversely, also in \figref{figure1}(d), the minimum time for near-unity first-state probability increases dramatically with more stringent constraints (red). Such constraints could not be incorporated into previous bound approaches.

\begin{figure}[thb]
    \centering
    \includegraphics[width=0.5\textwidth]{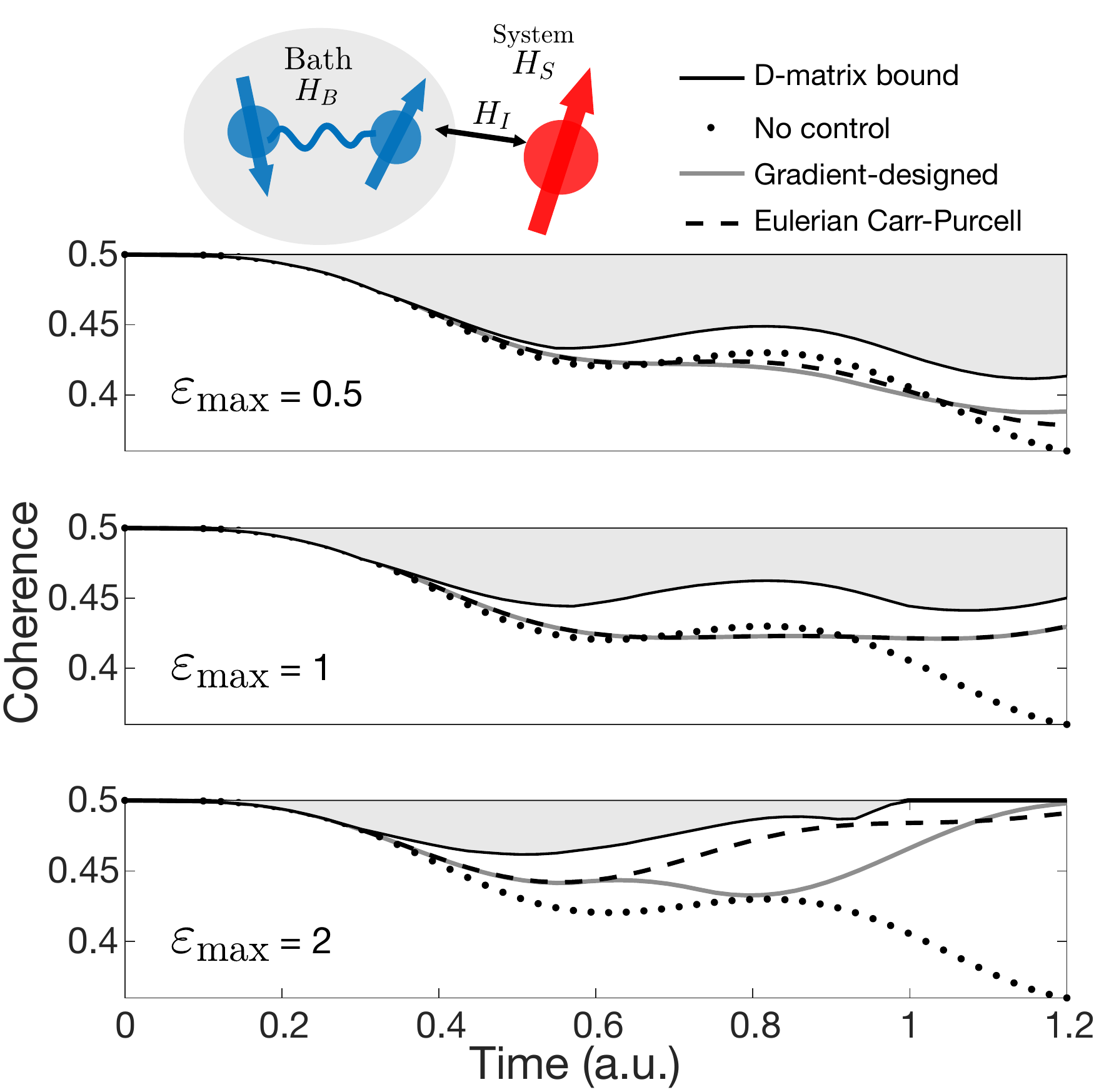}
    \caption{For a spin system interacting with a spin bath, the D-matrix approach enables bounds on maximum possible coherence as a function of time. The black solid line bounds the magnitude of the off-diagonal element of the system density matrix, $\magn{\rho^{S}_{12}}$, for varying maximum control amplitudes $\varepsilon_{\rm max}$. The time evolutions of $\magn{\rho^{S}_{12}}$ for pulses designed by gradient-ascent (solid grey) and finite Carr-Purcell (black dash lines) methods can closely approach the bounds.}
    \label{fig:figure2}
\end{figure}
A second example we consider is the extent to which one can prevent decoherence and dissipation due to interactions with the environment. The design of pulses to achieve such a goal has been studied extensively through semi-heuristic ``dynamic decoupling'' design schemes~\cite{viola1999dynamical,uhrig2009concatenated,west2010high,souza2011robust}, which may not be (and in many cases are not) globally optimal. A typical model of environmental effects is a spin system interacting with a spin bath. We consider a spin-bath system \cite{rossini2008bang} with Hamiltonian  $H_0 = H_S + H_E + H_{\rm int}$, where $H_S$ is the system Hamiltonian (two levels split by energy $\hbar\omega_0$), $H_E$ is the Hamiltonian of the environmental bath ($H_E = -J \sum_{j=1}^N \left( \sigma_j^x \sigma_{j+1}^x + \lambda \sigma_j^z \right)$), and $H_{\rm int}$ is the interaction between the system and the bath, $H_{\rm int} = -\nu \ket{\downarrow} \bra{\downarrow} \otimes \sum_j \sigma_j^z$, with $\omega_0 = \pi, J = 1, \lambda = 0.5$, and $\nu = 2$ here. The control Hamiltonian here is $H_c = \varepsilon(t) \sigma_x$ on the system only. Rather than use an approximation to the environmental coupling \cite{kosloff2019quantum}, we model the full dynamics of the wave function $\ket{\psi(t)}$. As a result, we only use a bath of size $N = 2$. Despite the bath being unrealistically small, it provide a qualitatively accurate description of the decoherence process~\cite{khodjasteh2005fault} and serves as a proof of principle. The system initial state is $\frac{1}{\sqrt{2}} \ket{\uparrow} + \frac{1}{\sqrt{2}} \ket{\downarrow}$, while the spin bath is in its ground state. The system density matrix $\rho^S$ is found by tracing out the bath part of the full density matrix, $\rho(t) = \ketbra{\psi(t)}{\psi(t)}$. The objective is to maximize $|\rho^S_{12}|$, the magnitude of the off diagonal elements of $\rho^S$, which represents the coherence of the system state. Instead of working with the absolute value (or its square, which is quartic in $\ket{\psi}$), we equivalently maximize $f = \Re\left(\rho^S_{12} e^{i\phi}\right)$ for a given $\phi$, and then iterate over possible values of $\phi$ between 0 and $2\pi$. \Figref{figure2} shows the bounds on maximal coherence as a function of time for three different bounded controls: $\varepsilon_{\rm max} = 0.5, 1$ and $2$. Also included are actual evolutions for three cases: without control, with a pulse designed by gradient ascent, and pulses designed by a bounded-control version of dynamical decoupling termed  ``Eulerian Carr-Purcell''~\cite{viola2003robust}. It is possible with strong controls to increase coherence at short times (as is particularly visible in \figref{figure2}(c)), but that would not be possible over longer time scales. We see that the bounds appear nearly tight, and provide information about what levels of coherence are possible as a function of time.

\begin{figure}[tbh]
    \centering
    \includegraphics[width=0.5\textwidth]{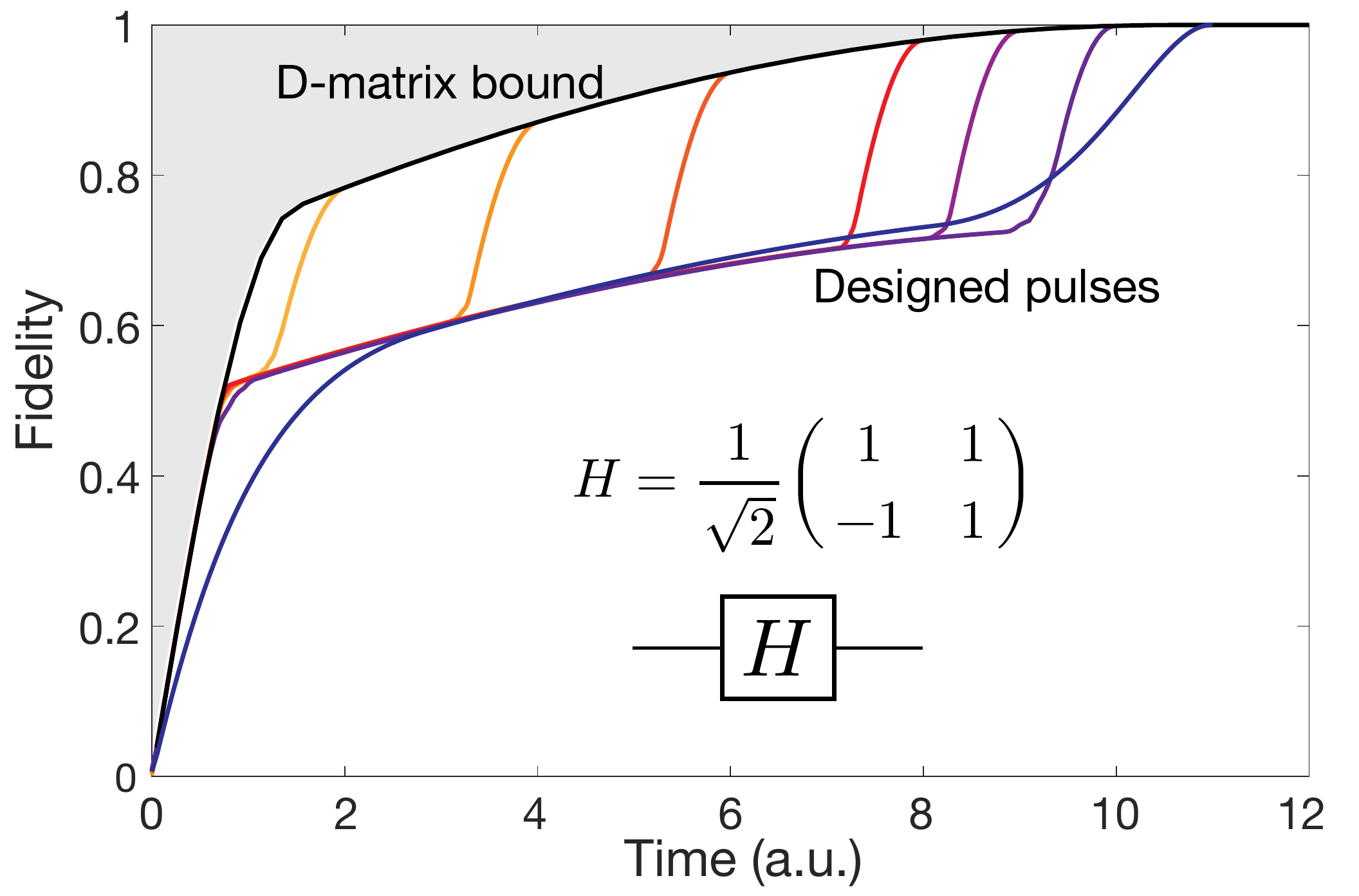}
    \caption{The black solid line bounds the maximum fidelity of a single-qubit Hadamard gate implemented in a two-level system with $H = \hbar \omega_0 \sigma_z  - \mu \varepsilon(t)\sigma_x$, and maximum control amplitude $\varepsilon_{\rm max} = 1$. Pulses optimized for different final times (colored lines) can achieve the upper bounds at all times.}
    \label{fig:figure3}
\end{figure}

For the third application, we consider the implementation of a single-qubit Hadamard gate. For a two-level system with Hamiltonian $H = \hbar \omega_0 \sigma_z  - \mu \varepsilon(t)\sigma_x$ ($\omega_0 =  0.0784$, $\mu = 1$) \cite{Werschnik2007}, the target time-evolution operator is given by $\frac{1}{\sqrt{2}} \bb{\mat{1 & 1\\-1 & 1}}$. The objective is to compute the maximal fidelity of a quantum gate at time $T$; for computational purposes, it is easier to work with the square of fidelity, $f^2 = \frac{1}{4} \left| \Tr \left\{ U_{\rm tar}^\dagger U(T) \right\} \right|^2 $. Identifying when the bound approaches 1 then indicates the minimum possible time to perform a gate operation. We consider a bounded control with $\varepsilon_{\rm max} = 1$. A crucial difference in the gate problem is that multiple inputs map to multiple outputs; the off-diagonal elements of the $D$ matrices in \eqref{Dconstr} inherently enforce the corresponding orthogonal-evolution requirement. \Figref{figure3} shows the fidelity bound as a function of time (solid black), along with time evolutions for locally optimized pulse sequences in the colored lines (optimized for different end times). The bound is tight, or very nearly so, across all times.

\emph{Conclusions}---Quadratic constraints representing generalized probability-conservation laws offer a framework for quantum control bounds. We have shown that this method can be significantly tighter than previous bounds and more widely applicable. There are further extensions that may be possible as well: in nanophotonic design problems, a hierarchy of bounds with varying analytical and semi-analytical complexity have been discovered as subsets of the $D$-matrix constraints~\cite{Miller2016, Yang2017, Shim2019, Molesky2019a, Ivanenko2019a, ShimChung2020, Shim2020, Molesky2020a, Gustafsson2020, Molesky2020b, Kuang2020,Kuang2020n2,Molesky2020c,Angeris2021}; the same may be possible in quantum control. In particular, environment-induced decoherence and dissipation are similar to material-absorption losses in electromagnetism, and may be amenable to general analytical bounds~\cite{Miller2016,Ivanenko2019a}. From an algorithmic perspective, \hl{there are significant computational speed-ups that should make the bound computations competitive with local optimizations, as a function of the number of degrees of freedom of the system, $N$ (the product of time steps and Hilbert-space dimensionality). Global optimization is presumably NP-hard; local optimizations require $O(N)$ time for each iteration and a number of iterations that may be large but independent of $N$. To find good local optima, however, requires restarting the search a number of times proportional to the number of local optima, which should scale at least as $O(N)$, for a total scaling of at least $O(N^2)$ (which is likely optimistic). For the bound computations, the simple implementation used for this work, using all possible constraints and interior-point-methods oblivious to the structure of the problem, scales as $O(N^{4.5})$~\cite{ma2008some}. Clever selection of the constraint matrices~\cite{Kuang2020n2} can reduce the scaling to $O(N^{3.5})$, while exploitation of the integral-operator's structure (e.g. via fast-multipole-type methods~\cite{greengard1987fast,coifman1993fast}) should further improve the scaling to $O(N^{2.5})$, making it highly competitive with local design methods.} More broadly, our approach and extensions thereof can be applied to problems across the quantum-control landscape, ranging from speed limits and gate fidelity to areas like NMR~\cite{Khaneja2005,Nielsen2010,Tovsner2018} and quantum complexity~\cite{Nielsen2006,Jefferson2017,Chapman2018}. 

\bibliography{qc_bib}

\begin{thebibliography}{87}%
\makeatletter
\providecommand \@ifxundefined [1]{%
 \@ifx{#1\undefined}
}%
\providecommand \@ifnum [1]{%
 \ifnum #1\expandafter \@firstoftwo
 \else \expandafter \@secondoftwo
 \fi
}%
\providecommand \@ifx [1]{%
 \ifx #1\expandafter \@firstoftwo
 \else \expandafter \@secondoftwo
 \fi
}%
\providecommand \natexlab [1]{#1}%
\providecommand \enquote  [1]{``#1''}%
\providecommand \bibnamefont  [1]{#1}%
\providecommand \bibfnamefont [1]{#1}%
\providecommand \citenamefont [1]{#1}%
\providecommand \href@noop [0]{\@secondoftwo}%
\providecommand \href [0]{\begingroup \@sanitize@url \@href}%
\providecommand \@href[1]{\@@startlink{#1}\@@href}%
\providecommand \@@href[1]{\endgroup#1\@@endlink}%
\providecommand \@sanitize@url [0]{\catcode `\\12\catcode `\$12\catcode
  `\&12\catcode `\#12\catcode `\^12\catcode `\_12\catcode `\%12\relax}%
\providecommand \@@startlink[1]{}%
\providecommand \@@endlink[0]{}%
\providecommand \url  [0]{\begingroup\@sanitize@url \@url }%
\providecommand \@url [1]{\endgroup\@href {#1}{\urlprefix }}%
\providecommand \urlprefix  [0]{URL }%
\providecommand \Eprint [0]{\href }%
\providecommand \doibase [0]{https://doi.org/}%
\providecommand \selectlanguage [0]{\@gobble}%
\providecommand \bibinfo  [0]{\@secondoftwo}%
\providecommand \bibfield  [0]{\@secondoftwo}%
\providecommand \translation [1]{[#1]}%
\providecommand \BibitemOpen [0]{}%
\providecommand \bibitemStop [0]{}%
\providecommand \bibitemNoStop [0]{.\EOS\space}%
\providecommand \EOS [0]{\spacefactor3000\relax}%
\providecommand \BibitemShut  [1]{\csname bibitem#1\endcsname}%
\let\auto@bib@innerbib\@empty
\bibitem [{\citenamefont {Kuang}\ and\ \citenamefont
  {Miller}(2020)}]{Kuang2020n2}%
  \BibitemOpen
  \bibfield  {author} {\bibinfo {author} {\bibfnamefont {Z.}~\bibnamefont
  {Kuang}}\ and\ \bibinfo {author} {\bibfnamefont {O.~D.}\ \bibnamefont
  {Miller}},\ }\bibfield  {title} {\bibinfo {title} {Computational bounds to
  light--matter interactions via local conservation laws},\ }\href
  {https://doi.org/10.1103/PhysRevLett.125.263607} {\bibfield  {journal}
  {\bibinfo  {journal} {Phys. Rev. Lett.}\ }\textbf {\bibinfo {volume} {125}},\
  \bibinfo {pages} {263607} (\bibinfo {year} {2020})},\ \Eprint
  {https://arxiv.org/abs/2008.13325} {arXiv:2008.13325} \BibitemShut {NoStop}%
\bibitem [{\citenamefont {Molesky}\ \emph
  {et~al.}(2020{\natexlab{a}})\citenamefont {Molesky}, \citenamefont {Chao},\
  and\ \citenamefont {Rodriguez}}]{Molesky2020c}%
  \BibitemOpen
  \bibfield  {author} {\bibinfo {author} {\bibfnamefont {S.}~\bibnamefont
  {Molesky}}, \bibinfo {author} {\bibfnamefont {P.}~\bibnamefont {Chao}},\ and\
  \bibinfo {author} {\bibfnamefont {A.~W.}\ \bibnamefont {Rodriguez}},\
  }\bibfield  {title} {\bibinfo {title} {{Hierarchical mean-field T operator
  bounds on electromagnetic scattering: Upper nounds on near-field radiative
  Purcell enhancement}},\ }\href {http://arxiv.org/abs/2008.08168} {\bibfield
  {journal} {\bibinfo  {journal} {Phys. Rev. Res.}\ }\textbf {\bibinfo {volume}
  {2}},\ \bibinfo {pages} {043398} (\bibinfo {year} {2020}{\natexlab{a}})},\
  \Eprint {https://arxiv.org/abs/2008.08168} {arXiv:2008.08168} \BibitemShut
  {NoStop}%
\bibitem [{\citenamefont {Laurent}\ and\ \citenamefont
  {Rendl}(2005)}]{Laurent2005}%
  \BibitemOpen
  \bibfield  {author} {\bibinfo {author} {\bibfnamefont {M.}~\bibnamefont
  {Laurent}}\ and\ \bibinfo {author} {\bibfnamefont {F.}~\bibnamefont
  {Rendl}},\ }\bibfield  {title} {\bibinfo {title} {{Semidefinite Programming
  and Integer Programming}},\ }\href
  {https://doi.org/10.1016/S0927-0507(05)12008-8} {\bibfield  {journal}
  {\bibinfo  {journal} {Handbooks Oper. Res. Manag. Sci.}\ }\textbf {\bibinfo
  {volume} {12}},\ \bibinfo {pages} {393} (\bibinfo {year} {2005})}\BibitemShut
  {NoStop}%
\bibitem [{\citenamefont {Luo}\ \emph {et~al.}(2010)\citenamefont {Luo},
  \citenamefont {Ma}, \citenamefont {So}, \citenamefont {Ye},\ and\
  \citenamefont {Zhang}}]{Luo2010}%
  \BibitemOpen
  \bibfield  {author} {\bibinfo {author} {\bibfnamefont {Z.~Q.}\ \bibnamefont
  {Luo}}, \bibinfo {author} {\bibfnamefont {W.~K.}\ \bibnamefont {Ma}},
  \bibinfo {author} {\bibfnamefont {A.}~\bibnamefont {So}}, \bibinfo {author}
  {\bibfnamefont {Y.}~\bibnamefont {Ye}},\ and\ \bibinfo {author}
  {\bibfnamefont {S.}~\bibnamefont {Zhang}},\ }\bibfield  {title} {\bibinfo
  {title} {{Semidefinite relaxation of quadratic optimization problems}},\
  }\href {https://doi.org/10.1109/MSP.2010.936019} {\bibfield  {journal}
  {\bibinfo  {journal} {IEEE Signal Process. Mag.}\ }\textbf {\bibinfo {volume}
  {27}},\ \bibinfo {pages} {20} (\bibinfo {year} {2010})}\BibitemShut {NoStop}%
\bibitem [{\citenamefont {Motzoi}\ \emph {et~al.}(2009)\citenamefont {Motzoi},
  \citenamefont {Gambetta}, \citenamefont {Rebentrost},\ and\ \citenamefont
  {Wilhelm}}]{motzoi2009simple}%
  \BibitemOpen
  \bibfield  {author} {\bibinfo {author} {\bibfnamefont {F.}~\bibnamefont
  {Motzoi}}, \bibinfo {author} {\bibfnamefont {J.~M.}\ \bibnamefont
  {Gambetta}}, \bibinfo {author} {\bibfnamefont {P.}~\bibnamefont
  {Rebentrost}},\ and\ \bibinfo {author} {\bibfnamefont {F.~K.}\ \bibnamefont
  {Wilhelm}},\ }\bibfield  {title} {\bibinfo {title} {Simple pulses for
  elimination of leakage in weakly nonlinear qubits},\ }\href@noop {}
  {\bibfield  {journal} {\bibinfo  {journal} {Physical Review Letters}\
  }\textbf {\bibinfo {volume} {103}},\ \bibinfo {pages} {110501} (\bibinfo
  {year} {2009})}\BibitemShut {NoStop}%
\bibitem [{\citenamefont {Khaneja}\ \emph {et~al.}(2005)\citenamefont
  {Khaneja}, \citenamefont {Reiss}, \citenamefont {Kehlet}, \citenamefont
  {Schulte-Herbr{\"{u}}ggen},\ and\ \citenamefont {Glaser}}]{Khaneja2005}%
  \BibitemOpen
  \bibfield  {author} {\bibinfo {author} {\bibfnamefont {N.}~\bibnamefont
  {Khaneja}}, \bibinfo {author} {\bibfnamefont {T.}~\bibnamefont {Reiss}},
  \bibinfo {author} {\bibfnamefont {C.}~\bibnamefont {Kehlet}}, \bibinfo
  {author} {\bibfnamefont {T.}~\bibnamefont {Schulte-Herbr{\"{u}}ggen}},\ and\
  \bibinfo {author} {\bibfnamefont {S.~J.}\ \bibnamefont {Glaser}},\ }\bibfield
   {title} {\bibinfo {title} {{Optimal control of coupled spin dynamics: Design
  of NMR pulse sequences by gradient ascent algorithms}},\ }\href
  {https://doi.org/10.1016/j.jmr.2004.11.004} {\bibfield  {journal} {\bibinfo
  {journal} {J. Magn. Reson.}\ }\textbf {\bibinfo {volume} {172}},\ \bibinfo
  {pages} {296} (\bibinfo {year} {2005})}\BibitemShut {NoStop}%
\bibitem [{\citenamefont {Nielsen}\ \emph {et~al.}(2010)\citenamefont
  {Nielsen}, \citenamefont {Kehlet}, \citenamefont {Glaser},\ and\
  \citenamefont {Khaneja}}]{Nielsen2010}%
  \BibitemOpen
  \bibfield  {author} {\bibinfo {author} {\bibfnamefont {N.~C.}\ \bibnamefont
  {Nielsen}}, \bibinfo {author} {\bibfnamefont {C.}~\bibnamefont {Kehlet}},
  \bibinfo {author} {\bibfnamefont {S.~J.}\ \bibnamefont {Glaser}},\ and\
  \bibinfo {author} {\bibfnamefont {N.}~\bibnamefont {Khaneja}},\ }\bibfield
  {title} {\bibinfo {title} {{Optimal Control Methods in NMR Spectroscopy}},\
  }\bibfield  {journal} {\bibinfo  {journal} {Encycl. Magn. Reson.}\ }\href
  {https://doi.org/10.1002/9780470034590.emrstm1043}
  {10.1002/9780470034590.emrstm1043} (\bibinfo {year} {2010})\BibitemShut
  {NoStop}%
\bibitem [{\citenamefont {To{\v{s}}ner}\ \emph {et~al.}(2018)\citenamefont
  {To{\v{s}}ner}, \citenamefont {Sarkar}, \citenamefont {Becker-Baldus},
  \citenamefont {Glaubitz}, \citenamefont {Wegner}, \citenamefont {Engelke},
  \citenamefont {Glaser},\ and\ \citenamefont {Reif}}]{Tovsner2018}%
  \BibitemOpen
  \bibfield  {author} {\bibinfo {author} {\bibfnamefont {Z.}~\bibnamefont
  {To{\v{s}}ner}}, \bibinfo {author} {\bibfnamefont {R.}~\bibnamefont
  {Sarkar}}, \bibinfo {author} {\bibfnamefont {J.}~\bibnamefont
  {Becker-Baldus}}, \bibinfo {author} {\bibfnamefont {C.}~\bibnamefont
  {Glaubitz}}, \bibinfo {author} {\bibfnamefont {S.}~\bibnamefont {Wegner}},
  \bibinfo {author} {\bibfnamefont {F.}~\bibnamefont {Engelke}}, \bibinfo
  {author} {\bibfnamefont {S.~J.}\ \bibnamefont {Glaser}},\ and\ \bibinfo
  {author} {\bibfnamefont {B.}~\bibnamefont {Reif}},\ }\bibfield  {title}
  {\bibinfo {title} {{Overcoming Volume Selectivity of Dipolar Recoupling in
  Biological Solid-State NMR Spectroscopy}},\ }\href@noop {} {\bibfield
  {journal} {\bibinfo  {journal} {Angew. Chemie Int. Ed.}\ }\textbf {\bibinfo
  {volume} {57}},\ \bibinfo {pages} {14514} (\bibinfo {year}
  {2018})}\BibitemShut {NoStop}%
\bibitem [{\citenamefont {Nielsen}\ \emph {et~al.}(2006)\citenamefont
  {Nielsen}, \citenamefont {Dowling}, \citenamefont {Gu},\ and\ \citenamefont
  {Doherty}}]{Nielsen2006}%
  \BibitemOpen
  \bibfield  {author} {\bibinfo {author} {\bibfnamefont {M.~A.}\ \bibnamefont
  {Nielsen}}, \bibinfo {author} {\bibfnamefont {M.~R.}\ \bibnamefont
  {Dowling}}, \bibinfo {author} {\bibfnamefont {M.}~\bibnamefont {Gu}},\ and\
  \bibinfo {author} {\bibfnamefont {A.~C.}\ \bibnamefont {Doherty}},\
  }\bibfield  {title} {\bibinfo {title} {{Optimal control, geometry, and
  quantum computing}},\ }\href {https://doi.org/10.1103/PhysRevA.73.062323}
  {\bibfield  {journal} {\bibinfo  {journal} {Phys. Rev. A}\ }\textbf {\bibinfo
  {volume} {73}},\ \bibinfo {pages} {1} (\bibinfo {year} {2006})},\ \Eprint
  {https://arxiv.org/abs/0603160} {arXiv:0603160 [quant-ph]} \BibitemShut
  {NoStop}%
\bibitem [{\citenamefont {Jefferson}\ and\ \citenamefont
  {Myers}(2017)}]{Jefferson2017}%
  \BibitemOpen
  \bibfield  {author} {\bibinfo {author} {\bibfnamefont {R.~A.}\ \bibnamefont
  {Jefferson}}\ and\ \bibinfo {author} {\bibfnamefont {R.~C.}\ \bibnamefont
  {Myers}},\ }\bibfield  {title} {\bibinfo {title} {{Circuit complexity in
  quantum field theory}},\ }\href {https://doi.org/10.1007/JHEP10(2017)107}
  {\bibfield  {journal} {\bibinfo  {journal} {J. High Energy Physics}\ }\textbf
  {\bibinfo {volume} {2017}},\ \bibinfo {pages} {107} (\bibinfo {year}
  {2017})}\BibitemShut {NoStop}%
\bibitem [{\citenamefont {Chapman}\ \emph {et~al.}(2018)\citenamefont
  {Chapman}, \citenamefont {Heller}, \citenamefont {Marrochio},\ and\
  \citenamefont {Pastawski}}]{Chapman2018}%
  \BibitemOpen
  \bibfield  {author} {\bibinfo {author} {\bibfnamefont {S.}~\bibnamefont
  {Chapman}}, \bibinfo {author} {\bibfnamefont {M.~P.}\ \bibnamefont {Heller}},
  \bibinfo {author} {\bibfnamefont {H.}~\bibnamefont {Marrochio}},\ and\
  \bibinfo {author} {\bibfnamefont {F.}~\bibnamefont {Pastawski}},\ }\bibfield
  {title} {\bibinfo {title} {{Toward a Definition of Complexity for Quantum
  Field Theory States}},\ }\href
  {https://doi.org/10.1103/PhysRevLett.120.121602} {\bibfield  {journal}
  {\bibinfo  {journal} {Phys. Rev. Lett.}\ }\textbf {\bibinfo {volume} {120}},\
  \bibinfo {pages} {121602} (\bibinfo {year} {2018})}\BibitemShut {NoStop}%
\bibitem [{\citenamefont {Peirce}\ \emph {et~al.}(1988)\citenamefont {Peirce},
  \citenamefont {Dahleh},\ and\ \citenamefont {Rabitz}}]{Peirce1988}%
  \BibitemOpen
  \bibfield  {author} {\bibinfo {author} {\bibfnamefont {A.~P.}\ \bibnamefont
  {Peirce}}, \bibinfo {author} {\bibfnamefont {M.~A.}\ \bibnamefont {Dahleh}},\
  and\ \bibinfo {author} {\bibfnamefont {H.}~\bibnamefont {Rabitz}},\
  }\bibfield  {title} {\bibinfo {title} {{Optimal control of quantum-mechanical
  systems: Existence, numerical approximation, and applications}},\ }\href@noop
  {} {\bibfield  {journal} {\bibinfo  {journal} {Phys. Rev. A}\ }\textbf
  {\bibinfo {volume} {37}},\ \bibinfo {pages} {4950} (\bibinfo {year}
  {1988})}\BibitemShut {NoStop}%
\bibitem [{\citenamefont {Werschnik}\ and\ \citenamefont
  {Gross}(2007)}]{Werschnik2007}%
  \BibitemOpen
  \bibfield  {author} {\bibinfo {author} {\bibfnamefont {J.}~\bibnamefont
  {Werschnik}}\ and\ \bibinfo {author} {\bibfnamefont {E.~K.}\ \bibnamefont
  {Gross}},\ }\bibfield  {title} {\bibinfo {title} {{Quantum optimal control
  theory}},\ }\bibfield  {journal} {\bibinfo  {journal} {J. Phys. B At. Mol.
  Opt. Phys.}\ }\textbf {\bibinfo {volume} {40}},\ \href
  {https://doi.org/10.1088/0953-4075/40/18/R01} {10.1088/0953-4075/40/18/R01}
  (\bibinfo {year} {2007}),\ \Eprint {https://arxiv.org/abs/0707.1883}
  {arXiv:0707.1883} \BibitemShut {NoStop}%
\bibitem [{\citenamefont {D'Alessandro}(2007)}]{dAlessandro2007}%
  \BibitemOpen
  \bibfield  {author} {\bibinfo {author} {\bibfnamefont {D.}~\bibnamefont
  {D'Alessandro}},\ }\href@noop {} {\emph {\bibinfo {title} {{Introduction to
  quantum control and dynamics}}}}\ (\bibinfo  {publisher} {CRC press},\
  \bibinfo {address} {Boca Raton, FL},\ \bibinfo {year} {2007})\BibitemShut
  {NoStop}%
\bibitem [{\citenamefont {Brif}\ \emph {et~al.}(2010)\citenamefont {Brif},
  \citenamefont {Chakrabarti},\ and\ \citenamefont {Rabitz}}]{Brif2010}%
  \BibitemOpen
  \bibfield  {author} {\bibinfo {author} {\bibfnamefont {C.}~\bibnamefont
  {Brif}}, \bibinfo {author} {\bibfnamefont {R.}~\bibnamefont {Chakrabarti}},\
  and\ \bibinfo {author} {\bibfnamefont {H.}~\bibnamefont {Rabitz}},\
  }\bibfield  {title} {\bibinfo {title} {{Control of quantum phenomena: Past,
  present and future}},\ }\bibfield  {journal} {\bibinfo  {journal} {New J.
  Phys.}\ }\textbf {\bibinfo {volume} {12}},\ \href
  {https://doi.org/10.1088/1367-2630/12/7/075008}
  {10.1088/1367-2630/12/7/075008} (\bibinfo {year} {2010}),\ \Eprint
  {https://arxiv.org/abs/0912.5121} {arXiv:0912.5121} \BibitemShut {NoStop}%
\bibitem [{\citenamefont {Glaser}\ \emph {et~al.}(2015)\citenamefont {Glaser},
  \citenamefont {Boscain}, \citenamefont {Calarco}, \citenamefont {Koch},
  \citenamefont {K{\"{o}}ckenberger}, \citenamefont {Kosloff}, \citenamefont
  {Kuprov}, \citenamefont {Luy}, \citenamefont {Schirmer}, \citenamefont
  {Schulte-Herbr{\"{u}}ggen}, \citenamefont {Sugny},\ and\ \citenamefont
  {Wilhelm}}]{Glaser2015}%
  \BibitemOpen
  \bibfield  {author} {\bibinfo {author} {\bibfnamefont {S.~J.}\ \bibnamefont
  {Glaser}}, \bibinfo {author} {\bibfnamefont {U.}~\bibnamefont {Boscain}},
  \bibinfo {author} {\bibfnamefont {T.}~\bibnamefont {Calarco}}, \bibinfo
  {author} {\bibfnamefont {C.~P.}\ \bibnamefont {Koch}}, \bibinfo {author}
  {\bibfnamefont {W.}~\bibnamefont {K{\"{o}}ckenberger}}, \bibinfo {author}
  {\bibfnamefont {R.}~\bibnamefont {Kosloff}}, \bibinfo {author} {\bibfnamefont
  {I.}~\bibnamefont {Kuprov}}, \bibinfo {author} {\bibfnamefont
  {B.}~\bibnamefont {Luy}}, \bibinfo {author} {\bibfnamefont {S.}~\bibnamefont
  {Schirmer}}, \bibinfo {author} {\bibfnamefont {T.}~\bibnamefont
  {Schulte-Herbr{\"{u}}ggen}}, \bibinfo {author} {\bibfnamefont
  {D.}~\bibnamefont {Sugny}},\ and\ \bibinfo {author} {\bibfnamefont {F.~K.}\
  \bibnamefont {Wilhelm}},\ }\bibfield  {title} {\bibinfo {title} {{Training
  Schr{\"{o}}dinger's cat: Quantum optimal control: Strategic report on current
  status, visions and goals for research in Europe}},\ }\bibfield  {journal}
  {\bibinfo  {journal} {Eur. Phys. J. D}\ }\textbf {\bibinfo {volume} {69}},\
  \href {https://doi.org/10.1140/epjd/e2015-60464-1}
  {10.1140/epjd/e2015-60464-1} (\bibinfo {year} {2015})\BibitemShut {NoStop}%
\bibitem [{\citenamefont {Bason}\ \emph {et~al.}(2012)\citenamefont {Bason},
  \citenamefont {Viteau}, \citenamefont {Malossi}, \citenamefont {Huillery},
  \citenamefont {Arimondo}, \citenamefont {Ciampini}, \citenamefont {Fazio},
  \citenamefont {Giovannetti}, \citenamefont {Mannella},\ and\ \citenamefont
  {Morsch}}]{Bason2012}%
  \BibitemOpen
  \bibfield  {author} {\bibinfo {author} {\bibfnamefont {M.~G.}\ \bibnamefont
  {Bason}}, \bibinfo {author} {\bibfnamefont {M.}~\bibnamefont {Viteau}},
  \bibinfo {author} {\bibfnamefont {N.}~\bibnamefont {Malossi}}, \bibinfo
  {author} {\bibfnamefont {P.}~\bibnamefont {Huillery}}, \bibinfo {author}
  {\bibfnamefont {E.}~\bibnamefont {Arimondo}}, \bibinfo {author}
  {\bibfnamefont {D.}~\bibnamefont {Ciampini}}, \bibinfo {author}
  {\bibfnamefont {R.}~\bibnamefont {Fazio}}, \bibinfo {author} {\bibfnamefont
  {V.}~\bibnamefont {Giovannetti}}, \bibinfo {author} {\bibfnamefont
  {R.}~\bibnamefont {Mannella}},\ and\ \bibinfo {author} {\bibfnamefont
  {O.}~\bibnamefont {Morsch}},\ }\bibfield  {title} {\bibinfo {title}
  {{High-fidelity quantum driving}},\ }\href
  {https://doi.org/10.1038/nphys2170} {\bibfield  {journal} {\bibinfo
  {journal} {Nat. Phys.}\ }\textbf {\bibinfo {volume} {8}},\ \bibinfo {pages}
  {147} (\bibinfo {year} {2012})},\ \Eprint {https://arxiv.org/abs/1111.1579}
  {arXiv:1111.1579} \BibitemShut {NoStop}%
\bibitem [{\citenamefont {{Van Frank}}\ \emph {et~al.}(2016)\citenamefont {{Van
  Frank}}, \citenamefont {Bonneau}, \citenamefont {Schmiedmayer}, \citenamefont
  {Hild}, \citenamefont {Gross}, \citenamefont {Cheneau}, \citenamefont
  {Bloch}, \citenamefont {Pichler}, \citenamefont {Negretti}, \citenamefont
  {Calarco},\ and\ \citenamefont {Montangero}}]{VanFrank2016}%
  \BibitemOpen
  \bibfield  {author} {\bibinfo {author} {\bibfnamefont {S.}~\bibnamefont {{Van
  Frank}}}, \bibinfo {author} {\bibfnamefont {M.}~\bibnamefont {Bonneau}},
  \bibinfo {author} {\bibfnamefont {J.}~\bibnamefont {Schmiedmayer}}, \bibinfo
  {author} {\bibfnamefont {S.}~\bibnamefont {Hild}}, \bibinfo {author}
  {\bibfnamefont {C.}~\bibnamefont {Gross}}, \bibinfo {author} {\bibfnamefont
  {M.}~\bibnamefont {Cheneau}}, \bibinfo {author} {\bibfnamefont
  {I.}~\bibnamefont {Bloch}}, \bibinfo {author} {\bibfnamefont
  {T.}~\bibnamefont {Pichler}}, \bibinfo {author} {\bibfnamefont
  {A.}~\bibnamefont {Negretti}}, \bibinfo {author} {\bibfnamefont
  {T.}~\bibnamefont {Calarco}},\ and\ \bibinfo {author} {\bibfnamefont
  {S.}~\bibnamefont {Montangero}},\ }\bibfield  {title} {\bibinfo {title}
  {{Optimal control of complex atomic quantum systems}},\ }\href
  {https://doi.org/10.1038/srep34187} {\bibfield  {journal} {\bibinfo
  {journal} {Sci. Rep.}\ }\textbf {\bibinfo {volume} {6}},\ \bibinfo {pages}
  {34187} (\bibinfo {year} {2016})},\ \Eprint
  {https://arxiv.org/abs/1511.02247} {arXiv:1511.02247} \BibitemShut {NoStop}%
\bibitem [{\citenamefont {Heeres}\ \emph {et~al.}(2017)\citenamefont {Heeres},
  \citenamefont {Reinhold}, \citenamefont {Ofek}, \citenamefont {Frunzio},
  \citenamefont {Jiang}, \citenamefont {Devoret},\ and\ \citenamefont
  {Schoelkopf}}]{Heeres2017}%
  \BibitemOpen
  \bibfield  {author} {\bibinfo {author} {\bibfnamefont {R.~W.}\ \bibnamefont
  {Heeres}}, \bibinfo {author} {\bibfnamefont {P.}~\bibnamefont {Reinhold}},
  \bibinfo {author} {\bibfnamefont {N.}~\bibnamefont {Ofek}}, \bibinfo {author}
  {\bibfnamefont {L.}~\bibnamefont {Frunzio}}, \bibinfo {author} {\bibfnamefont
  {L.}~\bibnamefont {Jiang}}, \bibinfo {author} {\bibfnamefont {M.~H.}\
  \bibnamefont {Devoret}},\ and\ \bibinfo {author} {\bibfnamefont {R.~J.}\
  \bibnamefont {Schoelkopf}},\ }\bibfield  {title} {\bibinfo {title}
  {{Implementing a universal gate set on a logical qubit encoded in an
  oscillator}},\ }\href {https://doi.org/10.1038/s41467-017-00045-1} {\bibfield
   {journal} {\bibinfo  {journal} {Nat. Commun.}\ }\textbf {\bibinfo {volume}
  {8}},\ \bibinfo {pages} {1} (\bibinfo {year} {2017})},\ \Eprint
  {https://arxiv.org/abs/1608.02430} {arXiv:1608.02430} \BibitemShut {NoStop}%
\bibitem [{\citenamefont {Goetz}\ \emph {et~al.}(2019)\citenamefont {Goetz},
  \citenamefont {Koch},\ and\ \citenamefont {Greenman}}]{Goetz2019}%
  \BibitemOpen
  \bibfield  {author} {\bibinfo {author} {\bibfnamefont {R.~E.}\ \bibnamefont
  {Goetz}}, \bibinfo {author} {\bibfnamefont {C.~P.}\ \bibnamefont {Koch}},\
  and\ \bibinfo {author} {\bibfnamefont {L.}~\bibnamefont {Greenman}},\
  }\bibfield  {title} {\bibinfo {title} {{Quantum Control of Photoelectron
  Circular Dichroism}},\ }\href
  {https://doi.org/10.1103/PhysRevLett.122.013204} {\bibfield  {journal}
  {\bibinfo  {journal} {Phys. Rev. Lett.}\ }\textbf {\bibinfo {volume} {122}},\
  \bibinfo {pages} {13204} (\bibinfo {year} {2019})},\ \Eprint
  {https://arxiv.org/abs/1809.04543} {arXiv:1809.04543} \BibitemShut {NoStop}%
\bibitem [{\citenamefont {Veps{\"{a}}l{\"{a}}inen}\ \emph
  {et~al.}(2019)\citenamefont {Veps{\"{a}}l{\"{a}}inen}, \citenamefont
  {Danilin},\ and\ \citenamefont {Paraoanu}}]{Vepsalainen2019}%
  \BibitemOpen
  \bibfield  {author} {\bibinfo {author} {\bibfnamefont {A.}~\bibnamefont
  {Veps{\"{a}}l{\"{a}}inen}}, \bibinfo {author} {\bibfnamefont
  {S.}~\bibnamefont {Danilin}},\ and\ \bibinfo {author} {\bibfnamefont {G.~S.}\
  \bibnamefont {Paraoanu}},\ }\bibfield  {title} {\bibinfo {title}
  {{Superadiabatic population transfer in a three-level superconducting
  circuit}},\ }\href {https://doi.org/10.1126/sciadv.aau5999} {\bibfield
  {journal} {\bibinfo  {journal} {Sci. Adv.}\ }\textbf {\bibinfo {volume}
  {5}},\ \bibinfo {pages} {eaau5999} (\bibinfo {year} {2019})}\BibitemShut
  {NoStop}%
\bibitem [{\citenamefont {Lam}\ \emph {et~al.}(2021)\citenamefont {Lam},
  \citenamefont {Peter}, \citenamefont {Groh}, \citenamefont {Alt},
  \citenamefont {Robens}, \citenamefont {Meschede}, \citenamefont {Negretti},
  \citenamefont {Montangero}, \citenamefont {Calarco},\ and\ \citenamefont
  {Alberti}}]{Lam2021}%
  \BibitemOpen
  \bibfield  {author} {\bibinfo {author} {\bibfnamefont {M.~R.}\ \bibnamefont
  {Lam}}, \bibinfo {author} {\bibfnamefont {N.}~\bibnamefont {Peter}}, \bibinfo
  {author} {\bibfnamefont {T.}~\bibnamefont {Groh}}, \bibinfo {author}
  {\bibfnamefont {W.}~\bibnamefont {Alt}}, \bibinfo {author} {\bibfnamefont
  {C.}~\bibnamefont {Robens}}, \bibinfo {author} {\bibfnamefont
  {D.}~\bibnamefont {Meschede}}, \bibinfo {author} {\bibfnamefont
  {A.}~\bibnamefont {Negretti}}, \bibinfo {author} {\bibfnamefont
  {S.}~\bibnamefont {Montangero}}, \bibinfo {author} {\bibfnamefont
  {T.}~\bibnamefont {Calarco}},\ and\ \bibinfo {author} {\bibfnamefont
  {A.}~\bibnamefont {Alberti}},\ }\bibfield  {title} {\bibinfo {title}
  {{Demonstration of Quantum Brachistochrones between Distant States of an
  Atom}},\ }\href {https://doi.org/10.1103/PhysRevX.11.011035} {\bibfield
  {journal} {\bibinfo  {journal} {Phys. Rev. X}\ }\textbf {\bibinfo {volume}
  {11}},\ \bibinfo {pages} {11035} (\bibinfo {year} {2021})},\ \Eprint
  {https://arxiv.org/abs/2009.02231} {arXiv:2009.02231} \BibitemShut {NoStop}%
\bibitem [{\citenamefont {{De Fouquieres}}\ \emph {et~al.}(2011)\citenamefont
  {{De Fouquieres}}, \citenamefont {Schirmer}, \citenamefont {Glaser},\ and\
  \citenamefont {Kuprov}}]{DeFouquieres2011}%
  \BibitemOpen
  \bibfield  {author} {\bibinfo {author} {\bibfnamefont {P.}~\bibnamefont {{De
  Fouquieres}}}, \bibinfo {author} {\bibfnamefont {S.~G.}\ \bibnamefont
  {Schirmer}}, \bibinfo {author} {\bibfnamefont {S.~J.}\ \bibnamefont
  {Glaser}},\ and\ \bibinfo {author} {\bibfnamefont {I.}~\bibnamefont
  {Kuprov}},\ }\bibfield  {title} {\bibinfo {title} {{Second order gradient
  ascent pulse engineering}},\ }\href
  {https://doi.org/10.1016/j.jmr.2011.07.023} {\bibfield  {journal} {\bibinfo
  {journal} {J. Magn. Reson.}\ }\textbf {\bibinfo {volume} {212}},\ \bibinfo
  {pages} {412} (\bibinfo {year} {2011})}\BibitemShut {NoStop}%
\bibitem [{\citenamefont {Dolde}\ \emph {et~al.}(2014)\citenamefont {Dolde},
  \citenamefont {Bergholm}, \citenamefont {Wang}, \citenamefont {Jakobi},
  \citenamefont {Naydenov}, \citenamefont {Pezzagna}, \citenamefont {Meijer},
  \citenamefont {Jelezko}, \citenamefont {Neumann}, \citenamefont
  {Schulte-Herbr{\"{u}}ggen}, \citenamefont {Biamonte},\ and\ \citenamefont
  {Wrachtrup}}]{Dolde2014}%
  \BibitemOpen
  \bibfield  {author} {\bibinfo {author} {\bibfnamefont {F.}~\bibnamefont
  {Dolde}}, \bibinfo {author} {\bibfnamefont {V.}~\bibnamefont {Bergholm}},
  \bibinfo {author} {\bibfnamefont {Y.}~\bibnamefont {Wang}}, \bibinfo {author}
  {\bibfnamefont {I.}~\bibnamefont {Jakobi}}, \bibinfo {author} {\bibfnamefont
  {B.}~\bibnamefont {Naydenov}}, \bibinfo {author} {\bibfnamefont
  {S.}~\bibnamefont {Pezzagna}}, \bibinfo {author} {\bibfnamefont
  {J.}~\bibnamefont {Meijer}}, \bibinfo {author} {\bibfnamefont
  {F.}~\bibnamefont {Jelezko}}, \bibinfo {author} {\bibfnamefont
  {P.}~\bibnamefont {Neumann}}, \bibinfo {author} {\bibfnamefont
  {T.}~\bibnamefont {Schulte-Herbr{\"{u}}ggen}}, \bibinfo {author}
  {\bibfnamefont {J.}~\bibnamefont {Biamonte}},\ and\ \bibinfo {author}
  {\bibfnamefont {J.}~\bibnamefont {Wrachtrup}},\ }\bibfield  {title} {\bibinfo
  {title} {{High-fidelity spin entanglement using optimal control}},\ }\href
  {https://doi.org/10.1038/ncomms4371} {\bibfield  {journal} {\bibinfo
  {journal} {Nat. Commun.}\ }\textbf {\bibinfo {volume} {5}},\ \bibinfo {pages}
  {1} (\bibinfo {year} {2014})},\ \Eprint {https://arxiv.org/abs/1309.4430}
  {arXiv:1309.4430} \BibitemShut {NoStop}%
\bibitem [{\citenamefont {Anderson}\ \emph {et~al.}(2015)\citenamefont
  {Anderson}, \citenamefont {Sosa-Martinez}, \citenamefont {Riofr{\'{i}}o},
  \citenamefont {Deutsch},\ and\ \citenamefont {Jessen}}]{Anderson2015}%
  \BibitemOpen
  \bibfield  {author} {\bibinfo {author} {\bibfnamefont {B.~E.}\ \bibnamefont
  {Anderson}}, \bibinfo {author} {\bibfnamefont {H.}~\bibnamefont
  {Sosa-Martinez}}, \bibinfo {author} {\bibfnamefont {C.~A.}\ \bibnamefont
  {Riofr{\'{i}}o}}, \bibinfo {author} {\bibfnamefont {I.~H.}\ \bibnamefont
  {Deutsch}},\ and\ \bibinfo {author} {\bibfnamefont {P.~S.}\ \bibnamefont
  {Jessen}},\ }\bibfield  {title} {\bibinfo {title} {{Accurate and Robust
  Unitary Transformations of a High-Dimensional Quantum System}},\ }\href
  {https://doi.org/10.1103/PhysRevLett.114.240401} {\bibfield  {journal}
  {\bibinfo  {journal} {Phys. Rev. Lett.}\ }\textbf {\bibinfo {volume} {114}},\
  \bibinfo {pages} {240401} (\bibinfo {year} {2015})}\BibitemShut {NoStop}%
\bibitem [{\citenamefont {Krotov}(1993)}]{Krotov1993}%
  \BibitemOpen
  \bibfield  {author} {\bibinfo {author} {\bibfnamefont {V.~F.}\ \bibnamefont
  {Krotov}},\ }\bibfield  {title} {\bibinfo {title} {{Global Methods in Optimal
  Control Theory}},\ }in\ \href {https://doi.org/10.1007/978-1-4612-0349-0_3}
  {\emph {\bibinfo {booktitle} {Adv. Nonlinear Dyn. Control A Rep. from
  Russ.}}}\ (\bibinfo  {publisher} {Birkh{\"{a}}user Boston},\ \bibinfo {year}
  {1993})\ pp.\ \bibinfo {pages} {74--121}\BibitemShut {NoStop}%
\bibitem [{\citenamefont {Soml{\'{o}}i}\ \emph {et~al.}(1993)\citenamefont
  {Soml{\'{o}}i}, \citenamefont {Kazakov},\ and\ \citenamefont
  {Tannor}}]{Somloi1993}%
  \BibitemOpen
  \bibfield  {author} {\bibinfo {author} {\bibfnamefont {J.}~\bibnamefont
  {Soml{\'{o}}i}}, \bibinfo {author} {\bibfnamefont {V.~A.}\ \bibnamefont
  {Kazakov}},\ and\ \bibinfo {author} {\bibfnamefont {D.~J.}\ \bibnamefont
  {Tannor}},\ }\bibfield  {title} {\bibinfo {title} {{Controlled dissociation
  of I2 via optical transitions between the X and B electronic states}},\
  }\href {https://doi.org/10.1016/0301-0104(93)80108-L} {\bibfield  {journal}
  {\bibinfo  {journal} {Chem. Phys.}\ }\textbf {\bibinfo {volume} {172}},\
  \bibinfo {pages} {85} (\bibinfo {year} {1993})}\BibitemShut {NoStop}%
\bibitem [{\citenamefont {Zhu}\ and\ \citenamefont {Rabitz}(1998)}]{Zhu1998}%
  \BibitemOpen
  \bibfield  {author} {\bibinfo {author} {\bibfnamefont {W.}~\bibnamefont
  {Zhu}}\ and\ \bibinfo {author} {\bibfnamefont {H.}~\bibnamefont {Rabitz}},\
  }\bibfield  {title} {\bibinfo {title} {{A rapid monotonically convergent
  iteration algorithm for quantum optimal control over the expectation value of
  a positive definite operator}},\ }\href {https://doi.org/10.1063/1.476575}
  {\bibfield  {journal} {\bibinfo  {journal} {J. Chem. Phys.}\ }\textbf
  {\bibinfo {volume} {109}},\ \bibinfo {pages} {385} (\bibinfo {year}
  {1998})}\BibitemShut {NoStop}%
\bibitem [{\citenamefont {Ohtsuki}\ \emph {et~al.}(2004)\citenamefont
  {Ohtsuki}, \citenamefont {Turinici},\ and\ \citenamefont
  {Rabitz}}]{Ohtsuki2004}%
  \BibitemOpen
  \bibfield  {author} {\bibinfo {author} {\bibfnamefont {Y.}~\bibnamefont
  {Ohtsuki}}, \bibinfo {author} {\bibfnamefont {G.}~\bibnamefont {Turinici}},\
  and\ \bibinfo {author} {\bibfnamefont {H.}~\bibnamefont {Rabitz}},\
  }\bibfield  {title} {\bibinfo {title} {{Generalized monotonically convergent
  algorithms for solving quantum optimal control problems}},\ }\href
  {https://doi.org/10.1063/1.1650297} {\bibfield  {journal} {\bibinfo
  {journal} {J. Chem. Phys.}\ }\textbf {\bibinfo {volume} {120}},\ \bibinfo
  {pages} {5509} (\bibinfo {year} {2004})}\BibitemShut {NoStop}%
\bibitem [{\citenamefont {Schirmer}\ and\ \citenamefont {{De
  Fouquieres}}(2011)}]{Schirmer2011}%
  \BibitemOpen
  \bibfield  {author} {\bibinfo {author} {\bibfnamefont {S.~G.}\ \bibnamefont
  {Schirmer}}\ and\ \bibinfo {author} {\bibfnamefont {P.}~\bibnamefont {{De
  Fouquieres}}},\ }\bibfield  {title} {\bibinfo {title} {{Efficient algorithms
  for optimal control of quantum dynamics: The Krotov method unencumbered}},\
  }\href {https://doi.org/10.1088/1367-2630/13/7/073029} {\bibfield  {journal}
  {\bibinfo  {journal} {New J. Phys.}\ }\textbf {\bibinfo {volume} {13}},\
  \bibinfo {pages} {73029} (\bibinfo {year} {2011})},\ \Eprint
  {https://arxiv.org/abs/1103.5435} {arXiv:1103.5435} \BibitemShut {NoStop}%
\bibitem [{\citenamefont {Goerz}\ \emph {et~al.}(2019)\citenamefont {Goerz},
  \citenamefont {Basilewitsch}, \citenamefont {Gago-Encinas}, \citenamefont
  {Krauss}, \citenamefont {Horn}, \citenamefont {Reich},\ and\ \citenamefont
  {Koch}}]{Goerz2019}%
  \BibitemOpen
  \bibfield  {author} {\bibinfo {author} {\bibfnamefont {M.~H.}\ \bibnamefont
  {Goerz}}, \bibinfo {author} {\bibfnamefont {D.}~\bibnamefont {Basilewitsch}},
  \bibinfo {author} {\bibfnamefont {F.}~\bibnamefont {Gago-Encinas}}, \bibinfo
  {author} {\bibfnamefont {M.~G.}\ \bibnamefont {Krauss}}, \bibinfo {author}
  {\bibfnamefont {K.~P.}\ \bibnamefont {Horn}}, \bibinfo {author}
  {\bibfnamefont {D.~M.}\ \bibnamefont {Reich}},\ and\ \bibinfo {author}
  {\bibfnamefont {C.~P.}\ \bibnamefont {Koch}},\ }\bibfield  {title} {\bibinfo
  {title} {{Krotov: A Python implementation of Krotov's method for quantum
  optimal control}},\ }\href {https://doi.org/10.21468/scipostphys.7.6.080}
  {\bibfield  {journal} {\bibinfo  {journal} {SciPost Phys.}\ }\textbf
  {\bibinfo {volume} {7}},\ \bibinfo {pages} {080} (\bibinfo {year} {2019})},\
  \Eprint {https://arxiv.org/abs/1902.11284} {arXiv:1902.11284} \BibitemShut
  {NoStop}%
\bibitem [{\citenamefont {Caneva}\ \emph {et~al.}(2011)\citenamefont {Caneva},
  \citenamefont {Calarco},\ and\ \citenamefont {Montangero}}]{Caneva2011}%
  \BibitemOpen
  \bibfield  {author} {\bibinfo {author} {\bibfnamefont {T.}~\bibnamefont
  {Caneva}}, \bibinfo {author} {\bibfnamefont {T.}~\bibnamefont {Calarco}},\
  and\ \bibinfo {author} {\bibfnamefont {S.}~\bibnamefont {Montangero}},\
  }\bibfield  {title} {\bibinfo {title} {{Chopped random-basis quantum
  optimization}},\ }\href {https://doi.org/10.1103/PhysRevA.84.022326}
  {\bibfield  {journal} {\bibinfo  {journal} {Phys. Rev. A}\ }\textbf {\bibinfo
  {volume} {84}},\ \bibinfo {pages} {022326} (\bibinfo {year}
  {2011})}\BibitemShut {NoStop}%
\bibitem [{\citenamefont {Doria}\ \emph {et~al.}(2011)\citenamefont {Doria},
  \citenamefont {Calarco},\ and\ \citenamefont {Montangero}}]{Doria2011}%
  \BibitemOpen
  \bibfield  {author} {\bibinfo {author} {\bibfnamefont {P.}~\bibnamefont
  {Doria}}, \bibinfo {author} {\bibfnamefont {T.}~\bibnamefont {Calarco}},\
  and\ \bibinfo {author} {\bibfnamefont {S.}~\bibnamefont {Montangero}},\
  }\bibfield  {title} {\bibinfo {title} {{Optimal control technique for
  many-body quantum dynamics}},\ }\href
  {https://doi.org/10.1103/PhysRevLett.106.190501} {\bibfield  {journal}
  {\bibinfo  {journal} {Phys. Rev. Lett.}\ }\textbf {\bibinfo {volume} {106}},\
  \bibinfo {pages} {1} (\bibinfo {year} {2011})}\BibitemShut {NoStop}%
\bibitem [{\citenamefont {Mandelstam}\ and\ \citenamefont
  {Tamm}(1945)}]{Mandelstam1945}%
  \BibitemOpen
  \bibfield  {author} {\bibinfo {author} {\bibfnamefont {L.}~\bibnamefont
  {Mandelstam}}\ and\ \bibinfo {author} {\bibfnamefont {I.}~\bibnamefont
  {Tamm}},\ }\bibfield  {title} {\bibinfo {title} {{The Uncertainty Relation
  Between Energy and Time in Non-relativistic Quantum Mechanics}},\ }\href
  {https://doi.org/10.1007/978-3-642-74626-0_8} {\bibfield  {journal} {\bibinfo
   {journal} {J. Phys. USSR}\ }\textbf {\bibinfo {volume} {9}},\ \bibinfo
  {pages} {249} (\bibinfo {year} {1945})}\BibitemShut {NoStop}%
\bibitem [{\citenamefont {Fleming}(1973)}]{Fleming1973}%
  \BibitemOpen
  \bibfield  {author} {\bibinfo {author} {\bibfnamefont {G.~N.}\ \bibnamefont
  {Fleming}},\ }\bibfield  {title} {\bibinfo {title} {{A unitarity bound on the
  evolution of nonstationary states}},\ }\href
  {https://doi.org/10.1007/BF02819419} {\bibfield  {journal} {\bibinfo
  {journal} {Nuovo Cim. A}\ }\textbf {\bibinfo {volume} {16}},\ \bibinfo
  {pages} {232} (\bibinfo {year} {1973})}\BibitemShut {NoStop}%
\bibitem [{\citenamefont {Vaidman}(1992)}]{Vaidman1992}%
  \BibitemOpen
  \bibfield  {author} {\bibinfo {author} {\bibfnamefont {L.}~\bibnamefont
  {Vaidman}},\ }\bibfield  {title} {\bibinfo {title} {{Minimum time for the
  evolution to an orthogonal quantum state}},\ }\href
  {https://doi.org/10.1119/1.16940} {\bibfield  {journal} {\bibinfo  {journal}
  {Am. J. Phys.}\ }\textbf {\bibinfo {volume} {60}},\ \bibinfo {pages} {182}
  (\bibinfo {year} {1992})}\BibitemShut {NoStop}%
\bibitem [{\citenamefont {Margolus}\ and\ \citenamefont
  {Levitin}(1998)}]{Margolus1998}%
  \BibitemOpen
  \bibfield  {author} {\bibinfo {author} {\bibfnamefont {N.}~\bibnamefont
  {Margolus}}\ and\ \bibinfo {author} {\bibfnamefont {L.~B.}\ \bibnamefont
  {Levitin}},\ }\bibfield  {title} {\bibinfo {title} {{The maximum speed of
  dynamical evolution}},\ }\href
  {https://doi.org/10.1016/S0167-2789(98)00054-2} {\bibfield  {journal}
  {\bibinfo  {journal} {Phys. D Nonlinear Phenom.}\ }\textbf {\bibinfo {volume}
  {120}},\ \bibinfo {pages} {188} (\bibinfo {year} {1998})},\ \Eprint
  {https://arxiv.org/abs/9710043} {arXiv:9710043 [quant-ph]} \BibitemShut
  {NoStop}%
\bibitem [{\citenamefont {Khaneja}\ \emph {et~al.}(2001)\citenamefont
  {Khaneja}, \citenamefont {Brockett},\ and\ \citenamefont
  {Glaser}}]{Khaneja2001}%
  \BibitemOpen
  \bibfield  {author} {\bibinfo {author} {\bibfnamefont {N.}~\bibnamefont
  {Khaneja}}, \bibinfo {author} {\bibfnamefont {R.}~\bibnamefont {Brockett}},\
  and\ \bibinfo {author} {\bibfnamefont {S.~J.}\ \bibnamefont {Glaser}},\
  }\bibfield  {title} {\bibinfo {title} {{Time optimal control in spin
  systems}},\ }\href {https://doi.org/10.1103/PhysRevA.63.032308} {\bibfield
  {journal} {\bibinfo  {journal} {Phys. Rev. A - At. Mol. Opt. Phys.}\ }\textbf
  {\bibinfo {volume} {63}},\ \bibinfo {pages} {032308} (\bibinfo {year}
  {2001})},\ \Eprint {https://arxiv.org/abs/0006114} {arXiv:0006114 [quant-ph]}
  \BibitemShut {NoStop}%
\bibitem [{\citenamefont {Khaneja}\ \emph {et~al.}(2002)\citenamefont
  {Khaneja}, \citenamefont {Glaser},\ and\ \citenamefont
  {Brockett}}]{Khaneja2002}%
  \BibitemOpen
  \bibfield  {author} {\bibinfo {author} {\bibfnamefont {N.}~\bibnamefont
  {Khaneja}}, \bibinfo {author} {\bibfnamefont {S.~J.}\ \bibnamefont
  {Glaser}},\ and\ \bibinfo {author} {\bibfnamefont {R.}~\bibnamefont
  {Brockett}},\ }\bibfield  {title} {\bibinfo {title} {{Sub-Riemannian geometry
  and time optimal control of three spin systems: Quantum gates and coherence
  transfer}},\ }\href {https://doi.org/10.1103/PhysRevA.65.032301} {\bibfield
  {journal} {\bibinfo  {journal} {Phys. Rev. A}\ }\textbf {\bibinfo {volume}
  {65}},\ \bibinfo {pages} {032301} (\bibinfo {year} {2002})},\ \Eprint
  {https://arxiv.org/abs/0106099} {arXiv:0106099 [quant-ph]} \BibitemShut
  {NoStop}%
\bibitem [{\citenamefont {Giovannetti}\ \emph
  {et~al.}(2003{\natexlab{a}})\citenamefont {Giovannetti}, \citenamefont
  {Lloyd},\ and\ \citenamefont {Maccone}}]{Giovannetti2003}%
  \BibitemOpen
  \bibfield  {author} {\bibinfo {author} {\bibfnamefont {V.}~\bibnamefont
  {Giovannetti}}, \bibinfo {author} {\bibfnamefont {S.}~\bibnamefont {Lloyd}},\
  and\ \bibinfo {author} {\bibfnamefont {L.}~\bibnamefont {Maccone}},\
  }\bibfield  {title} {\bibinfo {title} {{Quantum limits to dynamical
  evolution}},\ }\href {https://doi.org/10.1103/PhysRevA.67.052109} {\bibfield
  {journal} {\bibinfo  {journal} {Phys. Rev. A - At. Mol. Opt. Phys.}\ }\textbf
  {\bibinfo {volume} {67}},\ \bibinfo {pages} {8} (\bibinfo {year}
  {2003}{\natexlab{a}})},\ \Eprint {https://arxiv.org/abs/0210197}
  {arXiv:0210197 [quant-ph]} \BibitemShut {NoStop}%
\bibitem [{\citenamefont {Giovannetti}\ \emph
  {et~al.}(2003{\natexlab{b}})\citenamefont {Giovannetti}, \citenamefont
  {Lloyd},\ and\ \citenamefont {Maccone}}]{Giovannetti2003n2}%
  \BibitemOpen
  \bibfield  {author} {\bibinfo {author} {\bibfnamefont {V.}~\bibnamefont
  {Giovannetti}}, \bibinfo {author} {\bibfnamefont {S.}~\bibnamefont {Lloyd}},\
  and\ \bibinfo {author} {\bibfnamefont {L.}~\bibnamefont {Maccone}},\
  }\bibfield  {title} {\bibinfo {title} {{The role of entanglement in dynamical
  evolution}},\ }\href {https://doi.org/10.1209/epl/i2003-00418-8} {\bibfield
  {journal} {\bibinfo  {journal} {Europhys. Lett.}\ }\textbf {\bibinfo {volume}
  {62}},\ \bibinfo {pages} {615} (\bibinfo {year}
  {2003}{\natexlab{b}})}\BibitemShut {NoStop}%
\bibitem [{\citenamefont {Carlini}\ \emph {et~al.}(2006)\citenamefont
  {Carlini}, \citenamefont {Hosoya}, \citenamefont {Koike},\ and\ \citenamefont
  {Okudaira}}]{Carlini2006}%
  \BibitemOpen
  \bibfield  {author} {\bibinfo {author} {\bibfnamefont {A.}~\bibnamefont
  {Carlini}}, \bibinfo {author} {\bibfnamefont {A.}~\bibnamefont {Hosoya}},
  \bibinfo {author} {\bibfnamefont {T.}~\bibnamefont {Koike}},\ and\ \bibinfo
  {author} {\bibfnamefont {Y.}~\bibnamefont {Okudaira}},\ }\bibfield  {title}
  {\bibinfo {title} {{Time-optimal quantum evolution}},\ }\href
  {https://doi.org/10.1103/PhysRevLett.96.060503} {\bibfield  {journal}
  {\bibinfo  {journal} {Phys. Rev. Lett.}\ }\textbf {\bibinfo {volume} {96}},\
  \bibinfo {pages} {1} (\bibinfo {year} {2006})}\BibitemShut {NoStop}%
\bibitem [{\citenamefont {Carlini}\ \emph {et~al.}(2007)\citenamefont
  {Carlini}, \citenamefont {Hosoya}, \citenamefont {Koike},\ and\ \citenamefont
  {Okudaira}}]{Carlini2007}%
  \BibitemOpen
  \bibfield  {author} {\bibinfo {author} {\bibfnamefont {A.}~\bibnamefont
  {Carlini}}, \bibinfo {author} {\bibfnamefont {A.}~\bibnamefont {Hosoya}},
  \bibinfo {author} {\bibfnamefont {T.}~\bibnamefont {Koike}},\ and\ \bibinfo
  {author} {\bibfnamefont {Y.}~\bibnamefont {Okudaira}},\ }\bibfield  {title}
  {\bibinfo {title} {{Time-optimal unitary operations}},\ }\href
  {https://doi.org/10.1103/PhysRevA.75.042308} {\bibfield  {journal} {\bibinfo
  {journal} {Phys. Rev. A - At. Mol. Opt. Phys.}\ }\textbf {\bibinfo {volume}
  {75}},\ \bibinfo {pages} {1} (\bibinfo {year} {2007})},\ \Eprint
  {https://arxiv.org/abs/0608039} {arXiv:0608039 [quant-ph]} \BibitemShut
  {NoStop}%
\bibitem [{\citenamefont {Deffner}\ and\ \citenamefont
  {Lutz}(2013)}]{Deffner2013}%
  \BibitemOpen
  \bibfield  {author} {\bibinfo {author} {\bibfnamefont {S.}~\bibnamefont
  {Deffner}}\ and\ \bibinfo {author} {\bibfnamefont {E.}~\bibnamefont {Lutz}},\
  }\bibfield  {title} {\bibinfo {title} {{Quantum speed limit for non-Markovian
  dynamics}},\ }\href {https://doi.org/10.1103/PhysRevLett.111.010402}
  {\bibfield  {journal} {\bibinfo  {journal} {Phys. Rev. Lett.}\ }\textbf
  {\bibinfo {volume} {111}},\ \bibinfo {pages} {1} (\bibinfo {year} {2013})},\
  \Eprint {https://arxiv.org/abs/1302.5069} {arXiv:1302.5069} \BibitemShut
  {NoStop}%
\bibitem [{\citenamefont {{Del Campo}}\ \emph {et~al.}(2013)\citenamefont {{Del
  Campo}}, \citenamefont {Egusquiza}, \citenamefont {Plenio},\ and\
  \citenamefont {Huelga}}]{DelCampo2013}%
  \BibitemOpen
  \bibfield  {author} {\bibinfo {author} {\bibfnamefont {A.}~\bibnamefont {{Del
  Campo}}}, \bibinfo {author} {\bibfnamefont {I.~L.}\ \bibnamefont
  {Egusquiza}}, \bibinfo {author} {\bibfnamefont {M.~B.}\ \bibnamefont
  {Plenio}},\ and\ \bibinfo {author} {\bibfnamefont {S.~F.}\ \bibnamefont
  {Huelga}},\ }\bibfield  {title} {\bibinfo {title} {{Quantum speed limits in
  open system dynamics}},\ }\bibfield  {journal} {\bibinfo  {journal} {Phys.
  Rev. Lett.}\ }\textbf {\bibinfo {volume} {110}},\ \href
  {https://doi.org/10.1103/PhysRevLett.110.050403}
  {10.1103/PhysRevLett.110.050403} (\bibinfo {year} {2013}),\ \Eprint
  {https://arxiv.org/abs/1209.1737} {arXiv:1209.1737} \BibitemShut {NoStop}%
\bibitem [{\citenamefont {Poggi}\ \emph {et~al.}(2013)\citenamefont {Poggi},
  \citenamefont {Lombardo},\ and\ \citenamefont {Wisniacki}}]{Poggi2013}%
  \BibitemOpen
  \bibfield  {author} {\bibinfo {author} {\bibfnamefont {P.~M.}\ \bibnamefont
  {Poggi}}, \bibinfo {author} {\bibfnamefont {F.~C.}\ \bibnamefont
  {Lombardo}},\ and\ \bibinfo {author} {\bibfnamefont {D.~A.}\ \bibnamefont
  {Wisniacki}},\ }\bibfield  {title} {\bibinfo {title} {{Quantum speed limit
  and optimal evolution time in a two-level system}},\ }\href@noop {}
  {\bibfield  {journal} {\bibinfo  {journal} {EPL (Europhysics Lett.}\ }\textbf
  {\bibinfo {volume} {104}},\ \bibinfo {pages} {40005} (\bibinfo {year}
  {2013})}\BibitemShut {NoStop}%
\bibitem [{\citenamefont {Hegerfeldt}(2013)}]{Hegerfeldt2013}%
  \BibitemOpen
  \bibfield  {author} {\bibinfo {author} {\bibfnamefont {G.~C.}\ \bibnamefont
  {Hegerfeldt}},\ }\bibfield  {title} {\bibinfo {title} {{Driving at the
  quantum speed limit: Optimal control of a two-level system}},\ }\href
  {https://doi.org/10.1103/PhysRevLett.111.260501} {\bibfield  {journal}
  {\bibinfo  {journal} {Phys. Rev. Lett.}\ }\textbf {\bibinfo {volume} {111}},\
  \bibinfo {pages} {1} (\bibinfo {year} {2013})},\ \Eprint
  {https://arxiv.org/abs/1305.6403} {arXiv:1305.6403} \BibitemShut {NoStop}%
\bibitem [{\citenamefont {Hegerfeldt}(2014)}]{Hegerfeldt2014}%
  \BibitemOpen
  \bibfield  {author} {\bibinfo {author} {\bibfnamefont {G.~C.}\ \bibnamefont
  {Hegerfeldt}},\ }\bibfield  {title} {\bibinfo {title} {{High-speed driving of
  a two-level system}},\ }\href {https://doi.org/10.1103/PhysRevA.90.032110}
  {\bibfield  {journal} {\bibinfo  {journal} {Phys. Rev. A - At. Mol. Opt.
  Phys.}\ }\textbf {\bibinfo {volume} {90}},\ \bibinfo {pages} {1} (\bibinfo
  {year} {2014})},\ \Eprint {https://arxiv.org/abs/1407.6502} {arXiv:1407.6502}
  \BibitemShut {NoStop}%
\bibitem [{\citenamefont {Russell}\ and\ \citenamefont
  {Stepney}(2014)}]{Russell2014}%
  \BibitemOpen
  \bibfield  {author} {\bibinfo {author} {\bibfnamefont {B.}~\bibnamefont
  {Russell}}\ and\ \bibinfo {author} {\bibfnamefont {S.}~\bibnamefont
  {Stepney}},\ }\bibfield  {title} {\bibinfo {title} {{Zermelo navigation and a
  speed limit to quantum information processing}},\ }\href
  {https://doi.org/10.1103/PhysRevA.90.012303} {\bibfield  {journal} {\bibinfo
  {journal} {Phys. Rev. A - At. Mol. Opt. Phys.}\ }\textbf {\bibinfo {volume}
  {90}},\ \bibinfo {pages} {1} (\bibinfo {year} {2014})},\ \Eprint
  {https://arxiv.org/abs/1310.6731} {arXiv:1310.6731} \BibitemShut {NoStop}%
\bibitem [{\citenamefont {Brody}\ and\ \citenamefont
  {Meier}(2015)}]{Brody2015}%
  \BibitemOpen
  \bibfield  {author} {\bibinfo {author} {\bibfnamefont {D.~C.}\ \bibnamefont
  {Brody}}\ and\ \bibinfo {author} {\bibfnamefont {D.~M.}\ \bibnamefont
  {Meier}},\ }\bibfield  {title} {\bibinfo {title} {{Solution to the quantum
  Zermelo navigation problem}},\ }\href
  {https://doi.org/10.1103/PhysRevLett.114.100502} {\bibfield  {journal}
  {\bibinfo  {journal} {Phys. Rev. Lett.}\ }\textbf {\bibinfo {volume} {114}},\
  \bibinfo {pages} {1} (\bibinfo {year} {2015})},\ \Eprint
  {https://arxiv.org/abs/1409.3204} {arXiv:1409.3204} \BibitemShut {NoStop}%
\bibitem [{\citenamefont {Russell}\ and\ \citenamefont
  {Stepney}(2015)}]{Russell2015}%
  \BibitemOpen
  \bibfield  {author} {\bibinfo {author} {\bibfnamefont {B.}~\bibnamefont
  {Russell}}\ and\ \bibinfo {author} {\bibfnamefont {S.}~\bibnamefont
  {Stepney}},\ }\bibfield  {title} {\bibinfo {title} {{Zermelo navigation in
  the quantum brachistochrone}},\ }\href@noop {} {\bibfield  {journal}
  {\bibinfo  {journal} {J. Phys. A Math. Theor.}\ }\textbf {\bibinfo {volume}
  {48}},\ \bibinfo {pages} {115303} (\bibinfo {year} {2015})}\BibitemShut
  {NoStop}%
\bibitem [{\citenamefont {Frey}(2016)}]{Frey2016}%
  \BibitemOpen
  \bibfield  {author} {\bibinfo {author} {\bibfnamefont {M.~R.}\ \bibnamefont
  {Frey}},\ }\bibfield  {title} {\bibinfo {title} {{Quantum speed
  limits---primer, perspectives, and potential future directions}},\ }\href
  {https://doi.org/10.1007/s11128-016-1405-x} {\bibfield  {journal} {\bibinfo
  {journal} {Quantum Inf. Process.}\ }\textbf {\bibinfo {volume} {15}},\
  \bibinfo {pages} {3919} (\bibinfo {year} {2016})}\BibitemShut {NoStop}%
\bibitem [{\citenamefont {Deffner}\ and\ \citenamefont
  {Campbell}(2017)}]{Deffner2017}%
  \BibitemOpen
  \bibfield  {author} {\bibinfo {author} {\bibfnamefont {S.}~\bibnamefont
  {Deffner}}\ and\ \bibinfo {author} {\bibfnamefont {S.}~\bibnamefont
  {Campbell}},\ }\bibfield  {title} {\bibinfo {title} {{Quantum speed limits:
  From Heisenberg's uncertainty principle to optimal quantum control}},\ }\href
  {https://doi.org/10.1088/1751-8121/aa86c6} {\bibfield  {journal} {\bibinfo
  {journal} {J. Phys. A Math. Theor.}\ }\textbf {\bibinfo {volume} {50}},\
  \bibinfo {pages} {453001} (\bibinfo {year} {2017})}\BibitemShut {NoStop}%
\bibitem [{\citenamefont {Arenz}\ \emph {et~al.}(2017)\citenamefont {Arenz},
  \citenamefont {Russell}, \citenamefont {Burgarth},\ and\ \citenamefont
  {Rabitz}}]{Arenz2017}%
  \BibitemOpen
  \bibfield  {author} {\bibinfo {author} {\bibfnamefont {C.}~\bibnamefont
  {Arenz}}, \bibinfo {author} {\bibfnamefont {B.}~\bibnamefont {Russell}},
  \bibinfo {author} {\bibfnamefont {D.}~\bibnamefont {Burgarth}},\ and\
  \bibinfo {author} {\bibfnamefont {H.}~\bibnamefont {Rabitz}},\ }\bibfield
  {title} {\bibinfo {title} {{The roles of drift and control field constraints
  upon quantum control speed limits}},\ }\href
  {https://doi.org/10.1088/1367-2630/aa8242} {\bibfield  {journal} {\bibinfo
  {journal} {New J. Phys.}\ }\textbf {\bibinfo {volume} {19}},\ \bibinfo
  {pages} {103015} (\bibinfo {year} {2017})}\BibitemShut {NoStop}%
\bibitem [{\citenamefont {Lee}\ \emph {et~al.}(2018)\citenamefont {Lee},
  \citenamefont {Arenz}, \citenamefont {Rabitz},\ and\ \citenamefont
  {Russell}}]{Lee2018}%
  \BibitemOpen
  \bibfield  {author} {\bibinfo {author} {\bibfnamefont {J.}~\bibnamefont
  {Lee}}, \bibinfo {author} {\bibfnamefont {C.}~\bibnamefont {Arenz}}, \bibinfo
  {author} {\bibfnamefont {H.}~\bibnamefont {Rabitz}},\ and\ \bibinfo {author}
  {\bibfnamefont {B.}~\bibnamefont {Russell}},\ }\bibfield  {title} {\bibinfo
  {title} {{Dependence of the quantum speed limit on system size and control
  complexity}},\ }\href@noop {} {\bibfield  {journal} {\bibinfo  {journal} {New
  J. Phys.}\ }\textbf {\bibinfo {volume} {20}},\ \bibinfo {pages} {063002}
  (\bibinfo {year} {2018})}\BibitemShut {NoStop}%
\bibitem [{\citenamefont {Burgarth}\ \emph {et~al.}(2020)\citenamefont
  {Burgarth}, \citenamefont {Borggaard},\ and\ \citenamefont
  {Zimbor{\'a}s}}]{burgarth2020quantum}%
  \BibitemOpen
  \bibfield  {author} {\bibinfo {author} {\bibfnamefont {D.}~\bibnamefont
  {Burgarth}}, \bibinfo {author} {\bibfnamefont {J.}~\bibnamefont
  {Borggaard}},\ and\ \bibinfo {author} {\bibfnamefont {Z.}~\bibnamefont
  {Zimbor{\'a}s}},\ }\bibfield  {title} {\bibinfo {title} {Quantum distance to
  uncontrollability and quantum speed limits},\ }\href
  {https://arxiv.org/abs/2010.16156} {\bibfield  {journal} {\bibinfo  {journal}
  {arXiv preprint arXiv:2010.16156}\ } (\bibinfo {year} {2020})}\BibitemShut
  {NoStop}%
\bibitem [{\citenamefont {Knowles}(1981)}]{Knowles1981}%
  \BibitemOpen
  \bibfield  {author} {\bibinfo {author} {\bibfnamefont {G.}~\bibnamefont
  {Knowles}},\ }\href@noop {} {\emph {\bibinfo {title} {{An Introduction to
  Applied Optimal Control}}}}\ (\bibinfo  {publisher} {Academic Press, Inc.},\
  \bibinfo {address} {New York, NY},\ \bibinfo {year} {1981})\BibitemShut
  {NoStop}%
\bibitem [{\citenamefont {Englert}(2006)}]{Englert2006}%
  \BibitemOpen
  \bibfield  {author} {\bibinfo {author} {\bibfnamefont {B.-G.}\ \bibnamefont
  {Englert}},\ }\href@noop {} {\emph {\bibinfo {title} {{Lectures On Quantum
  Mechanics-Volume 3: Perturbed Evolution}}}}\ (\bibinfo  {publisher} {World
  Scientific Publishing Company},\ \bibinfo {year} {2006})\BibitemShut
  {NoStop}%
\bibitem [{\citenamefont {Sakurai}(1994)}]{Sakurai1994}%
  \BibitemOpen
  \bibfield  {author} {\bibinfo {author} {\bibfnamefont {J.~J.}\ \bibnamefont
  {Sakurai}},\ }\href@noop {} {\emph {\bibinfo {title} {{Modern Quantum
  Mechanics}}}},\ edited by\ \bibinfo {editor} {\bibfnamefont {S.~F.}\
  \bibnamefont {Tuan}}\ (\bibinfo  {publisher} {Addison-Wesley Publishing
  Co.},\ \bibinfo {address} {Reading, MA},\ \bibinfo {year} {1994})\BibitemShut
  {NoStop}%
\bibitem [{\citenamefont {Jackson}(1999)}]{Jackson1999}%
  \BibitemOpen
  \bibfield  {author} {\bibinfo {author} {\bibfnamefont {J.~D.}\ \bibnamefont
  {Jackson}},\ }\href@noop {} {\emph {\bibinfo {title} {{Classical
  Electrodynamics, 3rd Ed.}}}}\ (\bibinfo  {publisher} {John Wiley {\&} Sons},\
  \bibinfo {year} {1999})\BibitemShut {NoStop}%
\bibitem [{\citenamefont {Angeris}\ \emph {et~al.}(2021)\citenamefont
  {Angeris}, \citenamefont {Vu{\v{c}}kovi{\'{c}}},\ and\ \citenamefont
  {Boyd}}]{Angeris2021}%
  \BibitemOpen
  \bibfield  {author} {\bibinfo {author} {\bibfnamefont {G.}~\bibnamefont
  {Angeris}}, \bibinfo {author} {\bibfnamefont {J.}~\bibnamefont
  {Vu{\v{c}}kovi{\'{c}}}},\ and\ \bibinfo {author} {\bibfnamefont
  {S.}~\bibnamefont {Boyd}},\ }\bibfield  {title} {\bibinfo {title} {{Heuristic
  methods and performance bounds for photonic design}},\ }\href
  {https://doi.org/10.1364/OE.415052} {\bibfield  {journal} {\bibinfo
  {journal} {Opt. Express}\ }\textbf {\bibinfo {volume} {29}},\ \bibinfo
  {pages} {2827} (\bibinfo {year} {2021})},\ \Eprint
  {https://arxiv.org/abs/2011.08002} {arXiv:2011.08002} \BibitemShut {NoStop}%
\bibitem [{\citenamefont {Thijssen}(2012)}]{Thijssen2012}%
  \BibitemOpen
  \bibfield  {author} {\bibinfo {author} {\bibfnamefont {J.}~\bibnamefont
  {Thijssen}},\ }\href@noop {} {\emph {\bibinfo {title} {{Computational
  Physics}}}}\ (\bibinfo  {publisher} {Cambridge University Press},\ \bibinfo
  {year} {2012})\BibitemShut {NoStop}%
\bibitem [{\citenamefont {Vandenberghe}\ and\ \citenamefont
  {Boyd}(1996)}]{Vandenberghe1996}%
  \BibitemOpen
  \bibfield  {author} {\bibinfo {author} {\bibfnamefont {L.}~\bibnamefont
  {Vandenberghe}}\ and\ \bibinfo {author} {\bibfnamefont {S.}~\bibnamefont
  {Boyd}},\ }\bibfield  {title} {\bibinfo {title} {{Semidefinite
  Programming}},\ }\href {https://doi.org/10.1137/1038003} {\bibfield
  {journal} {\bibinfo  {journal} {SIAM Rev.}\ }\textbf {\bibinfo {volume}
  {38}},\ \bibinfo {pages} {49} (\bibinfo {year} {1996})}\BibitemShut {NoStop}%
\bibitem [{\citenamefont {Krantz}\ \emph {et~al.}(2019)\citenamefont {Krantz},
  \citenamefont {Kjaergaard}, \citenamefont {Yan}, \citenamefont {Orlando},
  \citenamefont {Gustavsson},\ and\ \citenamefont
  {Oliver}}]{krantz2019quantum}%
  \BibitemOpen
  \bibfield  {author} {\bibinfo {author} {\bibfnamefont {P.}~\bibnamefont
  {Krantz}}, \bibinfo {author} {\bibfnamefont {M.}~\bibnamefont {Kjaergaard}},
  \bibinfo {author} {\bibfnamefont {F.}~\bibnamefont {Yan}}, \bibinfo {author}
  {\bibfnamefont {T.~P.}\ \bibnamefont {Orlando}}, \bibinfo {author}
  {\bibfnamefont {S.}~\bibnamefont {Gustavsson}},\ and\ \bibinfo {author}
  {\bibfnamefont {W.~D.}\ \bibnamefont {Oliver}},\ }\bibfield  {title}
  {\bibinfo {title} {A quantum engineer's guide to superconducting qubits},\
  }\href {https://doi.org/10.1063/1.5089550} {\bibfield  {journal} {\bibinfo
  {journal} {Applied Physics Reviews}\ }\textbf {\bibinfo {volume} {6}},\
  \bibinfo {pages} {021318} (\bibinfo {year} {2019})}\BibitemShut {NoStop}%
\bibitem [{\citenamefont {Wood}\ and\ \citenamefont
  {Gambetta}(2018)}]{Wood2018}%
  \BibitemOpen
  \bibfield  {author} {\bibinfo {author} {\bibfnamefont {C.~J.}\ \bibnamefont
  {Wood}}\ and\ \bibinfo {author} {\bibfnamefont {J.~M.}\ \bibnamefont
  {Gambetta}},\ }\bibfield  {title} {\bibinfo {title} {{Quantification and
  characterization of leakage errors}},\ }\href
  {https://doi.org/10.1103/PhysRevA.97.032306} {\bibfield  {journal} {\bibinfo
  {journal} {Phys. Rev. A}\ }\textbf {\bibinfo {volume} {97}},\ \bibinfo
  {pages} {032306} (\bibinfo {year} {2018})},\ \Eprint
  {https://arxiv.org/abs/1704.03081} {arXiv:1704.03081} \BibitemShut {NoStop}%
\bibitem [{\citenamefont {Viola}\ \emph {et~al.}(1999)\citenamefont {Viola},
  \citenamefont {Knill},\ and\ \citenamefont {Lloyd}}]{viola1999dynamical}%
  \BibitemOpen
  \bibfield  {author} {\bibinfo {author} {\bibfnamefont {L.}~\bibnamefont
  {Viola}}, \bibinfo {author} {\bibfnamefont {E.}~\bibnamefont {Knill}},\ and\
  \bibinfo {author} {\bibfnamefont {S.}~\bibnamefont {Lloyd}},\ }\bibfield
  {title} {\bibinfo {title} {Dynamical decoupling of open quantum systems},\
  }\href {https://doi.org/10.1103/PhysRevLett.82.2417} {\bibfield  {journal}
  {\bibinfo  {journal} {Physical Review Letters}\ }\textbf {\bibinfo {volume}
  {82}},\ \bibinfo {pages} {2417} (\bibinfo {year} {1999})}\BibitemShut
  {NoStop}%
\bibitem [{\citenamefont {Uhrig}(2009)}]{uhrig2009concatenated}%
  \BibitemOpen
  \bibfield  {author} {\bibinfo {author} {\bibfnamefont {G.~S.}\ \bibnamefont
  {Uhrig}},\ }\bibfield  {title} {\bibinfo {title} {Concatenated control
  sequences based on optimized dynamic decoupling},\ }\href
  {https://doi.org/10.1103/PhysRevLett.102.120502} {\bibfield  {journal}
  {\bibinfo  {journal} {Physical Review Letters}\ }\textbf {\bibinfo {volume}
  {102}},\ \bibinfo {pages} {120502} (\bibinfo {year} {2009})}\BibitemShut
  {NoStop}%
\bibitem [{\citenamefont {West}\ \emph {et~al.}(2010)\citenamefont {West},
  \citenamefont {Lidar}, \citenamefont {Fong},\ and\ \citenamefont
  {Gyure}}]{west2010high}%
  \BibitemOpen
  \bibfield  {author} {\bibinfo {author} {\bibfnamefont {J.~R.}\ \bibnamefont
  {West}}, \bibinfo {author} {\bibfnamefont {D.~A.}\ \bibnamefont {Lidar}},
  \bibinfo {author} {\bibfnamefont {B.~H.}\ \bibnamefont {Fong}},\ and\
  \bibinfo {author} {\bibfnamefont {M.~F.}\ \bibnamefont {Gyure}},\ }\bibfield
  {title} {\bibinfo {title} {High fidelity quantum gates via dynamical
  decoupling},\ }\href {https://doi.org/10.1103/PhysRevLett.105.230503}
  {\bibfield  {journal} {\bibinfo  {journal} {Physical Review Letters}\
  }\textbf {\bibinfo {volume} {105}},\ \bibinfo {pages} {230503} (\bibinfo
  {year} {2010})}\BibitemShut {NoStop}%
\bibitem [{\citenamefont {Souza}\ \emph {et~al.}(2011)\citenamefont {Souza},
  \citenamefont {Alvarez},\ and\ \citenamefont {Suter}}]{souza2011robust}%
  \BibitemOpen
  \bibfield  {author} {\bibinfo {author} {\bibfnamefont {A.~M.}\ \bibnamefont
  {Souza}}, \bibinfo {author} {\bibfnamefont {G.~A.}\ \bibnamefont {Alvarez}},\
  and\ \bibinfo {author} {\bibfnamefont {D.}~\bibnamefont {Suter}},\ }\bibfield
   {title} {\bibinfo {title} {Robust dynamical decoupling for quantum computing
  and quantum memory},\ }\href {https://doi.org/10.1103/PhysRevLett.106.240501}
  {\bibfield  {journal} {\bibinfo  {journal} {Physical Review Letters}\
  }\textbf {\bibinfo {volume} {106}},\ \bibinfo {pages} {240501} (\bibinfo
  {year} {2011})}\BibitemShut {NoStop}%
\bibitem [{\citenamefont {Rossini}\ \emph {et~al.}(2008)\citenamefont
  {Rossini}, \citenamefont {Facchi}, \citenamefont {Fazio}, \citenamefont
  {Florio}, \citenamefont {Lidar}, \citenamefont {Pascazio}, \citenamefont
  {Plastina},\ and\ \citenamefont {Zanardi}}]{rossini2008bang}%
  \BibitemOpen
  \bibfield  {author} {\bibinfo {author} {\bibfnamefont {D.}~\bibnamefont
  {Rossini}}, \bibinfo {author} {\bibfnamefont {P.}~\bibnamefont {Facchi}},
  \bibinfo {author} {\bibfnamefont {R.}~\bibnamefont {Fazio}}, \bibinfo
  {author} {\bibfnamefont {G.}~\bibnamefont {Florio}}, \bibinfo {author}
  {\bibfnamefont {D.~A.}\ \bibnamefont {Lidar}}, \bibinfo {author}
  {\bibfnamefont {S.}~\bibnamefont {Pascazio}}, \bibinfo {author}
  {\bibfnamefont {F.}~\bibnamefont {Plastina}},\ and\ \bibinfo {author}
  {\bibfnamefont {P.}~\bibnamefont {Zanardi}},\ }\bibfield  {title} {\bibinfo
  {title} {Bang-bang control of a qubit coupled to a quantum critical spin
  bath},\ }\href {https://doi.org/10.1103/PhysRevA.77.052112} {\bibfield
  {journal} {\bibinfo  {journal} {Physical Review A}\ }\textbf {\bibinfo
  {volume} {77}},\ \bibinfo {pages} {052112} (\bibinfo {year}
  {2008})}\BibitemShut {NoStop}%
\bibitem [{\citenamefont {Kosloff}(2019)}]{kosloff2019quantum}%
  \BibitemOpen
  \bibfield  {author} {\bibinfo {author} {\bibfnamefont {R.}~\bibnamefont
  {Kosloff}},\ }\bibfield  {title} {\bibinfo {title} {Quantum thermodynamics
  and open-systems modeling},\ }\href {https://doi.org/10.1063/1.5096173}
  {\bibfield  {journal} {\bibinfo  {journal} {The Journal of Chemical Physics}\
  }\textbf {\bibinfo {volume} {150}},\ \bibinfo {pages} {204105} (\bibinfo
  {year} {2019})}\BibitemShut {NoStop}%
\bibitem [{\citenamefont {Khodjasteh}\ and\ \citenamefont
  {Lidar}(2005)}]{khodjasteh2005fault}%
  \BibitemOpen
  \bibfield  {author} {\bibinfo {author} {\bibfnamefont {K.}~\bibnamefont
  {Khodjasteh}}\ and\ \bibinfo {author} {\bibfnamefont {D.~A.}\ \bibnamefont
  {Lidar}},\ }\bibfield  {title} {\bibinfo {title} {Fault-tolerant quantum
  dynamical decoupling},\ }\href
  {https://doi.org/10.1103/PhysRevLett.95.180501} {\bibfield  {journal}
  {\bibinfo  {journal} {Physical Review Letters}\ }\textbf {\bibinfo {volume}
  {95}},\ \bibinfo {pages} {180501} (\bibinfo {year} {2005})}\BibitemShut
  {NoStop}%
\bibitem [{\citenamefont {Viola}\ and\ \citenamefont
  {Knill}(2003)}]{viola2003robust}%
  \BibitemOpen
  \bibfield  {author} {\bibinfo {author} {\bibfnamefont {L.}~\bibnamefont
  {Viola}}\ and\ \bibinfo {author} {\bibfnamefont {E.}~\bibnamefont {Knill}},\
  }\bibfield  {title} {\bibinfo {title} {Robust dynamical decoupling of quantum
  systems with bounded controls},\ }\href
  {https://doi.org/10.1103/PhysRevLett.90.037901} {\bibfield  {journal}
  {\bibinfo  {journal} {Physical Review Letters}\ }\textbf {\bibinfo {volume}
  {90}},\ \bibinfo {pages} {037901} (\bibinfo {year} {2003})}\BibitemShut
  {NoStop}%
\bibitem [{\citenamefont {Miller}\ \emph {et~al.}(2016)\citenamefont {Miller},
  \citenamefont {Polimeridis}, \citenamefont {Reid}, \citenamefont {Hsu},
  \citenamefont {Delacy}, \citenamefont {Joannopoulos}, \citenamefont {Solja{\v
  c}i\`c},\ and\ \citenamefont {Johnson}}]{Miller2016}%
  \BibitemOpen
  \bibfield  {author} {\bibinfo {author} {\bibfnamefont {O.~D.}\ \bibnamefont
  {Miller}}, \bibinfo {author} {\bibfnamefont {A.~G.}\ \bibnamefont
  {Polimeridis}}, \bibinfo {author} {\bibfnamefont {M.~T.~H.}\ \bibnamefont
  {Reid}}, \bibinfo {author} {\bibfnamefont {C.~W.}\ \bibnamefont {Hsu}},
  \bibinfo {author} {\bibfnamefont {B.~G.}\ \bibnamefont {Delacy}}, \bibinfo
  {author} {\bibfnamefont {J.~D.}\ \bibnamefont {Joannopoulos}}, \bibinfo
  {author} {\bibfnamefont {M.}~\bibnamefont {Solja{\v c}i\`c}},\ and\ \bibinfo
  {author} {\bibfnamefont {S.~G.}\ \bibnamefont {Johnson}},\ }\bibfield
  {title} {\bibinfo {title} {Fundamental limits to optical response in
  absorptive systems},\ }\href {https://doi.org/10.1364/oe.24.003329}
  {\bibfield  {journal} {\bibinfo  {journal} {Optics Express}\ }\textbf
  {\bibinfo {volume} {24}},\ \bibinfo {pages} {3329} (\bibinfo {year}
  {2016})}\BibitemShut {NoStop}%
\bibitem [{\citenamefont {Yang}\ \emph {et~al.}(2017)\citenamefont {Yang},
  \citenamefont {Miller}, \citenamefont {Christensen}, \citenamefont
  {Joannopoulos},\ and\ \citenamefont {Solja{\v c}i\`c}}]{Yang2017}%
  \BibitemOpen
  \bibfield  {author} {\bibinfo {author} {\bibfnamefont {Y.}~\bibnamefont
  {Yang}}, \bibinfo {author} {\bibfnamefont {O.~D.}\ \bibnamefont {Miller}},
  \bibinfo {author} {\bibfnamefont {T.}~\bibnamefont {Christensen}}, \bibinfo
  {author} {\bibfnamefont {J.~D.}\ \bibnamefont {Joannopoulos}},\ and\ \bibinfo
  {author} {\bibfnamefont {M.}~\bibnamefont {Solja{\v c}i\`c}},\ }\bibfield
  {title} {\bibinfo {title} {Low-loss plasmonic dielectric nanoresonators},\
  }\href {https://doi.org/10.1021/acs.nanolett.7b00852} {\bibfield  {journal}
  {\bibinfo  {journal} {Nano Letters}\ }\textbf {\bibinfo {volume} {17}},\
  \bibinfo {pages} {3238} (\bibinfo {year} {2017})}\BibitemShut {NoStop}%
\bibitem [{\citenamefont {Shim}\ \emph {et~al.}(2019)\citenamefont {Shim},
  \citenamefont {Fan}, \citenamefont {Johnson},\ and\ \citenamefont
  {Miller}}]{Shim2019}%
  \BibitemOpen
  \bibfield  {author} {\bibinfo {author} {\bibfnamefont {H.}~\bibnamefont
  {Shim}}, \bibinfo {author} {\bibfnamefont {L.}~\bibnamefont {Fan}}, \bibinfo
  {author} {\bibfnamefont {S.~G.}\ \bibnamefont {Johnson}},\ and\ \bibinfo
  {author} {\bibfnamefont {O.~D.}\ \bibnamefont {Miller}},\ }\bibfield  {title}
  {\bibinfo {title} {{Fundamental Limits to Near-Field Optical Response over
  Any Bandwidth}},\ }\href {https://doi.org/10.1103/PhysRevX.9.011043}
  {\bibfield  {journal} {\bibinfo  {journal} {Phys. Rev. X}\ }\textbf {\bibinfo
  {volume} {9}},\ \bibinfo {pages} {11043} (\bibinfo {year}
  {2019})}\BibitemShut {NoStop}%
\bibitem [{\citenamefont {Molesky}\ \emph {et~al.}(2019)\citenamefont
  {Molesky}, \citenamefont {Jin}, \citenamefont {Venkataram},\ and\
  \citenamefont {Rodriguez}}]{Molesky2019a}%
  \BibitemOpen
  \bibfield  {author} {\bibinfo {author} {\bibfnamefont {S.}~\bibnamefont
  {Molesky}}, \bibinfo {author} {\bibfnamefont {W.}~\bibnamefont {Jin}},
  \bibinfo {author} {\bibfnamefont {P.~S.}\ \bibnamefont {Venkataram}},\ and\
  \bibinfo {author} {\bibfnamefont {A.~W.}\ \bibnamefont {Rodriguez}},\
  }\bibfield  {title} {\bibinfo {title} {{T Operator Bounds on Angle-Integrated
  Absorption and Thermal Radiation for Arbitrary Objects}},\ }\href
  {https://doi.org/10.1103/PhysRevLett.123.257401} {\bibfield  {journal}
  {\bibinfo  {journal} {Phys. Rev. Lett.}\ }\textbf {\bibinfo {volume} {123}},\
  \bibinfo {pages} {257401} (\bibinfo {year} {2019})}\BibitemShut {NoStop}%
\bibitem [{\citenamefont {Ivanenko}\ \emph {et~al.}(2019)\citenamefont
  {Ivanenko}, \citenamefont {Gustafsson},\ and\ \citenamefont
  {Nordebo}}]{Ivanenko2019a}%
  \BibitemOpen
  \bibfield  {author} {\bibinfo {author} {\bibfnamefont {Y.}~\bibnamefont
  {Ivanenko}}, \bibinfo {author} {\bibfnamefont {M.}~\bibnamefont
  {Gustafsson}},\ and\ \bibinfo {author} {\bibfnamefont {S.}~\bibnamefont
  {Nordebo}},\ }\bibfield  {title} {\bibinfo {title} {{Optical theorems and
  physical bounds on absorption in lossy media}},\ }\href
  {https://doi.org/10.1364/oe.27.034323} {\bibfield  {journal} {\bibinfo
  {journal} {Opt. Express}\ }\textbf {\bibinfo {volume} {27}},\ \bibinfo
  {pages} {34323} (\bibinfo {year} {2019})},\ \Eprint
  {https://arxiv.org/abs/1908.09657} {arXiv:1908.09657} \BibitemShut {NoStop}%
\bibitem [{\citenamefont {Shim}\ \emph
  {et~al.}(2020{\natexlab{a}})\citenamefont {Shim}, \citenamefont {Chung},\
  and\ \citenamefont {Miller}}]{ShimChung2020}%
  \BibitemOpen
  \bibfield  {author} {\bibinfo {author} {\bibfnamefont {H.}~\bibnamefont
  {Shim}}, \bibinfo {author} {\bibfnamefont {H.}~\bibnamefont {Chung}},\ and\
  \bibinfo {author} {\bibfnamefont {O.~D.}\ \bibnamefont {Miller}},\ }\bibfield
   {title} {\bibinfo {title} {Maximal free-space concentration of
  electromagnetic waves},\ }\href
  {https://doi.org/10.1103/PhysRevApplied.14.014007} {\bibfield  {journal}
  {\bibinfo  {journal} {Physical Review Applied}\ }\textbf {\bibinfo {volume}
  {14}},\ \bibinfo {pages} {014007} (\bibinfo {year} {2020}{\natexlab{a}})},\
  \Eprint {https://arxiv.org/abs/1905.10500} {1905.10500} \BibitemShut
  {NoStop}%
\bibitem [{\citenamefont {Shim}\ \emph
  {et~al.}(2020{\natexlab{b}})\citenamefont {Shim}, \citenamefont {Kuang},\
  and\ \citenamefont {Miller}}]{Shim2020}%
  \BibitemOpen
  \bibfield  {author} {\bibinfo {author} {\bibfnamefont {H.}~\bibnamefont
  {Shim}}, \bibinfo {author} {\bibfnamefont {Z.}~\bibnamefont {Kuang}},\ and\
  \bibinfo {author} {\bibfnamefont {O.~D.}\ \bibnamefont {Miller}},\ }\bibfield
   {title} {\bibinfo {title} {Optical materials for maximal nanophotonic
  response ({I}nvited)},\ }\href {https://doi.org/10.1364/OME.396419}
  {\bibfield  {journal} {\bibinfo  {journal} {Optical Materials Express}\
  }\textbf {\bibinfo {volume} {10}},\ \bibinfo {pages} {1561} (\bibinfo {year}
  {2020}{\natexlab{b}})},\ \Eprint {https://arxiv.org/abs/2004.13132}
  {arXiv:2004.13132} \BibitemShut {NoStop}%
\bibitem [{\citenamefont {Molesky}\ \emph
  {et~al.}(2020{\natexlab{b}})\citenamefont {Molesky}, \citenamefont {Chao},
  \citenamefont {Jin},\ and\ \citenamefont {Rodriguez}}]{Molesky2020a}%
  \BibitemOpen
  \bibfield  {author} {\bibinfo {author} {\bibfnamefont {S.}~\bibnamefont
  {Molesky}}, \bibinfo {author} {\bibfnamefont {P.}~\bibnamefont {Chao}},
  \bibinfo {author} {\bibfnamefont {W.}~\bibnamefont {Jin}},\ and\ \bibinfo
  {author} {\bibfnamefont {A.~W.}\ \bibnamefont {Rodriguez}},\ }\bibfield
  {title} {\bibinfo {title} {{Global T operator bounds on electromagnetic
  scattering: Upper bounds on far-field cross sections}},\ }\href
  {https://doi.org/10.1103/physrevresearch.2.033172} {\bibfield  {journal}
  {\bibinfo  {journal} {Phys. Rev. Res.}\ }\textbf {\bibinfo {volume} {2}},\
  \bibinfo {pages} {033172} (\bibinfo {year} {2020}{\natexlab{b}})}\BibitemShut
  {NoStop}%
\bibitem [{\citenamefont {Gustafsson}\ \emph {et~al.}(2020)\citenamefont
  {Gustafsson}, \citenamefont {Schab}, \citenamefont {Jelinek},\ and\
  \citenamefont {Capek}}]{Gustafsson2020}%
  \BibitemOpen
  \bibfield  {author} {\bibinfo {author} {\bibfnamefont {M.}~\bibnamefont
  {Gustafsson}}, \bibinfo {author} {\bibfnamefont {K.}~\bibnamefont {Schab}},
  \bibinfo {author} {\bibfnamefont {L.}~\bibnamefont {Jelinek}},\ and\ \bibinfo
  {author} {\bibfnamefont {M.}~\bibnamefont {Capek}},\ }\bibfield  {title}
  {\bibinfo {title} {{Upper bounds on absorption and scattering}},\ }\href
  {https://doi.org/10.1088/1367-2630/ab83d3} {\bibfield  {journal} {\bibinfo
  {journal} {New J. Phys.}\ }\textbf {\bibinfo {volume} {22}},\ \bibinfo
  {pages} {073013} (\bibinfo {year} {2020})},\ \Eprint
  {https://arxiv.org/abs/1912.06699} {arXiv:1912.06699} \BibitemShut {NoStop}%
\bibitem [{\citenamefont {Molesky}\ \emph
  {et~al.}(2020{\natexlab{c}})\citenamefont {Molesky}, \citenamefont
  {Venkataram}, \citenamefont {Jin},\ and\ \citenamefont
  {Rodriguez}}]{Molesky2020b}%
  \BibitemOpen
  \bibfield  {author} {\bibinfo {author} {\bibfnamefont {S.}~\bibnamefont
  {Molesky}}, \bibinfo {author} {\bibfnamefont {P.~S.}\ \bibnamefont
  {Venkataram}}, \bibinfo {author} {\bibfnamefont {W.}~\bibnamefont {Jin}},\
  and\ \bibinfo {author} {\bibfnamefont {A.~W.}\ \bibnamefont {Rodriguez}},\
  }\bibfield  {title} {\bibinfo {title} {{Fundamental limits to radiative heat
  transfer: Theory}},\ }\href {https://doi.org/10.1103/PhysRevB.101.035408}
  {\bibfield  {journal} {\bibinfo  {journal} {Phys. Rev. B}\ }\textbf {\bibinfo
  {volume} {101}},\ \bibinfo {pages} {35408} (\bibinfo {year}
  {2020}{\natexlab{c}})},\ \Eprint {https://arxiv.org/abs/1907.03000}
  {arXiv:1907.03000} \BibitemShut {NoStop}%
\bibitem [{\citenamefont {Kuang}\ \emph {et~al.}(2020)\citenamefont {Kuang},
  \citenamefont {Zhang},\ and\ \citenamefont {Miller}}]{Kuang2020}%
  \BibitemOpen
  \bibfield  {author} {\bibinfo {author} {\bibfnamefont {Z.}~\bibnamefont
  {Kuang}}, \bibinfo {author} {\bibfnamefont {L.}~\bibnamefont {Zhang}},\ and\
  \bibinfo {author} {\bibfnamefont {O.~D.}\ \bibnamefont {Miller}},\ }\bibfield
   {title} {\bibinfo {title} {Maximal single-frequency electromagnetic
  response},\ }\href {https://doi.org/10.1364/OPTICA.398715} {\bibfield
  {journal} {\bibinfo  {journal} {Optica}\ }\textbf {\bibinfo {volume} {7}},\
  \bibinfo {pages} {1746} (\bibinfo {year} {2020})},\ \Eprint
  {https://arxiv.org/abs/2002.00521} {arXiv:2002.00521} \BibitemShut {NoStop}%
\bibitem [{\citenamefont {Ma}\ \emph {et~al.}(2008)\citenamefont {Ma},
  \citenamefont {Su}, \citenamefont {Jald{\'e}n},\ and\ \citenamefont
  {Chi}}]{ma2008some}%
  \BibitemOpen
  \bibfield  {author} {\bibinfo {author} {\bibfnamefont {W.-K.}\ \bibnamefont
  {Ma}}, \bibinfo {author} {\bibfnamefont {C.-C.}\ \bibnamefont {Su}}, \bibinfo
  {author} {\bibfnamefont {J.}~\bibnamefont {Jald{\'e}n}},\ and\ \bibinfo
  {author} {\bibfnamefont {C.-Y.}\ \bibnamefont {Chi}},\ }\bibfield  {title}
  {\bibinfo {title} {Some results on 16-qam mimo detection using semidefinite
  relaxation},\ }in\ \href {10.1109/ICASSP.2008.4518199} {\emph {\bibinfo
  {booktitle} {2008 IEEE International Conference on Acoustics, Speech and
  Signal Processing}}}\ (\bibinfo {organization} {IEEE},\ \bibinfo {year}
  {2008})\ pp.\ \bibinfo {pages} {2673--2676}\BibitemShut {NoStop}%
\bibitem [{\citenamefont {Greengard}\ and\ \citenamefont
  {Rokhlin}(1987)}]{greengard1987fast}%
  \BibitemOpen
  \bibfield  {author} {\bibinfo {author} {\bibfnamefont {L.}~\bibnamefont
  {Greengard}}\ and\ \bibinfo {author} {\bibfnamefont {V.}~\bibnamefont
  {Rokhlin}},\ }\bibfield  {title} {\bibinfo {title} {A fast algorithm for
  particle simulations},\ }\href {https://doi.org/10.1016/0021-9991(87)90140-9}
  {\bibfield  {journal} {\bibinfo  {journal} {Journal of Computational
  Physics}\ }\textbf {\bibinfo {volume} {73}},\ \bibinfo {pages} {325}
  (\bibinfo {year} {1987})}\BibitemShut {NoStop}%
\bibitem [{\citenamefont {Coifman}\ \emph {et~al.}(1993)\citenamefont
  {Coifman}, \citenamefont {Rokhlin},\ and\ \citenamefont
  {Wandzura}}]{coifman1993fast}%
  \BibitemOpen
  \bibfield  {author} {\bibinfo {author} {\bibfnamefont {R.}~\bibnamefont
  {Coifman}}, \bibinfo {author} {\bibfnamefont {V.}~\bibnamefont {Rokhlin}},\
  and\ \bibinfo {author} {\bibfnamefont {S.}~\bibnamefont {Wandzura}},\
  }\bibfield  {title} {\bibinfo {title} {The fast multipole method for the wave
  equation: A pedestrian prescription},\ }\href {10.1109/74.250128} {\bibfield
  {journal} {\bibinfo  {journal} {IEEE Antennas and Propagation magazine}\
  }\textbf {\bibinfo {volume} {35}},\ \bibinfo {pages} {7} (\bibinfo {year}
  {1993})}\BibitemShut {NoStop}%
\end{thebibliography}%


\begin{thebibliography}{11}%
\makeatletter
\providecommand \@ifxundefined [1]{%
 \@ifx{#1\undefined}
}%
\providecommand \@ifnum [1]{%
 \ifnum #1\expandafter \@firstoftwo
 \else \expandafter \@secondoftwo
 \fi
}%
\providecommand \@ifx [1]{%
 \ifx #1\expandafter \@firstoftwo
 \else \expandafter \@secondoftwo
 \fi
}%
\providecommand \natexlab [1]{#1}%
\providecommand \enquote  [1]{``#1''}%
\providecommand \bibnamefont  [1]{#1}%
\providecommand \bibfnamefont [1]{#1}%
\providecommand \citenamefont [1]{#1}%
\providecommand \href@noop [0]{\@secondoftwo}%
\providecommand \href [0]{\begingroup \@sanitize@url \@href}%
\providecommand \@href[1]{\@@startlink{#1}\@@href}%
\providecommand \@@href[1]{\endgroup#1\@@endlink}%
\providecommand \@sanitize@url [0]{\catcode `\\12\catcode `\$12\catcode
  `\&12\catcode `\#12\catcode `\^12\catcode `\_12\catcode `\%12\relax}%
\providecommand \@@startlink[1]{}%
\providecommand \@@endlink[0]{}%
\providecommand \url  [0]{\begingroup\@sanitize@url \@url }%
\providecommand \@url [1]{\endgroup\@href {#1}{\urlprefix }}%
\providecommand \urlprefix  [0]{URL }%
\providecommand \Eprint [0]{\href }%
\providecommand \doibase [0]{https://doi.org/}%
\providecommand \selectlanguage [0]{\@gobble}%
\providecommand \bibinfo  [0]{\@secondoftwo}%
\providecommand \bibfield  [0]{\@secondoftwo}%
\providecommand \translation [1]{[#1]}%
\providecommand \BibitemOpen [0]{}%
\providecommand \bibitemStop [0]{}%
\providecommand \bibitemNoStop [0]{.\EOS\space}%
\providecommand \EOS [0]{\spacefactor3000\relax}%
\providecommand \BibitemShut  [1]{\csname bibitem#1\endcsname}%
\let\auto@bib@innerbib\@empty
\bibitem [{\citenamefont {Angeris}\ \emph {et~al.}(2021)\citenamefont
  {Angeris}, \citenamefont {Vu{\v{c}}kovi{\'{c}}},\ and\ \citenamefont
  {Boyd}}]{Angeris2021}%
  \BibitemOpen
  \bibfield  {author} {\bibinfo {author} {\bibfnamefont {G.}~\bibnamefont
  {Angeris}}, \bibinfo {author} {\bibfnamefont {J.}~\bibnamefont
  {Vu{\v{c}}kovi{\'{c}}}},\ and\ \bibinfo {author} {\bibfnamefont
  {S.}~\bibnamefont {Boyd}},\ }\bibfield  {title} {\bibinfo {title} {{Heuristic
  methods and performance bounds for photonic design}},\ }\href
  {https://doi.org/10.1364/OE.415052} {\bibfield  {journal} {\bibinfo
  {journal} {Opt. Express}\ }\textbf {\bibinfo {volume} {29}},\ \bibinfo
  {pages} {2827} (\bibinfo {year} {2021})}\BibitemShut {NoStop}%
\bibitem [{\citenamefont {Werschnik}\ and\ \citenamefont
  {Gross}(2007)}]{Werschnik2007}%
  \BibitemOpen
  \bibfield  {author} {\bibinfo {author} {\bibfnamefont {J.}~\bibnamefont
  {Werschnik}}\ and\ \bibinfo {author} {\bibfnamefont {E.~K.}\ \bibnamefont
  {Gross}},\ }\bibfield  {title} {\bibinfo {title} {{Quantum optimal control
  theory}},\ }\bibfield  {journal} {\bibinfo  {journal} {J. Phys. B At. Mol.
  Opt. Phys.}\ }\textbf {\bibinfo {volume} {40}},\ \href
  {https://doi.org/10.1088/0953-4075/40/18/R01} {10.1088/0953-4075/40/18/R01}
  (\bibinfo {year} {2007})\BibitemShut {NoStop}%
\bibitem [{\citenamefont {Milton}(2002)}]{Milton2002}%
  \BibitemOpen
  \bibfield  {author} {\bibinfo {author} {\bibfnamefont {G.~W.}\ \bibnamefont
  {Milton}},\ }\href@noop {} {\emph {\bibinfo {title} {{The Theory of
  Composites}}}}\ (\bibinfo  {publisher} {Cambridge University Press},\
  \bibinfo {year} {2002})\BibitemShut {NoStop}%
\bibitem [{\citenamefont {Breuer}\ \emph {et~al.}(2002)\citenamefont {Breuer},
  \citenamefont {Petruccione} \emph {et~al.}}]{breuer2002theory}%
  \BibitemOpen
  \bibfield  {author} {\bibinfo {author} {\bibfnamefont {H.-P.}\ \bibnamefont
  {Breuer}}, \bibinfo {author} {\bibfnamefont {F.}~\bibnamefont {Petruccione}},
  \emph {et~al.},\ }\href@noop {} {\emph {\bibinfo {title} {The theory of open
  quantum systems}}}\ (\bibinfo  {publisher} {Oxford University Press on
  Demand},\ \bibinfo {year} {2002})\BibitemShut {NoStop}%
\bibitem [{\citenamefont {Horn}\ and\ \citenamefont
  {Johnson}(2012)}]{horn2012matrix}%
  \BibitemOpen
  \bibfield  {author} {\bibinfo {author} {\bibfnamefont {R.~A.}\ \bibnamefont
  {Horn}}\ and\ \bibinfo {author} {\bibfnamefont {C.~R.}\ \bibnamefont
  {Johnson}},\ }\href@noop {} {\emph {\bibinfo {title} {Matrix analysis}}}\
  (\bibinfo  {publisher} {Cambridge university press},\ \bibinfo {year}
  {2012})\BibitemShut {NoStop}%
\bibitem [{\citenamefont {Atkinson}(1967)}]{atkinson1967numerical}%
  \BibitemOpen
  \bibfield  {author} {\bibinfo {author} {\bibfnamefont {K.~E.}\ \bibnamefont
  {Atkinson}},\ }\bibfield  {title} {\bibinfo {title} {The numerical solution
  of fredholm integral equations of the second kind},\ }\href@noop {}
  {\bibfield  {journal} {\bibinfo  {journal} {SIAM Journal on Numerical
  Analysis}\ }\textbf {\bibinfo {volume} {4}},\ \bibinfo {pages} {337}
  (\bibinfo {year} {1967})}\BibitemShut {NoStop}%
\bibitem [{\citenamefont {Kreutz-Delgado}(2009)}]{kreutz2009complex}%
  \BibitemOpen
  \bibfield  {author} {\bibinfo {author} {\bibfnamefont {K.}~\bibnamefont
  {Kreutz-Delgado}},\ }\bibfield  {title} {\bibinfo {title} {The complex
  gradient operator and the cr-calculus},\ }\href
  {https://arxiv.org/abs/0906.4835} {\bibfield  {journal} {\bibinfo  {journal}
  {arXiv preprint arXiv:0906.4835}\ } (\bibinfo {year} {2009})}\BibitemShut
  {NoStop}%
\bibitem [{\citenamefont {Strang}(2007)}]{strang2007computational}%
  \BibitemOpen
  \bibfield  {author} {\bibinfo {author} {\bibfnamefont {G.}~\bibnamefont
  {Strang}},\ }\href@noop {} {\emph {\bibinfo {title} {Computational science
  and engineering}}}\ (\bibinfo  {publisher} {Wellesley-Cambridge Press},\
  \bibinfo {year} {2007})\BibitemShut {NoStop}%
\bibitem [{\citenamefont {Miller}(2013)}]{miller2013photonic}%
  \BibitemOpen
  \bibfield  {author} {\bibinfo {author} {\bibfnamefont {O.~D.}\ \bibnamefont
  {Miller}},\ }\bibfield  {title} {\bibinfo {title} {Photonic design: From
  fundamental solar cell physics to computational inverse design},\ }\href@noop
  {} {\bibfield  {journal} {\bibinfo  {journal} {arXiv preprint
  arXiv:1308.0212}\ } (\bibinfo {year} {2013})}\BibitemShut {NoStop}%
\bibitem [{\citenamefont {Arenz}\ \emph {et~al.}(2017)\citenamefont {Arenz},
  \citenamefont {Russell}, \citenamefont {Burgarth},\ and\ \citenamefont
  {Rabitz}}]{Arenz2017}%
  \BibitemOpen
  \bibfield  {author} {\bibinfo {author} {\bibfnamefont {C.}~\bibnamefont
  {Arenz}}, \bibinfo {author} {\bibfnamefont {B.}~\bibnamefont {Russell}},
  \bibinfo {author} {\bibfnamefont {D.}~\bibnamefont {Burgarth}},\ and\
  \bibinfo {author} {\bibfnamefont {H.}~\bibnamefont {Rabitz}},\ }\bibfield
  {title} {\bibinfo {title} {{The roles of drift and control field constraints
  upon quantum control speed limits}},\ }\href
  {https://doi.org/10.1088/1367-2630/aa8242} {\bibfield  {journal} {\bibinfo
  {journal} {New J. Phys.}\ }\textbf {\bibinfo {volume} {19}},\ \bibinfo
  {pages} {103015} (\bibinfo {year} {2017})}\BibitemShut {NoStop}%
\bibitem [{\citenamefont {Lee}\ \emph {et~al.}(2018)\citenamefont {Lee},
  \citenamefont {Arenz}, \citenamefont {Rabitz},\ and\ \citenamefont
  {Russell}}]{Lee2018}%
  \BibitemOpen
  \bibfield  {author} {\bibinfo {author} {\bibfnamefont {J.}~\bibnamefont
  {Lee}}, \bibinfo {author} {\bibfnamefont {C.}~\bibnamefont {Arenz}}, \bibinfo
  {author} {\bibfnamefont {H.}~\bibnamefont {Rabitz}},\ and\ \bibinfo {author}
  {\bibfnamefont {B.}~\bibnamefont {Russell}},\ }\bibfield  {title} {\bibinfo
  {title} {{Dependence of the quantum speed limit on system size and control
  complexity}},\ }\href@noop {} {\bibfield  {journal} {\bibinfo  {journal} {New
  J. Phys.}\ }\textbf {\bibinfo {volume} {20}},\ \bibinfo {pages} {063002}
  (\bibinfo {year} {2018})}\BibitemShut {NoStop}%
\end{thebibliography}%

\end{document}


\preprint{APS/123-QED}

\title{Supplementary Materials: \\Conservation-law-based global bounds to quantum optimal control}

\author{Hanwen Zhang}
\affiliation{Department of Applied Physics, Yale University, New Haven, Connecticut 06511, USA}
\affiliation{Energy Sciences Institute, Yale University, New Haven, Connecticut 06511, USA}
\author{Zeyu Kuang}
\affiliation{Department of Applied Physics, Yale University, New Haven, Connecticut 06511, USA}
\affiliation{Energy Sciences Institute, Yale University, New Haven, Connecticut 06511, USA}
\author{Shruti Puri}
\affiliation{Department of Applied Physics, Yale University, New Haven, Connecticut 06511, USA}
\affiliation{Yale Quantum Institute, Yale University, New Haven, Connecticut 06511, USA}
\author{Owen D. Miller}
\affiliation{Department of Applied Physics, Yale University, New Haven, Connecticut 06511, USA}
\affiliation{Energy Sciences Institute, Yale University, New Haven, Connecticut 06511, USA}

\date{\today}

\maketitle

\tableofcontents

\section{The QCQP is equivalent to the original control problem}
In the main text, we derive the conservation laws of Eq.~(3) as necessary conditions that must be satisfied by any solution of the control problem. In this section, we show that the conservation laws are also \emph{sufficient} conditions: any solution of the ultimate quadratically constrained quadratic program (QCQP) must also be a solution of the original control problem. Hence the optimal solution of the QCQP must also be the optimal solution of the original problem. The two problems are equivalent in this sense. (A related observation was made in \citeasnoun{Angeris2021}.)

The conservation-law constraints of Eq.~(3) in the main text are:
\begin{align}
                    \int_{t_0}^T \int_{t_0}^T \Phi^{\dagger}(t) D_i(t) \left[ \frac{H_c^{-1}(t)}{\epsmax} \delta(t-t') + \frac{i}{\hbar} G_0^{+}(t,t') \right] \Phi(t')\D t \,\D t' = \int_{t_0}^T \Phi^{\dagger}(t) D_i(t) U_0(t,t_0) \d{t},
                    \label{eq:Dconstr}
\end{align}
The QCQP arises from optimizing a linear or quadratic objective $f(\Phi)$ subject to \eqref{Dconstr} being satisfied for all possible $D_i$. Let us consider a solution $\Phi(t)$ of the QCQP that satisfies \eqref{Dconstr} for all $D_i(t)$. In particular, we can assume that it satisfied \eqref{Dconstr} for all $D_i(t)$ of the form
\begin{align}
    D^{i}_{jk}(t) = \delta_{j\ell}\delta_{km} \delta(t-t'),
\end{align}
where we have momentarily moved the $i$ to a superscript to explicitly consider the $jk$ element of $D^{(i)}$ in the Hilbert space, and $t'$ can be any time in the interval of interest. Given this choice of $D^{(i)}$, we can rewrite \eqref{Dconstr} as 
\begin{align}
    \Phi_{\ell i}^*(t') \left[\frac{H_c^{-1}(t')}{\epsmax}\Phi(t') + \frac{i}{\hbar}\int_{t_0}^T G_0^{+}(t',t'') \Phi(t'')\,{\rm d}t'' - U_0(t',t_0)\right]_{jm} = 0,
    \label{eq:prodform}
\end{align}
which must hold for all $i,j,\ell,m$ and $t'$. Since \eqref{prodform} must hold at all times, we can see that at any time $t'$ there are two possibilities for $\Phi(t')$: either $\Phi(t') = 0$, or $\Phi(t')$ satisfies the expression in square brackets. This dichotomy will dictate how to find a corresponding $U(t,t_0)$ that satisfied the original (Schrodinger) constraint. At times where $\Phi(t')=0$, we can take the control field $\varepsilon(t')=0$. We can also note that the integrand will be zero at such times, and that the domain of the integral can then be restricted to all times for which $\Phi$ is nonzero. \Eqref{prodform} can equivalently be written:
\begin{align}
    \Phi_{\ell i}^*(t') \left[\frac{H_c^{-1}(t')}{\epsmax}\Phi(t') + \frac{i}{\hbar}\int_{\left\{t''\in[t_0,T]|\Phi(t'')\neq0\right\}} G_0^{+}(t',t'') \Phi(t'')\,{\rm d}t'' - U_0(t',t_0)\right]_{jm} = 0,
    \label{eq:prodform2}
\end{align}
Now we consider all $t'$ for which $\Phi(t') \neq 0$. The term in square brackets must be zero at all such times. For such times, we can set $\varepsilon(t') = \epsmax$, and $U(t',t_0) = (H_c^{-1}(t')/\epsmax) \Phi(t')$. These $U(t',t_0)$ must satisfy the term in square brackets, which can now be written:
\begin{align}
    U(t',t_0) + \frac{i}{\hbar}\int_{\left\{t''\in[t_0,T]|\varepsilon(t'')\neq0\right\}} G_0^{+}(t',t'') \epsmax H_c(t'') U(t'',t_0)\,{\rm d}t'' - U_0(t',t_0) = 0.
    \label{eq:prodform3}
\end{align}
Finally, since $\varepsilon(t)$ is either $\epsmax$ or 0 at all times, we can rewrite the integral domain to take place over all possible times, with $\epsmax$ replaced by $\varepsilon(t)$, implying that $U(t,t_0)$ must satisfy
\begin{align}
    U(t',t_0) = U_0(t',t_0) - \frac{i}{\hbar}\int_{t_0}^T G_0^{+}(t',t'') \varepsilon(t'') H_c(t'') U(t'',t_0)\,{\rm d}t''.
    \label{eq:inteq}
\end{align}
\Eqref{inteq} is exactly the integral form of Eq.~(1) in the main text, which is equivalent to the Dyson equation and the differential Schrodinger equation! Hence we have shown that any solution $\Phi(t)$ that satisfies \eqref{Dconstr} for all $D_i$ implies a solution $U(t,t_0)$ of the original dynamical constraints. Therefore the optimal value of the QCQP will also be the optimal value of the original control problem.

\section{Three-level Hamiltonians}
The asymmetric double-well case is taken from Sec.~2.8 of \cite{Werschnik2007},
\begin{eqnarray}
	H &=& \bb{\mat{\omega_0 & 0&0\\ 0 & \omega_1 & 0\\ 0& 0& \omega_2 }} -\epsilon(t) \bb{\mat{\mu_{00} & \mu_{01}&\mu_{02}\\ \mu_{10} & \mu_{11} & \mu_{12}\\ \mu_{20}& \mu_{21}&\mu_{22}}}\nonumber\\
	 &=& \bb{\mat{0 & 0&0\\ 0 & 0.1568 & 0\\ 0& 0& 0.7022 }} -\epsilon(t) \bb{\mat{-2.5676 &0.3921& 0.6382\\ 0.3921 &2.3242& -0.7037\\  0.6382 & -0.7037 &-0.5988}}\,.
\end{eqnarray}
We consider a maximum control amplitude of $\magn{\epsilon} \le 0.15$ in this example.

For the transmon example, the exact Hamiltonian is
\begin{eqnarray}
	H &=& \bb{\mat{\omega_0 & 0&0\\ 0 & \omega_1 & 0\\ 0& 0& \omega_2 }} - \epsilon(t)  \bb{\mat{0 & \mu_{01}& 0 \\ \mu_{10} & 0 & \mu_{12}\\ 0& \mu_{21}& 0}}\nonumber\\
	&=&\bb{\mat{0 & 0&0\\ 0 & 1.9 & 0\\ 0& 0& 3.7 }} + \epsilon(t)  \bb{\mat{0 & 1 & 0 \\ 1 & 0 & \sqrt{2}\\ 0& \sqrt{2} & 0}}\,,
\end{eqnarray}
with a maximum control amplitude $\magn{\epsilon} \le 0.3$ in this example.

\section{Finding a binary pulse equivalent to a continuous one by local averaging}
Given a Hamiltonian $H(t) = H_0 (t) + \varepsilon(t)H_c(t)$ with some control $\varepsilon(t)$, designed controls might be of ``bang--bang'' type, taking only two discrete values, or they might be smooth, continuous controls, possibly bounded in magnitude. From a theoretical bound perspective, the latter case is subsumed by the former: any smooth, continuous control can be approximated to arbitrarily high accuracy with a particular bang--bang control. Intuitively, one can see that this might be true: if you oscillate a bang--bang control at high enough frequency, oscillating over smaller time scales than any transitions (real or virtual) induced by the Hamiltonian, then the wave function will not respond to the particular high-frequency details of the control but rather to a homogenized average (which will lie between the two extremes). Mathematically, there is a rich literature on the field of ``homogenization'' theory~\cite{Milton2002}, for applications from material science to optics. Here, we prove that any smooth, continuous control can be approximated by a bang--bang control, which then allows us to use bang--bang controls in the formulation of our bounds. (Note the key point that such bang--bang controls would not ever need to be implemented; they simply inform us of the generality of the bounds.)

Suppose that we are interested in the time evolution of the system from $t_0$ to $t_f$, with continuous control $\varepsilon_c(t)$ and propagator $U_c(t_f,t_0)$. We can divide $[t_0,t_f]$ into $N$ intervals of equal length $\tau = (t_f - t_0)/N$: $[t_0,t_1], [t_1,t_2], \ldots, [t_{N-1},t_N]$ ($t_f = t_N$), such that each interval is shorter than any timescale of the Hamiltonian. We can then write $U_c (t_f,t_0) = U_c (t_N,t_{N-1})\cdots U_c (t_2,t_1)U_c (t_1,t_0)$. Since each interval is smaller than any transition time scale, we can apply first-order perturbation theory to each $U(t_{i+1},t_i)$ to analyze the accuracy of a bang--bang-control approximation. First, we analyze the accuracy of using a first-order approximation to the time-evolution operator over each interval:
\begin{eqnarray}
	U_c(t_{i+1},t_i)& =& U_0 (t_{i+1},t_i) - \frac{\I}{\hbar}\int_{t_i}^{t_{i+1}}\D t\, U_0 (t_{i+1},t)H_I U_0 (t,t_i) \epsilon_c (t) + e'_i\\
	&&( \mbox{Taylor expand} \quad U_0 (t_{i+1},t)H_I U_0 (t,t_i) \quad \mathrm{around} \quad t = t_i )\nonumber \\
	&=& \equals{U_0 (t_{i+1},t_i) - \frac{\I}{\hbar}U_0 (t_{i+1},t_i)H_I \int_{t_i}^{t_{i+1}}\D t\, \epsilon_c (t)}{\tilde{U}_c(t_{i+1},t_i)} + e_{i}\label{eq:perturb}
\end{eqnarray}
where $U_0$ is the propagator of $H_0$ and $e'_{i}$ is the error term of the perturbation series of $\BigO{\tau^2}$. Together with the error due to Taylor expansion of $O(\tau^2)$, the overall local error $e_i$ is of $O(\tau^2)$. If we stitch these local approximations back to the global propagator, i.e. $U_c (t_f,t_0) = U_c (t_N,t_{N-1})\cdots U_c (t_2,t_1)U_c (t_1,t_0)$, we see that the global error is $\sum_{i=0}^{N-1} U_0 (t_N,t_{i+1}) e_i U_0 (t_{i},t_0)$, which is at most $N$ ($\sim 1/\tau$) copies of $O(\tau^2)$, which is of $O(\tau)$. Hence the global error goes to zero as $\tau \rightarrow 0$, and we can approximate $U_c$ with the local apprxomations $\tilde{U}_c$:
\begin{equation}
	U_c (t_f,t_0) \sim \tilde{U}_c (t_N,t_{N-1})\cdots \tilde{U}_c (t_2,t_1) \tilde{U}_c (t_1,t_0)\quad \mathrm{as} \quad \tau \rightarrow 0.
\end{equation}

The above perturbation analysis shows that if we can find a bang--bang control $\epsilon_b(t)$ with propagator $U_b(t_f,t_0)$ whose local approximation $\tilde{U}_b(t_{i+1},t_i)$ agrees with  $\tilde{U}_c(t_{i+1},t_i)$ up to the first order in $\tau$, then we will have  
\begin{equation}
	U_c (t_f,t_0) \sim \tilde{U}_c (t_N,t_{N-1})\cdots \tilde{U}_c (t_1,t_0) \sim \tilde{U}_b (t_N,t_{N-1})\cdots \tilde{U}_b (t_1,t_0) \sim U_b (t_f,t_0) 
\end{equation}
as $\tau \rightarrow 0$. This will show the global equivalence between $U_c (t_f,t_0)$ and $U_b (t_f,t_0)$. Such an $\epsilon_b (t)$ is easy to find: by the form of the perturbation in \eqref{perturb}, any choice of $\epsilon_b$ is valid as long as $\int_{t_i}^{t_{i+1}}\D t \, \epsilon_b (t) = \int_{t_i}^{t_{i+1}}\D t \, \epsilon_c (t)$ for all $i$, which can be achieve simply by requiring the average of the bang--bang control equaling the average of the continuous control. (This choice of $\epsilon_b(t) $ over $[t_i,t_{i+1}]$ has $\epsilon_b = \varepsilon_{\rm min}$ for a duration $t'$ and then switch to $\epsilon_b = \varepsilon_{\rm max}$ for the rest of the time to $t_{i+1}$, for $t' = \frac{\varepsilon_{\rm max} \tau - M_i}{\varepsilon_{\rm max} - \varepsilon_{\rm min}}$, where $M_i =  \int_{t_i}^{t_{i+1}}\D t \, \epsilon_c (t)$. Certainly, there are other choices as well.) 

\begin{figure}[thh]
    \centering
    \includegraphics[width=0.9\textwidth]{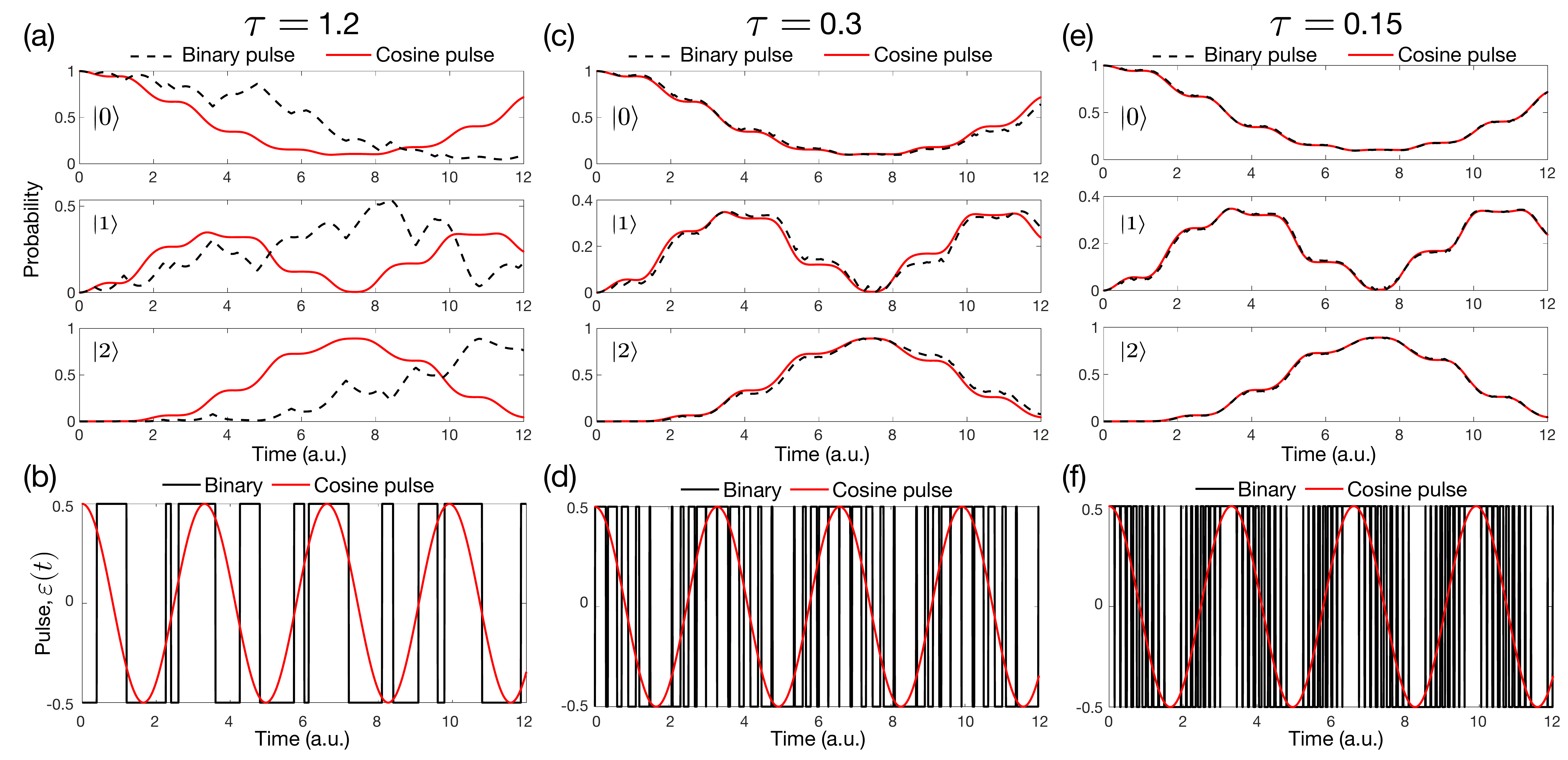}
    \caption{(a) Comparisons between probability time evolution under $\varepsilon_c (t)$(red) and $\varepsilon_b (t)$(black dash) for all three levels, from $t_0 = 0$ to $T = 12$. (b) $\varepsilon_c (t) = E\cos(\omega t)$ is plotted in red and the $\varepsilon_b (t)$ in black, obtained through averaging for $\tau = 1.2$. (c) and (d), (e) and (f) are similar to (a) and (b), but the former pair is for $\tau = 0.3$ and the latter is for $\tau = 0.15$\,. One can see that as $\tau$ gets smaller, the $\varepsilon_b(t)$ oscillates more rapidly and the evolution gets closer to the one produce by $\varepsilon_c (t)$. For $\tau = 0.15$, the effect of $\varepsilon_b(t)$ and $\varepsilon_c(t)$ are almost identical.}
    \label{fig:averaging}
\end{figure}

\begin{figure}[thb]
    \centering
    \includegraphics[width=0.5\textwidth]{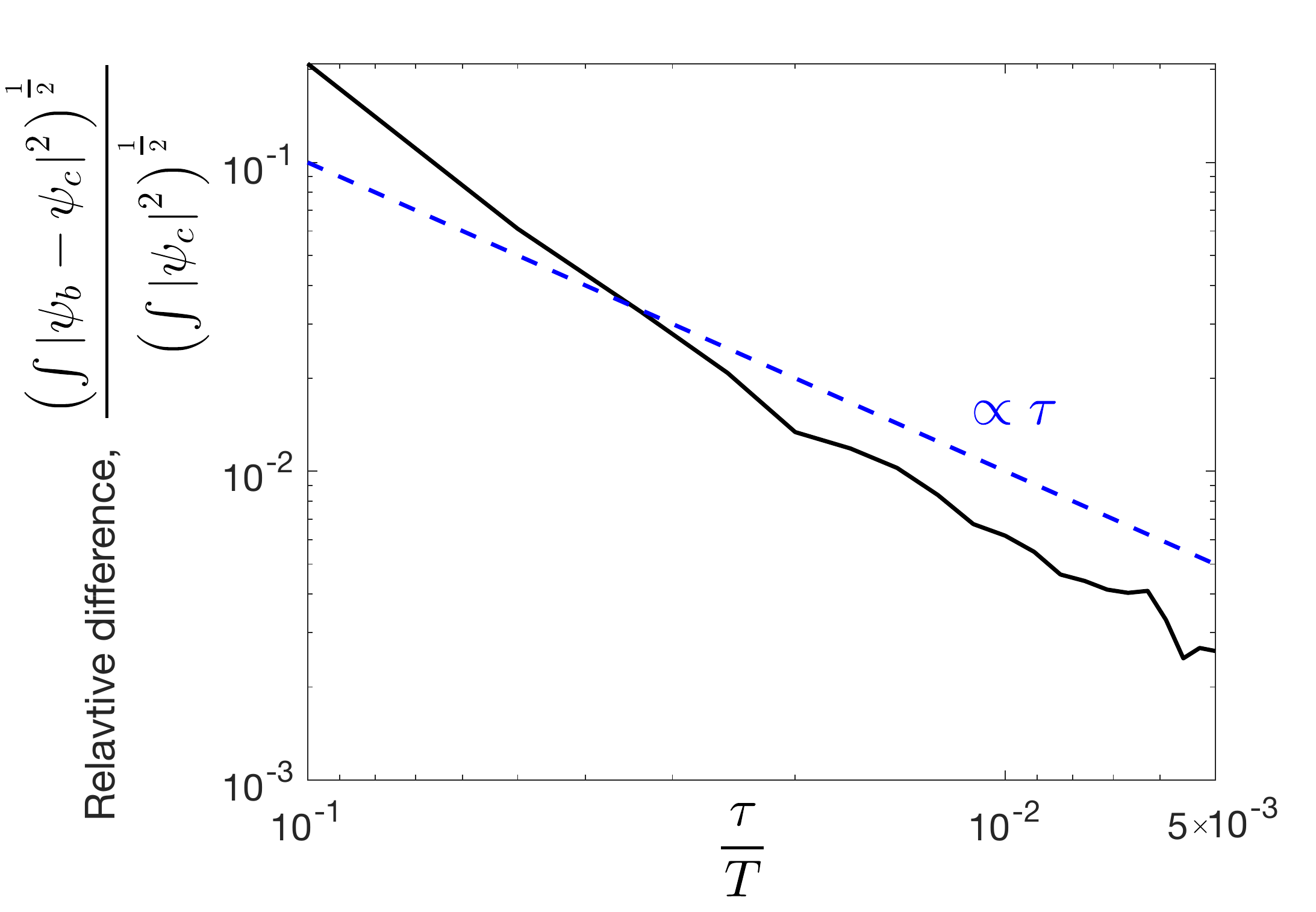}
   \caption{The relative difference between evolution of $\varepsilon_c$ and $\varepsilon_b$, measured by $\frac{(\int_{t_0}^T |\psi_{b}(t) - \psi_{c}(t)|^2)^{\frac{1}{2}}}{(\int_{t_0}^T |\psi_{c}(t)|^2)^{\frac{1}{2}}}$ is plotted against $\frac{\tau}{T}$. It shows the $O(\tau)$ convergence for small $\tau$.}
    \label{fig:converge}
\end{figure}

We illustrate the above proof numerically through the transmon example in the main text. We apply a continuous control $\varepsilon_c (t) = E \cos(\omega t)$ with $\omega = \omega_1$ and $E = 0.5$ from $t_0=0$ to $T = 12$. We want to find a binary pulse with $\varepsilon_{\rm max} = - \varepsilon_{\rm min} = E$,  such that the dynamics $\psi_{b}(t)$ approaches $\psi_{c}(t)$ under the continuous pulse when $\tau$ goes to zeros. We construct such $\varepsilon_b (t)$ according to the method described in the previous paragraph. In \figref{averaging}, we compare $\psi_b$ with $\psi_c$ for $\tau = 1.2$ for (a), $0.3$ for (c) and $0.15$ for (e) and show their respective pulses in (b), (d) and (f).  We see from In \figref{averaging} the the evolution under each $\varepsilon_b (t)$ (red) gets closer to the evolution under $\varepsilon_c (t)$ (black dash) as $\tau$ decreases. More explicitly, we plot in \figref{converge} the relative difference between evolution of $\varepsilon_c$ and $\varepsilon_b$, measured by $\frac{(\int_{t_0}^T |\psi_{b}(t) - \psi_{c}(t)|^2)^{\frac{1}{2}}}{(\int_{t_0}^T |\psi_{c}(t)|^2)^{\frac{1}{2}}}$, against $\frac{\tau}{T}$. We can see that the convergence demonstrates the $O(\tau)$ behavior proved above.
We emphasize again that the purpose of numerical results here is to show the equivalence between continuous and binary pulses when $\tau$ goes to zero. In the bound computation, we do not need to find such $\varepsilon_b(t)$ and the bound converges automatically and faster than $O(\tau)$ in the proof\,.

To conclude, we have proven that for any continuously valued control $\varepsilon_c(t)$ there is a corresponding binary control $\varepsilon_b(t)$ producing the same time evolution as $\epsilon_c(t)$, when the local averaging time $\tau$ approaches zero. Hence, a bound derived for bang--bang controls will include all possible bounded continuous controls as well.

\clearpage
\section{Local probability conservation laws as a subset of D-matrix constraints }
In this section, we show explicitly that Eq. 4 in the main text, which represents local probability conservation, is a subset of constraints  of Eq. 3  in the main text with $D_i (t) = \eta(t_1 - t) \mathcal{I} $ with $t\ge t_0$, i.e. identity operator $\mathcal{I}$ supported from $t_0$ to $t_1$. 

Now consider Eq. 2 in the main text\,,
\begin{eqnarray}
                    \int_{t=t_0}^T \int_{t'=t_0}^T \Phi^\dagger (t) D_i(t) \sb{\frac{H^{-1}_c }{\varepsilon_{\rm max}} \delta(t-t') + \frac{i}{\hbar}G_0^+ (t,t')}\Phi (t') \D t \, \D t' = \int_{t_0}^T \Phi^\dagger (t)D_i(t) U_0 (t,t_0) \D t\,,
                    \label{eq:mt3}
\end{eqnarray}
where $G_0^+ (t,t') = U_0 (t,t') \eta(t-t')$.
Now with $D_i (t) = \eta(t_1 - t) \mathcal{I}$, it becomes
\begin{eqnarray}
               \int_{t_0}^{t_1} \Phi^\dagger (t) \frac{H^{-1}_c }{\varepsilon_{\rm max}}   \Phi (t) \D t +  \frac{i}{\hbar} \int_{t=t_0}^{t_1} \int_{t'=t_0}^t \Phi^\dagger (t) U_0(t,t')\Phi (t') \D t \, \D t' = \int_{t_0}^{t_1} \Phi^\dagger (t)U_0 (t,t_0) \D t\,.
                    \label{eq:mt3t1}
\end{eqnarray}
Taking Hermitian conjugation both side of \eqref{mt3t1}, and using $U^\dagger_0 (t,t') = U_0 (t',t)$ and $(H^{-1}_c)^\dagger = H^{-1}_c$\,, produces
\begin{eqnarray}
               \int_{t_0}^{t_1} \Phi^\dagger (t) \frac{H^{-1}_c }{\varepsilon_{\rm max}}   \Phi (t) \D t -  \frac{i}{\hbar} \int_{t=t_0}^{t_1} \int_{t'=t_0}^t \Phi^\dagger (t') U_0(t',t)\Phi (t) \D t \, \D t' = \int_{t_0}^{t_1} U^\dagger_0 (t,t_0) \Phi(t)\D t\,
                    \label{eq:mt3t1cg}
\end{eqnarray}
Note we can exchange the $t,t'$ integration in the the following integral
\begin{eqnarray*}
	\int_{t = t_0}^{t_1} 	\int_{t' = t_0}^{t} \ldots \D t' \D t &=&  \int_{t' = t_0}^{t_1} 	\int_{t = t'}^{t_1} \ldots \D t \D t' \,.
\end{eqnarray*}
As a result, the double integral in \eqref{mt3t1cg} becomes
\begin{eqnarray*}
	 && \int_{t=t_0}^{t_1} \int_{t'=t_0}^t \Phi^\dagger (t') U_0(t',t)\Phi (t) \D t \, \D t'  \quad (\mbox{exchange $t, t'$ integrations})\\
 	 &=& \int_{t' =t_0}^{t_1} \int_{t = t'}^{t_1} \Phi^\dagger (t') U_0(t',t)\Phi (t) \D t \, \D t' \quad (\mbox{relabeling by} \,t\rightarrow t', t' \rightarrow t)\\
 	 &=& \int_{t =t_0}^{t_1} \int_{t' = t}^{t_1} \Phi^\dagger (t) U_0(t,t')\Phi (t') \D t \, \D t'\,.
\end{eqnarray*}
Now we subtract \eqref{mt3t1} by \eqref{mt3t1cg} to obtain the imaginary part of \eqref{mt3t1}, and we get
\begin{equation}
	\frac{1}{\hbar} \int_{t_0}^{t_1} \int_{t_0}^{t_1}\Phi^\dagger (t) U_0(t,t')\Phi (t') \D t \, \D t' = 2\Im \int_{t_0}^{t_1} \Phi^\dagger (t)U_0 (t,t_0) \D t\,,
\end{equation}
which is exactly  Eq. 4 in the main text, since $U^\dagger U = \mathcal{I}$ by probability conservation. $ \int_{t_0}^{t_1} \Phi^\dagger (t) \frac{H^{-1}_c }{\varepsilon_{\rm max}}   \Phi (t) \D t$ disappears since its imaginary part is zero. This makes sense since $H_c$ is a Hermitian matrix and no dissipation is present.

Often $D_i (t) = \delta(t-t')D$ is localized at time $t'$ ($D$ is some constant matrix). To connect such $D_i(t)$ with the above result, we notice that $\int_{t_0} ^{t_1} \delta (t-t') \D t' = \eta(t_1-t)$ for $t \ge t_0$. As a result, we can interpret the above conservation law as the imaginary part of a linear combination  of constraints with $D_i (t) =  \delta(t-t') \mathcal{I}$ for $t_0 \le t' \le t_1$, a subset of all D-matrix constraints. More generally, $D$ is not necessarily $\mathcal{I}$ and the real part is also a valid D-matrix constraint. We thus can interpret the entire set of D-matrix constraints as generalized local probability conservation laws.

\newpage
\section{Bound formulation for problems with multiple controllable terms}
In the main text, we only consider problems with a single control. We show here that our framework extends naturally to problems with multiple controls. Consider a Hamiltonian $H(t) = H_0(t) + \sum_{i=1}^{N}  \varepsilon_i(t) H^i_{c}(t)$, where $H_0(t)$ is the non-controllable part and  $\sum_{i=1}^{N} \varepsilon_i(t) H^i_{c}$ is the controllable part. To avoid overloading the notation, we here formulate the bound problem for a two-term problem with $H(t) = H_0(t) + \varepsilon_1(t) H^1_{c}  +  \varepsilon_2(t) H^2_{c}$, which can be extended to more terms in a straightforward way. 

The first step is to write the integral-equation form (analog of Eq.~(1) of the main text) for multiple controls. To do this, we will assume the two controls cannot be turned on at the same times. (Note that two controls that can be turned on at the same time is equivalent to three controls that cannot be turned on at the same time, so there is no loss of generality here.) With two independent controls, the integral equation for $U(t,t_0)$ is:
\begin{equation}
    U(t,t_0) = U_0(t,t_0) -\frac{i}{\hbar}\int_{t_0}^{T} G_0^+(t,t') H^1_c(t) \varepsilon_1(t')U(t',t_0) - \frac{i}{\hbar}\int_{t_0}^{T} G_0^+(t,t') H^2_c(t) \varepsilon_2(t')U(t',t_0) 
\end{equation}
For times where $\varepsilon_2(t)$ is zero, we have
\begin{equation}
	U(t,t_0) = U_0(t,t_0) -\frac{i}{\hbar}\int_{\cb{\varepsilon_1(t) \ne  0}} G_0^+(t,t') H^1_c \varepsilon_1(t')U(t',t_0) \D t'\,,
	\label{eq:precon1}
\end{equation}
 and for times where $\varepsilon_1(t)$ is zero, 
\begin{equation}
	U(t,t_0) = U_0(t,t_0) -\frac{i}{\hbar}\int_{\cb{\varepsilon_2(t) \ne  0}} G_0^+(t,t') H^2_c \varepsilon_2(t')U(t',t_0) \D t'\,.
	\label{eq:precon2}
\end{equation} 
We then left multiply $U^\dagger(t,t_0)\varepsilon_1(t)H^1_c D_i(t)$ to \eqref{precon1} and $U^\dagger(t,t_0)\varepsilon_2(t)H^2_cD_j(t) $ to \eqref{precon2}, and integrate over $t$ from $t_0$ to $T$. Then \eqref{precon1}  and \eqref{precon2} become
\begin{eqnarray}
	&&\int_{t_0}^{T}\D t \, U^\dagger(t,t_0)\varepsilon_1(t)H^1_c D_i(t) U(t,t_0) 
	= \int_{t_0}^{T}\D t \, U^\dagger(t,t_0)\varepsilon_1(t)H^1_c D_i(t)U_0(t,t_0) \nonumber\\
	&& -\frac{i}{\hbar}\int_{t_0}^{T}\D t \int_{\cb{\varepsilon_1(t) \ne 0}} \D t' \, U^\dagger(t,t_0)\varepsilon_1(t)H^1_cD_i(t)G_0^+(t,t') H^1_c \varepsilon_1(t')U(t',t_0)\,,
\end{eqnarray} 
and 
\begin{eqnarray}
	&&\int_{t_0}^{T}\D t \, U^\dagger(t,t_0)\varepsilon_2(t)H^2_c D_j(t)U(t,t_0) 
	= \int_{t_0}^{T}\D t \, U^\dagger(t,t_0)\varepsilon_2(t)H^2_c D_j(t) U_0(t,t_0) \nonumber\\
	&& -\frac{i}{\hbar}\int_{t_0}^{T}\D t \int_{\cb{\varepsilon_2(t) \ne 0}} \D t' \, U^\dagger(t,t_0)\varepsilon_2(t)H^2_c D_j(t)G_0^+(t,t') H^2_c \varepsilon_2(t')U(t',t_0)\,,
\end{eqnarray} 
We define $\Phi_1(t) = \varepsilon_1(t)H^1_c U(t,t_0)$ and $\Phi_2(t) = \varepsilon_2(t)H^2_c U(t,t_0)$.  The above equations are quadratic in $\Phi_1$ and $\Phi_2$, thus they hold for $t'$ from $t_0$ and $T$. Consequently, we can replace the integral range of $t'$ by $t_0$ and $T$ and obtain constraints
\begin{equation}
	\int_{t_0}^{T}\D t \int_{t_0}^{T}\D t' \,  \Phi^\dagger_1(t)D_i(t)\bb{ \delta(t-t')\frac{(H^1_c)^{-1}}{\varepsilon_{1\mathrm{max}}} + \frac{i}{\hbar}G_0^+(t,t')}  \Phi_1(t') = \int_{t_0}^{T}\D t \,  \Phi^\dagger_1(t) D_i(t)U_0(t,t_0) \,,
	\label{eq:con1}
\end{equation} 
and 
\begin{equation}
	\int_{t_0}^{T}\D t \int_{t_0}^{T}\D t' \,  \Phi^\dagger_2(t)D_j(t)\bb{ \delta(t-t')\frac{(H^2_c)^{-1}}{\varepsilon_{2\mathrm{max}}} + \frac{i}{\hbar}G_0^+(t,t')}  \Phi_2(t') = \int_{t_0}^{T}\D t \,  \Phi^\dagger_2(t)D_j(t) U_0(t,t_0) \,,
	\label{eq:con2}
\end{equation} 
similar to Eq. (3) in the main text, but now with two different components. In addition to these two sets of constraints, we also need to impose that only one of $\varepsilon_1(t)$ and $\varepsilon_2(t)$ can be nonzero at any given time, which is represented in the constraint
\begin{equation}
    \Phi_1(t)\Phi_2^\dagger(t) = 0\, \quad \mbox{for all components}, 
	\label{eq:exclude}
\end{equation}
for all times $t$. Hence bounds for two controls can be formulated as
                \begin{equation}
                    \begin{aligned}
                        & \underset{\Phi_1,\Phi_2}{\text{max.}} & & f(\Phi_1,\Phi_2) \\
                        & \text{s.t.} & & \textrm{\Eqref{con1} satisfied for all } D_i(t)\,,\nonumber\\
					  &  & & \textrm{\Eqref{con2} satisfied for all } D_j(t)\,,\\
					  &&& \Phi_1(t)\Phi_2^\dagger(t)= 0 \quad \mbox{for all } t\,, \mbox{for all components}\,.
                    \end{aligned}
                    \label{eq:opt_prob}
                \end{equation}
                One can see immediately how to generalize to multiterm pulse designs. For $H(t) = H_0(t) + \sum_{i=1}^{N}  \varepsilon_i(t) H^i_{c}$, we will have $N$ different $\Phi_i$ with its corresponding sets of quadratic constraints. We also need to impose that $\Phi_i(t) \Phi_j^\dagger (t) = 0$ for all $i\ne j$.

Finally, note that if both $\varepsilon_1(t)$ and $\varepsilon_2(t)$ can be nonzero at the same time, it is equivalent to considering a problem with three controllable terms: $\varepsilon_1(t)H^1_c$, $\varepsilon_2(t)H^2_c$, and $\varepsilon_1(t)H^1_c + \varepsilon_2(t)H^2_c$, only one of which can be nonzero at any given time.

\section{Bound formulation for open systems governed by master equations}
In the main text, we only consider problems whose dynamics is governed by the Schr\"odinger equation. Now we show how to apply our bound framework to open systems described by differential equations for the density matrix $\rho$. In principle, this framework applies to any equation that is linear in $\rho$. Here we specifically consider the Lindblad master equation \cite{breuer2002theory}:

\begin{equation}
	\frac{\partial }{\partial t}\rho = -\frac{i}{\hbar}[H,\rho] + \sum_{i}\gamma_i \bb{A_i\rho A_i^\dagger - \frac{1}{2}\cb{A_i^\dagger A_i, \rho }}\,,
\end{equation}
where we also assume the Hamiltonian has the form $H = H_0 + \varepsilon(t)H_c$, and the second term is the dissipative part. A common trick is to vectorize the equation to form ``superoperators''~\cite{breuer2002theory},  through flattening the $n^2$ by $n^2$ density matrix $\rho$ into a column vector $\vec{r}$ by ordering the columns of $\rho$ in a single vector~\cite{horn2012matrix}. Then all operators are transformed according to this new ordering (basis):
\begin{eqnarray}
	&&[H,\rho] \longrightarrow \bb{I\otimes H - H^T\otimes I}\vec{r}\nonumber\\
	&&\quad\quad\quad\quad =  \bb{I\otimes H_0 - H_0^T\otimes I}\vec{r} +  \bb{I\otimes H_c - H_c^T\otimes I}\vec{r}\\
	&& A_i \rho A_i^\dagger \longrightarrow  \bb{(A_i^\dagger)^T \otimes A_i}\vec{r} \\
     && A_i A_i^\dagger \rho  \longrightarrow  \bb{I \otimes A_i^\dagger A_i}\vec{r} \\
     && \rho A_i^\dagger  A_i \longrightarrow  \bb{(A_i^\dagger A_i)^T \otimes I}\vec{r}\,.
\end{eqnarray}
If we define the vectorized  $-\frac{i}{\hbar}[H_0,\rho]$ by $L_0$, the vectorized dissipative term by $L_D$, and vectorized  $-\frac{i}{\hbar}[H_c,\rho]$ by $L_c$, we turn the original equation into
\begin{equation}
	\frac{\partial }{\partial t}\vec{r} = (L_0+ L_D)\vec{r} + L_c \vec{r}\,.
\end{equation}
This equation again can be turned into the integral form
\begin{equation}
	\vec{r}(t) = \vec{r}_0(t) + \int_{t_0}^T \D t' G^+(t,t')L_c(t’) \vec{r}(t’),
	\label{eq:int_r}
\end{equation}
where $\vec{r_0}(t)$ is the evolution of $\vec{r}$ without the presence of $L_c$, and $G^+(t,t')$ is the retarded Green's function of $\frac{\partial }{\partial t}\vec{r} =(L_0+ L_D)\vec{r}$. The dissipative nature of the problem is encoded in $G^+(t,t')$.

\eqref{int_r} is the starting point of the bound formulation, just as Eq. (1) in the main text for the Schr\"odinger equation. Carrying out the same manipulations as in the main text gives the constraints
\begin{eqnarray}
	&&\int_{t_0}^{T} \D t \,\vec{r}^\dagger(t) L_c (t) D_i(t) \vec{r}(t) + \int_{t_0}^{T} \D t \int_{t_0}^{T} \D t' \, \vec{r}^\dagger(t) L_c (t) D_i(t) G^+(t,t') L_c (t') \vec{r}(t') \nonumber \\
	&=&   \int_{t_0}^{T} \D t \vec{r}^\dagger(t) L_c (t) D_i(t) \vec{r}(t)\,,
\end{eqnarray}
which are analogous to Eq. (2) in the main text. The quadratic constraints can then be formed analogously.

\section{Numerical details}
In this section, we provide details for the numerical scheme used in the main text. Unlike conventional time-stepping methods to solve initial-value differential equations, we directly use the integral equation (Eq. (1) of the main text) to solve for the dynamics. We do this for three reasons: (1) In our bound method, constraints come from the integral equation, so the discretized form of the integral equation is directly available after solving the bound problem. (2) Pulse design is straightforward to perform with methods based on the integral equation. (3) Methods based on the integral equation are equally viable as those based on the differential equation. 

\subsection{Nystr\"om discretization}
We consider the integral equation in Eq. (1). We consider systems with $L$ levels, modeling either discrete levels or a subset of continuous states, over times $[t_0,T]$. Here we isolate a single column of the time-evolution operator, essentially selecting an initial state $\ket{\psi_0}$ and the time-dependent state $\ket{\psi(t)} = U(t,t_0) \ket{\psi_0}$. Then the integral equation for $\ket{\psi(t)}$ is
\begin{equation}
	\ket{\psi(t)}= \ket{\psi_0(t)} -\frac{\I}{\hbar} \int_{t_0}^{t} U_0(t,t')H_c \varepsilon(t')\ket{\psi(t')}\D t'\,,
	\label{eq:VIE}
\end{equation}
where we have replaced the retarded Green's function with the time-evolution operator multiplied by a step function (changing the upper limit of the integral to $t$). This is in the form of a Volterra integral equation of the second kind.

The Nystr\"om method \cite{atkinson1967numerical} for discretizing this equation consistent of defining quadrature nodes $\{t_i\}_{i=0}^N\in[t_0,T]$ and corresponding weights\,, and enforcing
\eqref{VIE} to hold at $\{t_i\}_{i=0}^N$
\begin{equation}
	\ket{\psi(t_i)} = \ket{\psi_0(t_i)} -\frac{\I}{\hbar} \int_{t_0}^{t_i} U_0(t_i,t')H_c \varepsilon(t') \ket{\psi(t')} \D t'\,, \quad i = 0,1,2,\ldots, N\,,
\end{equation}
and then approximating the integral operator by
\begin{equation}
	\frac{\I}{\hbar} \int_{t_0}^{t_i} U_0(t_i,t')H_c \varepsilon(t')\ket{ \psi(t')} \D t' \approx \sum_{j = 1}^{t_i} w_{ij} \equals{\frac{\I}{\hbar}U_0(t_i,t_j)H_c \varepsilon(t_j)}{K_{ij}} \ket{\psi(t_j)}\,.
	 \label{eq:quad}
\end{equation}
for some weights $w_{ij}$. Then \eqref{VIE} becomes a linear system 
\begin{equation}
	 \ket{\psi(t_i)} = \psi_0(t_i) -\sum_{j = 0}^{i} w_{ij} K_{ij}\ket{\psi(t_j)}\,, \quad i = 0,1,2,\ldots, N\,,
	 \label{eq:pre_linsys}
\end{equation}
or  more compactly as 
\begin{equation}
	\vec{\psi}  = \vec{\psi}_0 - \vec{A}\vec{\psi}\,,
	\label{eq:linsys}
\end{equation}
where $\vec{\psi}$ and $\vec{\psi}_0$ are column vector of length $L(N+1)$ with the $i$th column block $\vec{\psi}_i = \ket{\psi(t_i)}$ and $(\vec{\psi}_0)_i = \ket{\psi_0(t_i)}$ respectively. And $\vec{A}$ is a matrix of size $L(N+1)\times L(N+1)$ whose explicit form we will show later. Then $\vec{\psi}$ can be solved as $(\vec{I} + \vec{A})^{-1}\vec{\psi}_0$\,. 

We note that the convergence rate of $\vec{\psi}$ to the exact value $\psi$ is the same as the order of quadrature used in \eqref{quad}. The matrix norm of $\vec{A}$ is provably small, so $\vec{I} + \vec{A}$ is always well-conditioned. To conclude, the scheme we described above is stable and convergences quickly for good choice of $\{t_i\}_{i}^N$ and weights\,.

In this work, we use the trapezoidal rule
\begin{equation}
	\int_{a}^{b} f(x) \D x \approx \frac{h}{2} \bb{f(a) +2f(a+h) + \ldots + 2f(b-h)+f(b)} + O(h^2)
	\label{eq:trap}
\end{equation}
where $h$ is the spacing between the equally spaced nodes from $a$ to $b$. As a result, the method we are using has convergence rate equivalent to Runge Kutta of order 2. Without special techniques for the matrix inversion process, this method is not as efficient as Runge Kutta of order 2. Despite of this drawback and the fact that higher order methods are available, the method here is sufficient for the examples we consider. If we apply the trapezoidal rule to \eqref{pre_linsys}, then the nodes $\{t_i\}_{i=0}^N\in[t_0,T]$ are $t_0, t_1,\ldots,t_N = T$ with $t_i = t_0 + ih$ and $h = \frac{T-t_0}{N}$\,. \eqref{pre_linsys} then becomes
\begin{eqnarray}
	\ket{\psi(t_0)} &=& \ket{\psi_0(t_0)} \nonumber\\
	\ket{\psi(t_1)} &=& \ket{\psi_0(t_1)} - \frac{h}{2}(K_{10} \ket{\psi(t_0)}+ K_{11}\ket{\psi(t_1)} ) \nonumber\\
	\ket{\psi(t_2)} &=& \ket{\psi_0(t_2)} - \frac{h}{2}(K_{20}\ket{ \psi(t_0)} + 2K_{21}\ket{\psi(t_1)} + K_{22} \ket{\psi(t_2)}  ) \nonumber\\
	&\vdots& \nonumber\\
	\ket{\psi(t_N)} &=& \ket{\psi_0(t_N)} - \frac{h}{2}(K_{N0} \ket{\psi(t_0)} + 2\sum_{j =1}^{N-1}K_{Nj}\ket{\psi(t_j)} + K_{NN}\ket{\psi(t_N)}  )\nonumber\,,\\
	\label{eq:linsys_exp}
\end{eqnarray}
which is \eqref{linsys} in component form with the trapezoidal rule.

To solve for $U$\,, we simply replace  the vector $\vec{\psi}$ by the matrix $\vec{U} = \bb{\mat{\vec{\psi}^1 & \vec{\psi}^2 & \ldots} \vec{\psi}^L }$ and $\vec{\psi}_0$ by the matrix $\vec{U}_0 = \bb{\mat{\vec{\psi}_0^1 & \vec{\psi}_0^2 & \ldots \vec{\psi}_0^L}}$\,, both of size $L(N+1)\times L$\,, where $\vec{\psi}^i $ represents the discretized version of  different states wave function $\ket{\psi^i (t)} = U(t,t_0)\ket{\psi_0^i (t_0)}$. 
By comparing \eqref{linsys_exp} and \eqref{pre_linsys}, we can see that the weight $w_{ij} = h$ for $i \ne j$ and $w_{ij} = h/2$ for $i=j$ and $w_{ij} = 0$ for $i<j$.
Then the discretized version of Eq. (1) in the main text is given by 
\begin{equation}
	\vec{U}  = \vec{U}_0 - \vec{A}\vec{U}\,, 
	\label{eq:linsys_U}
\end{equation}
and $\vec{U} = (\vec{I} + \vec{A})^{-1}\vec{U}_0$\,.
\subsection{Discretization for quadratic constraints}
With the understanding of how the integral equation is discretized, we are ready to give the discretization form of Eq. (3), which constitutes the constraints for computing the bound. 
Constraints in Eq. (3) are given by 
\begin{eqnarray}
	&& \int_{t_0}^{T}\D t \int_{t_0}^{T}\D t' \,  \Phi^\dagger (t)D_i(t)\bb{ \delta(t-t')\frac{H^{-1}_c}{\varepsilon_{\mathrm{max}}} + \frac{i}{\hbar}G_0^+(t,t')}  \Phi(t')
	\nonumber\\
	&& = \int_{t_0}^{T}\D t \,  \Phi^\dagger(t)D_i(t) U_0(t,t_0) \,,
\end{eqnarray}
or 
\begin{eqnarray}
		&& \int_{t_0}^{T}\D t \int_{t_0}^{t}\D t' \,  \Phi^\dagger (t)D_i(t) \frac{i}{\hbar} U_0(t,t') \Phi(t') +  \int_{t_0}^{T}\D t \, \Phi^\dagger (t)D_i(t)\frac{H^{-1}_c}{\varepsilon_{\mathrm{max}}}   \Phi(t) 
	\nonumber\\
	&& = \int_{t_0}^{T}\D t \,  \Phi^\dagger(t)D_i(t) U_0(t,t_0) \,,
	\label{eq:pre_dis_con}
\end{eqnarray}
We have seen in the previous section on how to discretize the $t'$ integral. The $t$ integral is in fact much easier with fixed integration limit, and we can apply the trapezoidal rule directly to the integral $\int_{t_0}^T \D t f(t) \approx \sum_{i=0}^{N} w_i f(t_i)$ with the same equally spaced nodes $\{t_i \}_{i=0}^N$ and weights $w_i$ in \eqref{trap}\,. 
Then \eqref{pre_dis_con} becomes
\begin{equation}
	\vec{\Phi}^\dagger \vec{D}_i \vec{W}\vec{G}\vec{\Phi} + \vec{\Phi}^\dagger \vec{D}_i \vec{W}\frac{\vec{H}^{-1}_c}{\varepsilon_{\rm max}}\vec{\Phi} = \vec{\Phi}^\dagger \vec{D}_i \vec{W}\vec{U}_0\,,
	\label{eq:QC_con}
\end{equation}
where $\vec{\Phi}$ and $\vec{U}_0$ are matrix of size $L(N+1)\times L$. 
$\vec{G}$ is the discretized version of the integral operator $\int_{t_0}^t \D t' U(t,t')$. Similar to the way we discretized \eqref{quad}\,, we have the $ij$ block of $\vec{G}$
\begin{equation}
	\vec{G}_{ij} = w_{ij} \frac{\I}{\hbar} U_0(t_i,t_j)
	\label{eq:G}
\end{equation}
with $w_{ij}$ the weight in \eqref{linsys_exp}.
$\vec{W}$ is the matrix contains the effect of weights $w_i$ from the $t$ integral, and it is of the from
\begin{equation}
	\vec{W} = \bb{\mat{w_0 &&&\\&w_2&&\\ &&\ddots&\\&&&w_N}}\otimes \vec{I}_{L\times L}\,,
\end{equation}
where $\otimes$ is the kroncker product for matrices and $ \vec{I}_{L\times L}$ represents an identity matrix of dimension $L$.
Here $\vec{H}^{-1}_c$ is
\begin{equation}
	\vec{H}^{-1}_c = \vec{I}_{(N+1)\times(N+1)}\otimes H^{-1}_c 
\end{equation}
Also, since $D_i = \delta(t-t_i)D$ and $D$ is some constant matrix of dimension $L$ by $L$, we have 
\begin{equation}
	\vec{D}_i =  \bb{\mat{0 &&&&\\&\ddots&&&\\ &&1 & &\\&&&\ddots&\\&&&&0}}\otimes D\,,
\end{equation}
where the first diagonal matrix is the discretized $\delta(t-t_i)$ and is nonzero only in the $i$th diagonal entry. 
$D$ of dimension $L$ by $L$ consists of $L^2$ unit basis matrices. For a two level system, all possible independent $D$ consists of $\bb{\mat{1 &0\\ 0& 0}}$, $\bb{\mat{0 &1\\ 0& 0}}$, $\bb{\mat{0 &0\\1& 0}}$ and $\bb{\mat{0 &0\\0& 1}}$\,. Higher dimension systems can be generalized accordingly.

By varying $D_i$ over all possible $t_i$ and $D$, we obtain all possible constraints in our problem in the form of a $L$ by $L$ matrix equality, which must hold componentwise.
$\vec{\Phi}$ can be flattened column by column into a vector, and the equality constraints for each components can then be transformed  into a quadratic constraint of the flattened $\vec{\Phi}$. 
Then semidefinite relaxation can be carried out to form the semidefinite program, which can be solved by standard convex optimization package.

\subsection{Pulse designs via local optimization}
In this section, we describe how to efficiently compute the gradient  $\frac{\partial f(U)}{\partial \varepsilon(t)}$ of the objective $f(U)$ for all $t$ from $[t_0, T]$ by solving for the dynamics twice only. Now, we return to \eqref{linsys_U}, the discretized version of Eq. (1) in the main text. To make the variables  $\varepsilon(t_i)$ here more prominent, we separate  $\varepsilon(t_i)$ from $\vec{A}$ in \eqref{linsys_U} and write it as
\begin{equation}
	\vec{U} = \vec{U}_0 - \equals{\vec{G}\vec{H}_c \vec{\varepsilon}}{\vec{A}}\vec{U}\,,
	\label{eq:VIE_design}
\end{equation}
where 
\begin{equation}
	\vec{H}_c = \vec{I}_{(N+1)\times(N+1)}\otimes H_c\,,
\end{equation}
\begin{equation}
	\vec{\varepsilon} =  \bb{\mat{\varepsilon(t_0) &&&\\&\varepsilon(t_1) &&\\ &&\ddots&\\&&&\varepsilon(t_N) }}\otimes \vec{I}_{L\times L}\,,
\end{equation}
and $\vec{G}$ is given in \eqref{G}. We also denote the column vector $\bb{\mat{\varepsilon(t_0) & \varepsilon(t_1)&\ldots&\varepsilon(t_N)}}^T$ by $\vec{p}$\,.

The goal here is to compute the discretized gradient $\frac{\partial f(\vec{U})}{\partial \vec{p}}$ in a way that the work required is independent of the number of variable $\varepsilon(t_i)$\,. To do so, we first use the CR calculus to formally treat $\vec{U}$ and $\vec{U}^*$ as independent variables \cite{kreutz2009complex}; the chain rule yields
\begin{equation}
		\frac{\partial f(\vec{U})}{\partial \vec{p}} = \frac{\partial f(\vec{U})}{\partial \vec{U}} \frac{\partial \vec{U}}{\partial \vec{p}} + \frac{\partial f(\vec{U})}{\partial \vec{U}^*} \frac{\partial \vec{U}^*}{\partial \vec{p}}  = 2\Re{\frac{\partial f(\vec{U})}{\partial \vec{U}} \frac{\partial \vec{U}}{\partial \vec{p}} }\,.
		\label{eq:grad}
\end{equation}
Then we differentiate \eqref{VIE_design} with respect to $\vec{p}$ and obtain
\begin{equation}
	\vec{G}\vec{H}_c \frac{\partial \vec{\varepsilon}}{\partial \vec{p}}\vec{U} = -(\vec{I} + \vec{A})\frac{\partial \vec{U}}{\partial \vec{p}}\,,
\end{equation}
and we can solve for $\frac{\partial \vec{U}}{\partial \vec{p}} = -(\vec{I} + \vec{A})^{-1}\vec{G}\vec{H}_c \frac{\partial \vec{\varepsilon}}{\partial \vec{p}}\vec{U} $\,. Substituting this into \eqref{grad} gives
\begin{equation}
		\frac{\partial f(\vec{U})}{\partial \vec{p}}  = -2\Re{\frac{\partial f(\vec{U})}{\partial \vec{U}}(\vec{I} + \vec{A})^{-1}\vec{G}\vec{H}_c \frac{\partial \vec{\varepsilon}}{\partial \vec{p}}\vec{U} }.
\end{equation}
Since $\vec{G}$, $\vec{H}_c$ are known $\frac{\partial \vec{\varepsilon}}{\partial \vec{p}}$ can be calculated analytically, then if we define the so-called ``adjoint'' solution \cite{strang2007computational,miller2013photonic}
\begin{equation}
    (\vec{I} + \vec{A})^T\vec{U}_{\rm adj} =\bb{ \frac{\partial f(\vec{U})}{\partial \vec{U}}}^T\,,\
	\label{eq:adjoint}
\end{equation}
we can obtain the gradient as
\begin{equation}
    \frac{\partial f(\vec{U})}{\partial \vec{p}} = -2\Re{\vec{U}_{\rm adj}^T \vec{G}\vec{H}_c \frac{\partial \vec{\varepsilon}}{\partial \vec{p}}\vec{U} }\,,
\end{equation}
by solving for $\vec{U}_{\rm adj} = (\vec{I} + \vec{A}^T)^{-1}\bb{ \frac{\partial f(\vec{U})}{\partial \vec{U}}}^T$  from \eqref{adjoint} and $\vec{U} = (\vec{I}+\vec{A})^{-1} \vec{U}_0$ from \eqref{VIE_design} only.
This allows rapid computation of $\frac{\partial f(\vec{U})}{\partial \vec{p}}$. Then standard optimization methods, such as gradient ascent, can be applied to maximize or minimize $f(\vec{U})$\,. All designed pulses in the main text are obtained by this approach described here.

\section{Guide to computing prior-literature bounds}
In this section, we describe how we computed the bounds from prior literature in Fig.~1 of the main text: Mandelstam--Tamm (MT), Margolus--Levitin (ML), and the bounds from Refs.~\cite{Arenz2017,Lee2018}. We start with the MT bound: the minimum time, which we denote $\tau_{\rm MT}$, for a given evolution is given by
\begin{align}
    \tau_{\rm MT} = \frac{\pi}{2\Delta H},
\end{align}
where the standard deviation $\Delta H$ is given by $\Delta H = \sqrt{\langle H^2 \rangle - \langle H \rangle^2}$. We do not know the mean of the Hamiltonian over the optimal trajectory, but can bound its square below by zero, which gives $\Delta H \leq \sqrt{\langle H^2 \rangle}$. We don't know the value of $\langle H^2 \rangle$ over the optimal trajectory either, but it can be bounded above by the square of the largest possible eigenvalue of the Hamiltonian, which we denote $\lambda_{\rm max}$. This eigenvalue can be found by a grid-search-based optimization over all possible instantaneous Hamiltonians (assuming bounded controls), and we are left with the bound:
\begin{align}
    \tau_{\rm MT} \geq \frac{\pi}{2\lambda_{\rm max}}.
\end{align}
The ML bound, which we denote $\tau_{\rm ML}$, follows similar reasoning. Now given by $\tau_{\rm ML} = \pi / 2\langle H \rangle$, we again do not know the mean energy, but can now bound it above by $\lambda_{\rm max}$, which implies that
\begin{align}
    \tau_{\rm ML} \geq \frac{\pi}{2\lambda_{\rm max}}.
\end{align}
By this reasoning, the MT and ML bounds coincides when presented with uncertainty about the optimal trajectory and bounded in this way.

Next, we consider the bound of \citeasnoun{Arenz2017}. In this case, one defines two functions incorporating information about the controllable and non-controllable Hamiltonians, $H_c$ and $H_0$, respectively:
\begin{align}
    C(U_g,H_c) = \frac{\sqrt{2\left(d - \sum_{j=1}^d \left| \langle \phi_j^{(c)} | U_g | \phi_j^{(c)} \rangle \right| \right)}}{2}, \\
    C(U_g,H_0) = \frac{\sqrt{2\left(d - \sum_{j=1}^d \left| \langle \phi_j^{(0)} | U_g | \phi_j^{(0)} \rangle \right| \right)}}{2},
\end{align}
where $U_g$ is a target evolution operator, $d$ is the dimensionality, and $\phi_j^{(c)}$ and $\phi_j^{(0)}$ are the eigenfunctions of the controllable and non-controllable Hamiltonians, respectively. From these functions, the time bounds of \citeasnoun{Arenz2017} can be written:
\begin{align}
    T \geq {\rm max} \left\{ \frac{2C(U_g,H_c)}{\|H_0\|}, \frac{2C(U_g,H_0)}{|f_{\rm max}|\|H_c\|} \right\},
    \label{eq:Tmax}
\end{align}
where $|f_{\rm max}|$ is the maximum amplitude of the bounded control parameter. \Eqref{Tmax} is straightforward to compute, though hard to apply to the maximum transition probability in a multilevel system, as there is no specific target unitary evolution. In this case, to compute a meaningful bound at all (for the sake of comparison), we instead consider the best numerically-optimized control sequence, and use its evolution operator evaluated at the final time as the target unitary. In this way, we take a known achievable propagator and can generate a meaningful bound to compare it to. It should be noted that if one were to want to derive a strict bound in this way, it would be required to find the smallest bound over all possible unitaries that correspond to unity probability in the desired state; hence, our computation of a bound is actually an over-estimate of the ``true'' bound using this approach, which is necessarily smaller (and thereby looser). 

The bound of \citeasnoun{Lee2018} is similarly for a target unitary, but now uses a different expression. Now, the minimum time is given by:
\begin{align}
    T \geq \max_{V\in {\rm Stab}(iH_c)} \frac{\|[U_g,V]\|}{\|[H_0,V]\|}.
    \label{eq:Tmax2}
\end{align}
This bound arises from maximizing the ratio of norms on the right-hand side of \eqref{Tmax2} over all matrices that commute with $H_c$. Generically, there is no simple way to optimize over all such matrices. To compute a reasonable bound, then, we choose $H_c$ for $V$, as it is guaranteed to commute with itself. Then we evaluate the ratio in \eqref{Tmax2}, again using a single unitary as described above (instead of minimizing over all possible unitaries). In both \eqref{Tmax} and \eqref{Tmax2}, the Frobenius norm is the norm that is used.
\bibliography{qc_bib}